\documentclass[%
reprint,
superscriptaddress,
nofootinbib,
amsmath,
amssymb,
prx,
floatfix,
]{revtex4-2}
\usepackage[utf8]{inputenc}
\usepackage{graphicx}
\usepackage{soul}
\usepackage[table,xcdraw]{xcolor}
\usepackage{etoolbox}
\usepackage{amsmath}
\usepackage{enumitem}
\usepackage{makecell}
\usepackage{hyperref}
\usepackage{amssymb}
\usepackage{siunitx} % si stuff
\usepackage{multirow}
\usepackage{subfiles}
\usepackage{bbm}
\usepackage{verbatim}
\usepackage{booktabs}
\usepackage{cancel}
\usepackage{lineno}
\usepackage[caption=false]{subfig}
\usepackage{dcolumn}% Align table columns on decimal point
\usepackage{bm}% bold math
\usepackage{fullpage}
\usepackage{colortbl}

\makeatletter

    \def\CT@@do@color{%
      \global\let\CT@do@color\relax
            \@tempdima\wd\z@
            \advance\@tempdima\@tempdimb
            \advance\@tempdima\@tempdimc
    \advance\@tempdimb\tabcolsep
    \advance\@tempdimc\tabcolsep
    \advance\@tempdima2\tabcolsep
            \kern-\@tempdimb
            \leaders\vrule
                    \hskip\@tempdima\@plus  1fill
            \kern-\@tempdimc
            \hskip-\wd\z@ \@plus -1fill }
    \makeatother
\newcolumntype{?}{!{\vrule width 0.1pt}}
\apptocmd{\sloppy}{\hbadness 10000\relax}{}{}

\def\ket#1{| #1 \rangle}
\def\bra#1{\langle #1 |}
\def\bracket#1#2{\langle #1 | #2 \rangle}
\def\ketbra#1#2{| #1 \rangle\!\langle #2 |}
\def\vev#1{\langle #1 \rangle}

\newcommand{\sixj}[6]{ \begin{Bmatrix}
		#1 & #2 & #3 \\
		#4 & #5 & #6 
\end{Bmatrix}}

\graphicspath{{Figures/}}
\date{\today}

\begin{document}
%\linenumbers
\title{Practical trapped-ion protocols for universal qudit-based quantum computing}
\author{Pei Jiang Low}
\thanks{These authors contributed equally}
\affiliation{Department of Physics and Astronomy, University of Waterloo, Waterloo, N2L 3R1, Canada}
\affiliation{Institute for Quantum Computing, University of Waterloo, Waterloo, N2L 3R1, Canada}
\author{Brendan M. White}
\thanks{These authors contributed equally}
\affiliation{Department of Physics and Astronomy, University of Waterloo, Waterloo, N2L 3R1, Canada}
\affiliation{Institute for Quantum Computing, University of Waterloo, Waterloo, N2L 3R1, Canada}
\author{Andrew A. Cox}
\affiliation{Department of Physics and Astronomy, University of Waterloo, Waterloo, N2L 3R1, Canada}
\author{Matthew L. Day}
\affiliation{Department of Physics and Astronomy, University of Waterloo, Waterloo, N2L 3R1, Canada}
\affiliation{Institute for Quantum Computing, University of Waterloo, Waterloo, N2L 3R1, Canada}
\author{Crystal Senko}
\email{csenko@uwaterloo.ca}
\affiliation{Department of Physics and Astronomy, University of Waterloo, Waterloo, N2L 3R1, Canada}
\affiliation{Institute for Quantum Computing, University of Waterloo, Waterloo, N2L 3R1, Canada}
\begin{abstract}
      The notion of universal quantum computation can be generalized to multi-level qudits, which offer advantages in resource usage and algorithmic efficiencies. Trapped ions, which are pristine and well-controlled quantum systems, offer an ideal platform to develop qudit-based quantum information processing. Previous work has not fully explored the practicality of implementing trapped-ion qudits accounting for known experimental error sources. Here, we describe a universal set of protocols for state preparation, single-qudit gates, a new generalization of the M\o{}lmer-S\o{}rensen gate for two-qudit gates, and a measurement scheme which utilizes shelving to a meta-stable state. We numerically simulate known sources of error from previous trapped ion experiments, and show that there are no fundamental limitations to achieving fidelities above \(99\%\) for three-level qudits encoded in \(^{137}\mathrm{Ba}^+\) ions. Our methods are extensible to higher-dimensional qudits, and our measurement and single-qudit gate protocols can achieve \(99\%\) fidelities for five-level qudits. We identify avenues to further decrease errors in future work. Our results suggest that three-level trapped ion qudits will be a useful technology for quantum information processing.
\end{abstract}
\maketitle

\section{Introduction}
\label{Sec:Intro}
In current approaches to developing quantum computing hardware, each constituent building block -- such as a trapped ion, superconducting resonator, etc. -- is typically used to encode a two-level qubit. However, in contrast to classical computing hardware using binary transistors, it is less obviously an optimal choice to encode only two states within a unit of quantum information. Trapped ions, superconducting transmons, and many other quantum technologies typically feature many possible physical states, and must be artificially restricted to the two states used as a qubit. A natural question is whether we are optimizing the resources extracted from our quantum building blocks by choosing to use only two of these levels \cite{Cerf-et-al-2002, Thew-et-al-2002, Bartlett-Guise-Sanders-2002, Drub-Macchiavello-2002, Kaszlikowski-et-al-2003, O'Leary-Brennen-Bullock-2006, Mohammadi-Niknafs-Eshghi-2011, Luo-Wang-2014}. Experimentalists have developed sufficient control to envision that a quantum processor could benefit from using more of the physical states afforded by the quantum system. Making use of multi-level quantum building blocks, or qudits, presents clear challenges. More controls will be needed to fully exploit the new degrees of freedom, while at the same time, more opportunities arise for errors during a computation. However, there could be substantial benefits if these challenges are overcome.

In this paper, we propose methods to perform quantum information processing with multi-level qudits, using trapped atomic ions as the hardware. Trapped ions are one of the leading qubit technologies owing to their superb coherence and controllability \cite{Haffner-Roos-Blatt-2008, Ozeri-2011, Schindler-et-al-2013, Bruzewics-et-al-2019}, and as such are an attractive hardware choice for developing a qudit-based technology. Within some sections in this paper, we deviate from a generalized discussion to specific implementations using our chosen ion species, \({^{137}\mathrm{Ba}^+}\).

To determine whether qudit-based quantum processors could be more scalable than qubit-based processors, several lines of inquiry are needed. One, it must be determined whether idealized qudits offer advantages over idealized qubits; two, it must be shown that the necessary qudit operations can be practically achieved in experiments; and three, it must be investigated whether the advantages offered by idealized qudits are outweighed by tradeoffs with increased experimental complexity and more potential sources of error. The remainder of this introduction addresses the first question. The main work of this article seeks to answer the second inquiry, on whether qudit-based quantum computation can be practically realized. The third inquiry is left for future works.

Previous efforts have synthesized the motivations for pursuing qudit-based computation and described possible experimental toolkits \cite{Muthukrishnan-Stroud-Jr.-2000,Klimov-et-al-2003,Mischuck-Molmer-2013,Luo-Wang-2014}, while other efforts have implemented limited amounts of control over three-level trapped ion qudits (i.e. qutrits) \cite{Senko-et-al-2014, Randall-et-al-2015, Leupold-et-al-2018}. Platforms other than trapped ions have also been considered as qudits \cite{Neeley-et-al-2009, Strauch-2011, Svetitsky-et-al-2014, Kues-et-al-2017, Wang-et-al-2018, Moro-et-al-2019, Mashhadi-2019, Soltamov-et-al-2019, Sawant-et-al-2020}. In trapped ions, most proposals for two-qudit gates have utilized the Cirac-Zoller \cite{Cirac-Zoller-1995} entangling scheme, which has been found to be less practical than the M\o{}lmer-S\o{}rensen (MS) scheme \cite{Sorensen-Molmer-2000}. The novelty of our work is the development of a new and more practical qudit entangling gate which is a generalization of the MS gate, and the demonstration that high-fidelity qudit operations could be performed with existing technologies, even when accounting for realistic error sources. Scaling the number of ions in trapped ion systems represents a significant challenge to useful quantum computation, and using qudits could help by increasing the information capacity of each ion. The straightforward methods we describe are directly generalized from current ion trap qubit techniques, and so are intuitive to implement for experimentalists already working with ion qubits. 

Current understanding of practical quantum computing rests on several early theoretical discoveries, including the notion of a universal gate set. Namely, there exists a finite set of operations that suffice to implement any ``algorithm'' (i.e. any unitary operation) with arbitrary precision, if these operations can be performed on any single qubit or two-qubit pair within a large enough collection of qubits \cite{Solovay-1995,Kitaev-1997,Nielsen-Chuang-2000}. Furthermore, error mechanisms in physical hardware are pervasive enough to require fault-tolerant error correction protocols, in which logical qubits are encoded using multiple physical qubits. If physical error rates can be made sufficiently small, the structure of the error-correcting code guarantees that errors can be detected and corrected. Both of these notions are extensible to qudits \cite{Gottesman-1998, Grassl-Rotteler-Beth-2003, Brennen-et-al-2005, Brennen-Bullock-O'Leary-2006}. Therefore, by encoding qudits rather than qubits, a larger Hilbert space is accessible with the same physical information carriers, and there is no sacrifice of universality or of the potential for fault tolerant implementations. 

It is not \textit{a priori} clear that the larger Hilbert space accessed by using qudits instead of qubits should translate to a computational advantage. However, some algorithms can be shown to require fewer qudits of higher dimension to achieve comparable results to a qubit-based algorithm, suggesting that there will be computational advantages \cite{Gokhale-et-al-2019}. In particular, the quantum phase estimation algorithm, which forms the basis of Shor’s factoring algorithm \cite{Shor-1995} and of many quantum chemistry calculations \cite{Aspuru-et-al-2005}, is known to benefit from an increase in the dimension of the qudits used. For example, as seen in Table 1 from Reference \cite{Parasa-Perkowski-2011}, making use of 5-level qudits roughly halves the number of atoms required to perform quantum phase estimation with the same precision as compared to qubits. There are also indications that simulations of higher-dimensional quantum systems, such as spins with \({S>1/2}\), will be more efficient when performed on qudit-based processors \cite{Bullock-O'Leary-Brennen-2005}. This suggests that qudits could be useful for understanding questions from fundamental particle physics (where higher dimensions are necessary to simulate color charge) to exotic quantum material properties \cite{Cohen-et-al-2015}.

Qudits may also offer advantages for quantum error correction \cite{Lanyon-et-al-2008, Ralph-Resch-Gilchrist-2007, Campbell-2014, Campbell-Anwar-Browne-2012, Andrist-Wootton-Katzgraber-2015}, which is a more important long-term concern compared to algorithmic advantages. Current estimates for the number of qubits required to perform practical calculations, such as simulating reaction mechanisms for nitrogen fixation, or factoring numbers of the size used in RSA encryption, are in the range of 500-1000 logical qubits \cite{Lekitsch-et-al-2015,Reiher-et-al-2016}. However, millions to billions of physical qubits will be required to perform these calculations. Qudits can ameliorate the difficulty of resource overheads for quantum error correction in several contexts. In existing qubit codes, a frequently used construction called the Toffoli gate requires half as many physical operations when introducing a third level to the qubits \cite{Lanyon-et-al-2008, Ralph-Resch-Gilchrist-2007}. Qudit error correcting codes offer more favorable error thresholds than equivalent qubit codes \cite{Campbell-2014}, as well as improved efficiency in magic-state distillation \cite{Campbell-Anwar-Browne-2012}, which in many cases is the most resource-intensive aspect of error correction \cite{Reiher-et-al-2016}. Work by Andrist et. al. \cite{Andrist-Wootton-Katzgraber-2015} and Campbell et. al. \cite{Campbell-2014,Campbell-Anwar-Browne-2012} indicates that the error thresholds to successfully implement error correction increase with the number of qudit levels, indicating that fault-tolerant quantum computing for qudits may be able to sustain a higher error rate. For surface code quantum computing, this could mean that the fidelity needed for fault tolerance is lower than the \({99.25\%}\) threshold \cite{Fowler-et-al-2012} given for qubits. 

There is one main class of qudits that we are interested in: small prime-dimensional qudits (\({d=3,5,7}\)).
These are interesting because a set of single-qudit Pauli group gates and
the generalized \({\pi/8}\) gate, along with a two-qudit gate, form a universal gate set \cite{Howard-Vala-2012, Gottesman-1998} for the prime-dimensional Hilbert space. Furthermore, many qudit error correction codes are based on prime dimensions \cite{Gottesman-1998,Campbell-2014,Campbell-Anwar-Browne-2012,Aharanov-Ben-Or-2018}. In this paper, up to 7-level qudits are studied for measurement, and up to 5-level qudits are studied for single-qudit and two-qudit gates. In our analysis, error sources pertaining to hardware control imperfections are excluded as they can arguably be improved as technology advances, thus not posing a fundamental limit to the fidelities of qudit operations. Environmental noises are, however, considered difficult to be removed, and are thus included in our assessment.

\section{Qudit Requirements and Encoding}
\label{Sec:Desiderata}
There are many requirements to perform quantum computation with qudits. In this section, we first describe these requirements. Then, we describe our rationale for using \(^{137}\mathrm{Ba}^+\) for our numeric calculations. Finally, we discuss the coherence limitations for the different available encoding schemes within this platform, and describe our specific chosen encoding scheme.
\subsection{Qudit Requirements}

%We choose to encode as in (b), picking levels with lowest \(m\) quantum numbers first. This encoding gives qudit state transitions which are the most well resolved from each other. 
\begin{figure*}[!htb]%
    \subfloat[\hspace{-1.1cm}\label{sfig:testa}]{\includegraphics[height=120pt]{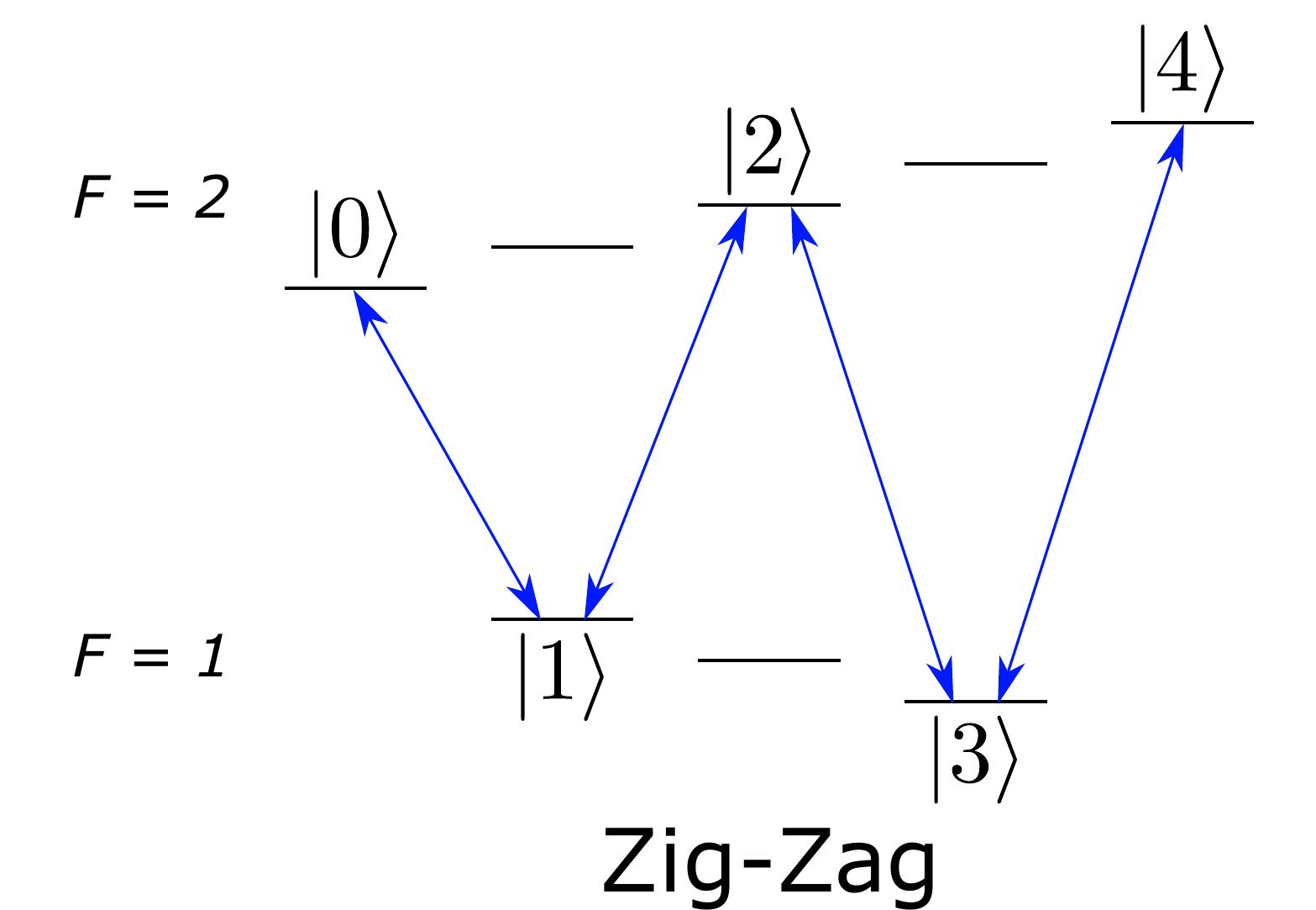}}\hfill%
    \subfloat[\hspace{0cm}\label{sfig:testa2}]{\includegraphics[height=120pt]{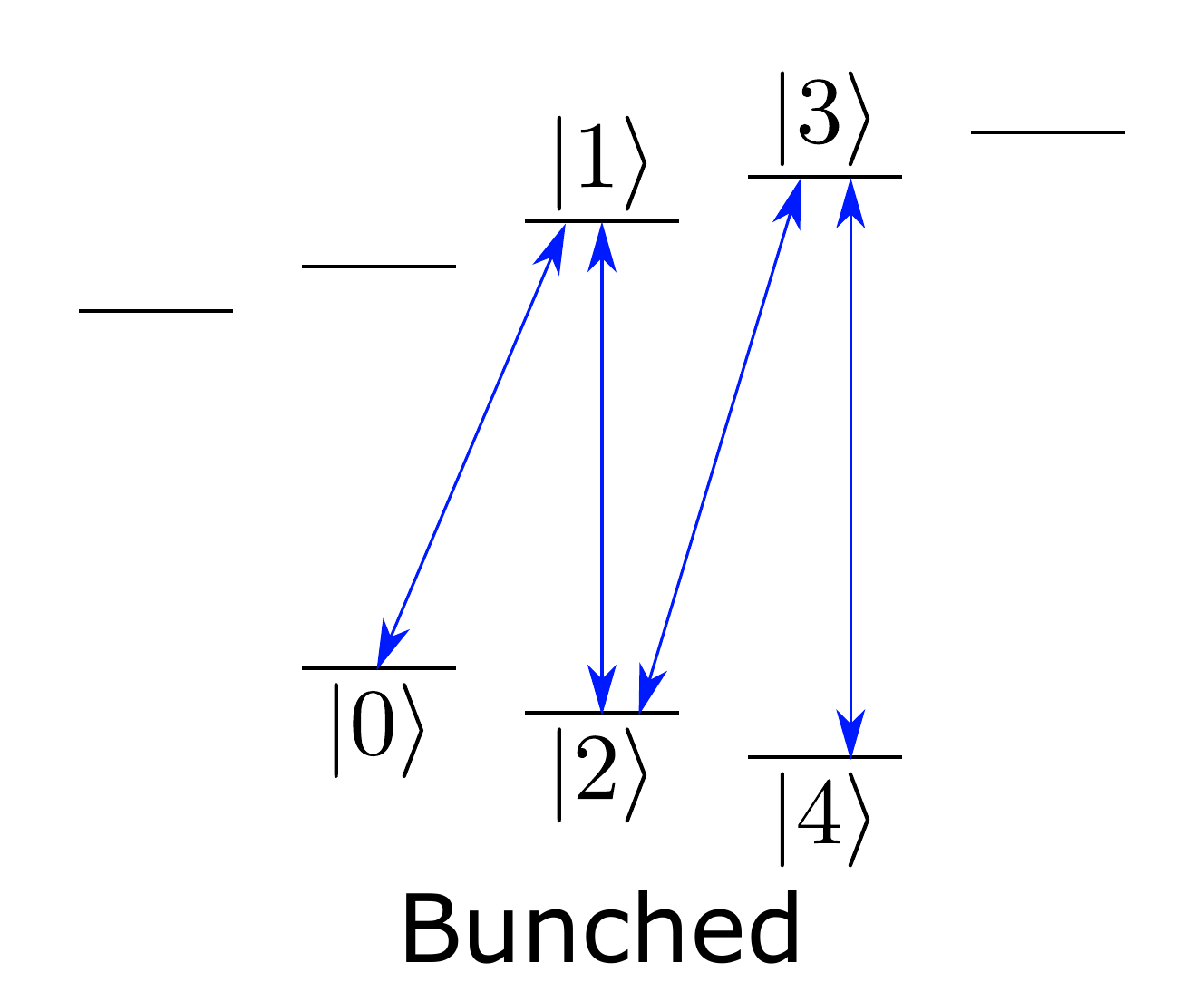}}\hfill%
    \subfloat[\label{sfig:testa3}]{\includegraphics[height=120pt]{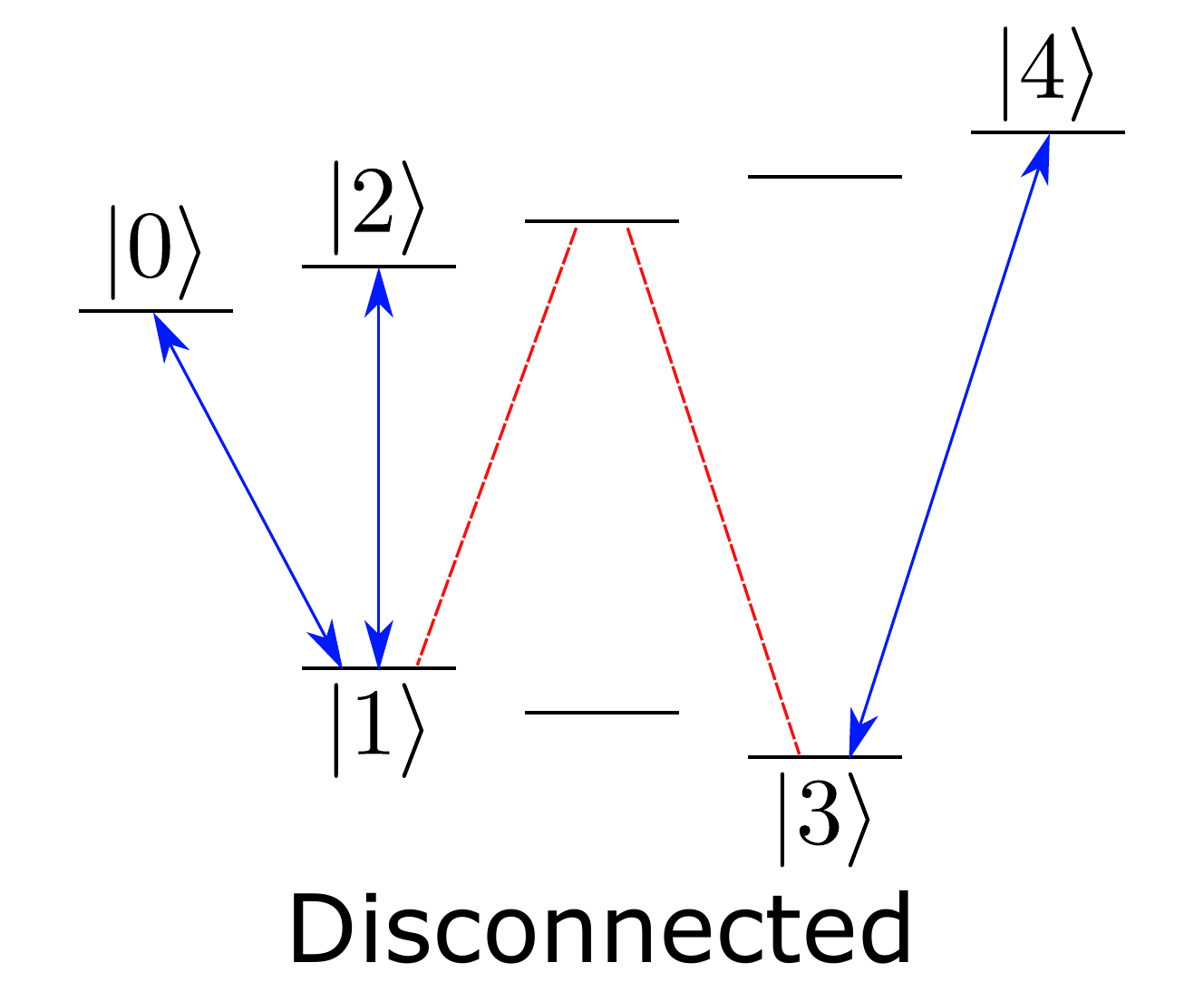}}\hfill
    \caption{Examples of qudit encodings. (a) We prefer a zigzag configuration because it simplifies laser manipulations. (b) A bunched configuration minimizes decoherence due to magnetic field fluctuations but requires more polarization control of lasers. (c) Disconnected configurations are not preferred due to the experimental complexity of transferring population among all possible states.}
    \label{fig:Qudit-Encoding}
\end{figure*}

The requirements to consider for qudit-based information processing are as follows:
\begin{enumerate}
    \item Ability to encode multiple basis states.
    \item Stability of basis states against decoherence processes.
    \item Ability to prepare a fiducial initial state.
    \item A method to reliably measure in the physical basis.
    \item Ability to perform arbitrary single-qudit gates.
    \item Ability to apply an entangling operation between qudit pairs.
\end{enumerate}
Each of these desiderata may be accomplished in an ion trap context, using straightforward generalizations of existing techniques. For specificity, we focus on the case of encoding qudit states in hyperfine sublevels of a ground \({S_{1/2}}\) electronic manifold of a hydrogenic atomic ion. A weak magnetic field is used to lift the degeneracy, and standard optical pumping techniques are used to prepare a fiducial initial state \cite{Dietrich-2010}.
\subsection{Ions}
\setlength\tabcolsep{5pt}
\begin{table*}[!htb]
\centering
\begin{tabular}{cccccccc}
\toprule\toprule
\multicolumn{1}{c|}{\textbf{\begin{tabular}[c]{@{}c@{}}Ion \\ Species\end{tabular}}}
& \multicolumn{1}{c|}{\textbf{\begin{tabular}[c]{@{}c@{}}Nuclear \\ spin,   \(I\)\end{tabular}}} 
& \multicolumn{1}{c|}{\textbf{\begin{tabular}[c]{@{}c@{}}States \\ in \(^2S_{1/2}\)\end{tabular}}}
& \multicolumn{1}{c|}{\textbf{\begin{tabular}[c]{@{}c@{}}Metastable \\ Lifetime \cite{Werth-1989,Roberts-et-al-2000} \end{tabular}}}
& \multicolumn{1}{c|}{\textbf{\begin{tabular}[c]{@{}c@{}}Primary \\ Transition\end{tabular}}}
& \multicolumn{1}{c|}{\textbf{\begin{tabular}[c]{@{}c@{}}Metastable \\
Transition\end{tabular}}} 
& \multicolumn{1}{c|}{\textbf{\begin{tabular}[c]{@{}c@{}}3-level \\
\(\tau\)\end{tabular}}}
& \multicolumn{1}{c}{\textbf{\begin{tabular}[c]{@{}c@{}}5-level \\
\(\tau\)\end{tabular}}}\\  \midrule
\multicolumn{8}{c}{\(\mathbf{D_{5/2}}\) \textbf{Metastable State}} \\  \midrule
\(^{43}\mathrm{Ca}^+\) & 7/2 & 16  & \(\SI{1}{s}\)  & \SI{397}{nm} & \SI{729}{nm} & \SI{8.4}{s} & \SI{4.2}{s}\\
\(^{87}\mathrm{Sr}^+\) & 9/2 & 20 &  \(\SI{0.345}{s}\) & \SI{422}{nm} & \SI{674}{nm} & \SI{10.5}{s} & \SI{5.3}{s}\\
\(^{133}\mathrm{Ba}^+\) & 1/2 & 4 &  \(\SI{35}{s}\) & \SI{493}{nm} & \SI{1762}{nm} & \SI{2.1}{s} & n/a \\
\(^{137}\mathrm{Ba}^+\) & 3/2 & 8 &  \(\SI{35}{s}\) & \SI{493}{nm} & \SI{1762}{nm} & \SI{4.2}{s} & \SI{2.1}{s}\\ \midrule
\multicolumn{8}{c}{\(\mathbf{F_{7/2}}\) \textbf{Metastable State}}\\ \midrule
\(^{171}\mathrm{Yb}^+\) & 1/2 & 4 &  \(\SI{5.4}{y}\) & \SI{369.5}{nm} & \SI{467}{nm} & \SI{2.1}{s} & \SI{1.1}{s}\\
\(^{173}\mathrm{Yb}^+\) & 5/2 & 12 &  \(\sim\SI{10}{d}\) & \SI{369.5}{nm} & \SI{467}{nm} & \SI{6.3}{s} & \SI{3.2}{s}\\ \bottomrule\bottomrule
\end{tabular}
\caption{A list of different qudit candidates and their relevant properties. In Yb, the octupole transition to the \({F_{7/2}}\) state would be required because the lifetime of the \({D_{5/2}}\) state is only \(\SI{7}{ms}\), which would substantially limit the attainable readout fidelity.
Next, we list the relevant transitions in each ion species: the primary transition is used for Doppler cooling, optical pumping, and fluorescence measurement, and the metastable transition allows us to measure qudits using the shelving technique described in Section \ref{Sec:Measurement}.
The last two columns are the dephasing/coherence times for a \({3}\)- and \({5}\)-level qudit with \({\SI{2.7}{pT}}\) magnetic field fluctuations, calculated from Equation \ref{Eq:DephasingTime}; we assume a zig-zag encoding centered at \(m_F = 0\).}
\label{tab:Qudit-Ions1}
\end{table*}
Table \ref{tab:Qudit-Ions1} compares many possible ion species options for encoding qudits. Atomic structure data for the selected ion species are presented, while the final columns show the coherence decay time in 3- and 5-level qudits from magnetic field fluctuations. Having a metastable state is an important requirement in order to implement the shelving scheme described in Section \ref{Sec:Measurement}. The longer the lifetime of this metastable state, the more fiducious the shelving procedure becomes. We chose \({^{137}\mathrm{Ba}^+}\) to encode our qudits because it features the longest \({D_{5/2}}\) lifetime, and does not require an octupole transition for the shelving operation (an octupole transition requires more laser power than a quadrupole transition to a \({D_{5/2}}\) state). With this species, there are enough hyperfine ground states to implement up to 8-level qudits. For the calculations in this paper, a quantization field of \({\SI{470}{\mu T}}\) is selected. Furthermore, as can be seen in Figure \ref{fig:Ba137} of the appendix, most of the lasers required are in the visible range, simplifying the optical technology required to build an experiment \cite{Dietrich-et-al-2009, Dietrich-2010}.

\subsection{Encoding and coherence}
Depending on the qudit dimension and the hyperfine structure of the candidate ion, there may be multiple ways to encode a qudit. We consider only encodings where the basis states form the nodes of a connected graph whose edges represent frequency-resolved transitions allowed under magnetic dipole selection rules (see Figure \ref{fig:Qudit-Encoding}). The encodings which satisfy these requirements are the zig-zag and the bunched encodings in Figure \ref{fig:Qudit-Encoding}.

Qudits will experience first-order sensitivity to magnetic field noise, which is the most common source of dephasing in ion trap experiments. Two-level qubit states can be chosen to share the same magnetic field sensitivity, but that solution does not generalize to more than two states. Technological solutions have been found to stabilize magnetic fields to the order of \(\SI{1}{pT}\) by utilizing magnetic shielding and applying quantization fields with permanent magnet arrays, resulting in coherence times of order one second for magnetic-field-sensitive qubits \cite{Ruster-et-al-2016,Hakelberg-et-al-2018}. 

The error from magnetic field noise of a qudit in the arbitrary initial state
\begin{equation}
    \ket{\psi_0} = \sum_l a_l \ket{l}
\end{equation}
can be obtained following a similar derivation as outlined in \cite{Monz-2011}. We assume that the deviation of the magnetic field from the ideal field, \({\Delta B(t)}\), is a stationary Gaussian process; for the case \( t \ll 1/\gamma \), the fidelity as a function of time is
\begin{align}
    \label{Eq:Decoherence-Fidelity}
    \mathcal{F}&(t) = \sum_l \lvert a_l \rvert^4 +
    \nonumber\\
    &\sum_l \sum_{l'>l} 2 \lvert a_l \rvert^2 \lvert a_{l'} \rvert^2 e^{-\frac{\mu_B^2}{2 \hbar^2} \left( g_{Fl}m_{Fl} - g_{Fl'}m_{Fl'} \right)^2 \langle \Delta B^2 \rangle t^2 },
\end{align}
where \(t_c = 1/\gamma\) is the magnetic field noise correlation decay time, \({a_l}\) is the wavefunction amplitude in the state \( \ket{l} \), \({g_{Fl}}\) and \({g_{Fl'}}\) are the two levels’ respective hyperfine g-factors, \(m_{Fl}\) are the magnetic quantum numbers of the hyperfine sublevels, \( \hbar = h/2\pi \) is the reduced Planck constant, and \({\mu_B \approx h\times\SI{14}{GHz/T}}\) is the Bohr magneton. When making use of magnetic sub-levels in the ground state, the g-factor is \({g_F = \pm\frac{1}{(I + 1/2)}}\) for \({F = I \pm 1/2}\). The result is a series of terms with different dephasing times 
\begin{equation}
    \tau_{l,l'} = \frac{\hbar}{\lvert g_{Fl}m_{Fl} - g_{Fl'}m_{Fl'} \rvert \mu_B \sqrt{\langle \Delta B^2 \rangle}},
    \label{Eq:DephasingTime2}
\end{equation}
where we have defined dephasing time as the time taken for an off-diagonal element in the density matrix of a state to decay by a factor of $e^{-1/2}$. To obtain a single parameter characterizing the qudit decoherence, we choose the pair of qudit states with the largest value of \({\tau}\), which corresponds to the shortest dephasing time in the series of terms:
\begin{equation}
    \label{Eq:DephasingTime}
    \tau = \frac{\hbar}{\mu\sqrt{\vev{\Delta B(t)^2}}}.
\end{equation}
For example, for the zigzag encoding in Figure \ref{fig:Qudit-Encoding}, the estimate for the coherence time is the relative dephasing of the \({\ket{0}\equiv \ket{F=2,m_F=-2}}\) and \({\ket{4}\equiv \ket{F=2,m_F=2}}\) levels.
In this expression, if the two most sensitive levels of the qudit are \({\ket{F_1,m_{F1}}}\) and \({\ket{F_2,m_{F2}}}\), we obtain \({\mu=\mu_B(g_{F1}m_{F1}-g_{F2}m_{F2})}\). Using the estimate of an achievable magnetic field noise from \cite{Ruster-et-al-2016}, where \(\sqrt{\vev{\Delta B(t)^2}}\approx\SI{2.7}{pT}\), we may thus calculate a lower bound dephasing rate for any qudit encoding.

Coherence times will be maximized when the relative sensitivity of the entire set of states is minimized, as in the bunched configuration shown in Figure \ref{fig:Qudit-Encoding}. Using this encoding scheme, the largest relative sensitivity for two states within a d-level qudit is \({\mu = \mathrm{max}(\frac{d-1}{2})\frac{\mu_B}{I+1/2}}\), where \({\mathrm{max}(x)}\) denotes the largest integer smaller than or equal to \({x}\).

In practice, encoding a qudit in the least sensitive levels is difficult to implement because of the great deal of polarization control necessary. We choose instead to use the zigzag encoding exemplified in Figure \ref{fig:Qudit-Encoding}(a), where each consecutive pair of states obey \({ \Delta F=1\text{, }\Delta m_F =1}\), for the \({d=3}\) and \({d=5}\) qudits. This encoding can be manipulated with Raman transitions that have straightforward laser polarization requirements, as described in Section \ref{Sec:Single-Qudit}. In this case, the largest relative sensitivity for two states within the qudit is \({\mu = \frac{d\mu_B}{I + 1/2}}\). For comparison, the relative sensitivity of a Zeeman qubit used in reference \cite{Ruster-et-al-2016}, which uses the two states of a single electron spin, is \({2\mu_B}\). This means that for any of the ions listed in Table \ref{tab:Qudit-Ions1}, the coherence time \({\tau}\) we estimate is greater than or equal to that in Reference \cite{Ruster-et-al-2016}, depending on how many levels we are using; if we use all states, the lifetimes are equal. As pointed out in reference \cite{Brown-Brown-2018}, this coherence time is already long enough to
envision implementing error correcting codes with existing techniques.

\section{Qudit Measurements}
\label{Sec:Measurement}
The typical method for measuring qubits must be modified for qudits in order to account for the higher number of states encoded. This section describes a protocol for measuring all of these encoded states. An analytical description for the error of this measurement is developed and realistic error estimations are presented using this model.

State measurement for trapped ion qubits is accomplished by exposing the ion to laser radiation, configured so that only one of the qubit states fluoresces, and the fluorescence is collected on a detector such as a charge-coupled device or photomultiplier tube. In generalizing the fluorescence technique to multiple levels, one must produce a signal that differs for each physical basis state. A straightforward way to accomplish this goal is to sequentially check each basis state separately:
\begin{enumerate}
    \item Engineer a situation where only one of the basis states (e.g. \({\ket{0}}\)) produces fluorescence when exposed to laser radiation
    \item If no fluorescence is detected, engineer a situation where another state (e.g. \({\ket{1}}\)) fluoresces
    \item Etc,
\end{enumerate}
repeating this process until the presence of fluorescence has indicated which of the basis states is occupied. The criterion that only a single basis state respond to the detection lasers at any given step is crucial. If two or more states are induced to fluoresce simultaneously, then the information about which state was occupied will be lost.

\begin{figure}
    \centering
    \includegraphics[width=\linewidth]{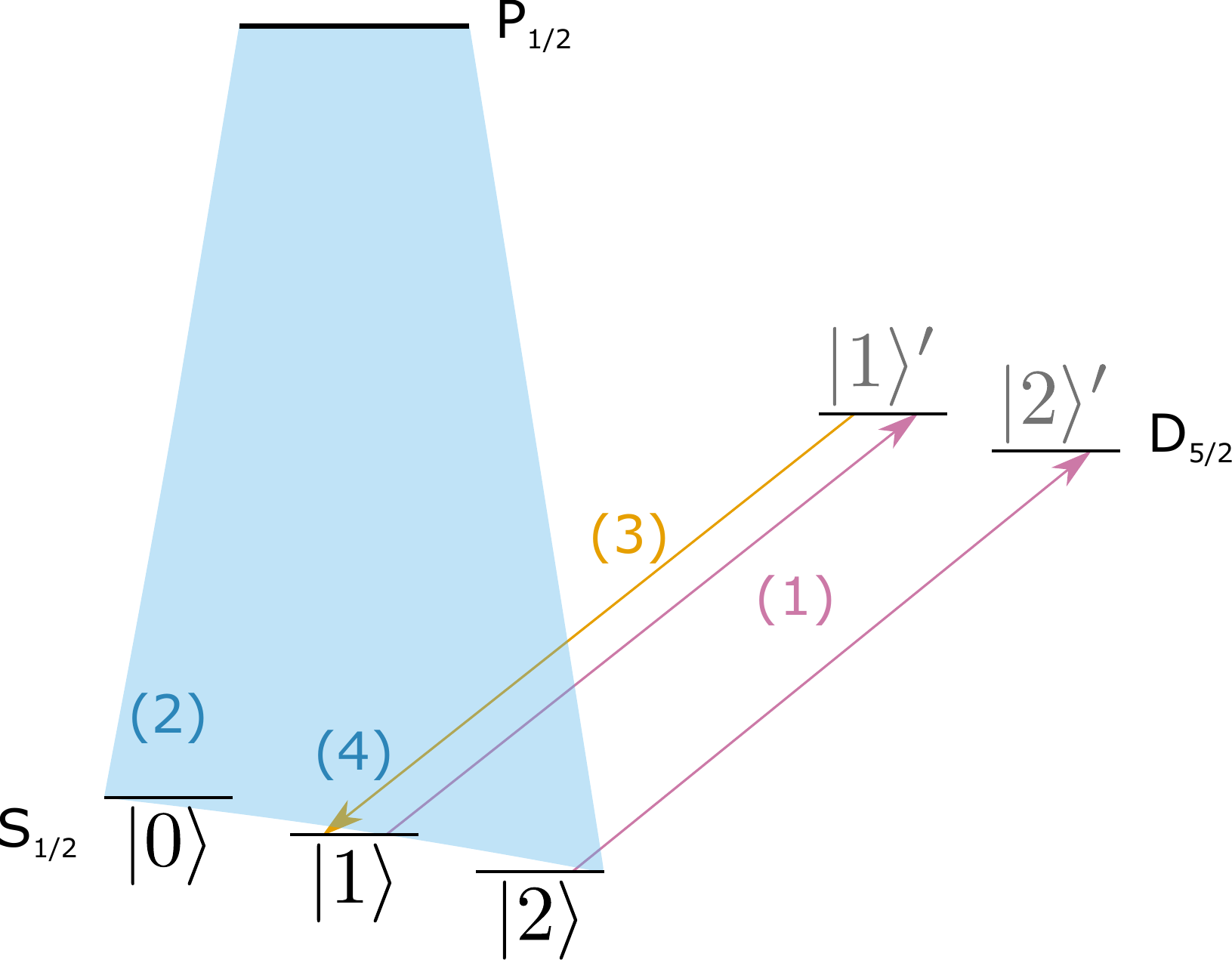}
    \caption{The shelving procedure for a 3-level qudit measurement. (1): Map states \({\ket{1}, \ket{2}}\) to the metastable state. (2): Fluoresce on the cycling transition \({S_{1/2} \leftrightarrow P_{1/2}}\). (3): If no is fluorescence detected, return one state from the metastable state to the ground state and (4): measure it with fluorescence.}
\label{fig:ShelvingProcedure}
\end{figure}

Many ions used for QIP feature metastable states, which can be exploited for state readout. The metastable state chosen should not be part of the closed-cycle transition used for Doppler cooling. The ``shelving” approach to measuring a qudit encoded in such an ion is illustrated in Figure \ref{fig:ShelvingProcedure} for 3-levels. It consists of shelving all but one state in the metastable state, measuring the remaining state, then repeatedly de-shelving and measuring states until the overall state of the qudit is known.
This approach assumes that the transitions between each qudit state and its corresponding metastable shelf state are resolved in frequency, so that each state can be checked independently during the fluorescence step. 

Here, we estimate fundamental limitations on the measurement speed and fidelity for our chosen ion, \({^{137}\mathrm{Ba}^+}\). The finite lifetime of the metastable state \({D_{5/2}}\) imposes a fundamental limitation on the measurement fidelity as a function of the measurement duration. In this article, we assess the feasibility of implementing rapid adiabatic passage with rectangular driving pulses for qudit measurements. There are more advanced schemes for rapid adiabatic passage that reduces error by careful pulse shaping of the driving field \cite{Wunderlich-et-al-2007}, but a simpler scheme is sufficient for evaluating the practicality of our qudit measurement protocol. A similar scheme was used in Reference \cite{Noel-et-al-2012} for \({^{138}\mathrm{Ba}^+}\) shelving, and in Reference \cite{Christensen-et-al-2019} for readout of \({^{133}\mathrm{Ba}^+}\) ions.

For a 2-level system, if one couples a field detuned from the transition by \({\Delta(t) = \omega(t) - \omega_0}\) (where \({\omega(t)}\) is the laser frequency and \({\omega_0}\) is the transition frequency) and uses the rotating wave approximation (RWA), the Hamiltonian in the rotating frame of the laser frequency is found to be
\begin{equation}
    \mathcal{H} = \begin{bmatrix}0 & \Omega/2 \\ \Omega/2 & -\Delta(t)\end{bmatrix},
\end{equation}
where \({\Omega}\) is the resulting Rabi frequency of the transition. The eigenstates of the system are called the adiabatic or dressed states. The important result is that if one sweeps the frequency of the field adiabatically from some detuning \({\Delta_0}\) across resonance, stopping at \(-\Delta_0\), then the system will remain in the adiabatic state it was initialized in. While the adiabatic state does not change, the composition of this adiabatic state in terms of the diabatic (undressed) states changes so that one has near perfect fidelity population transfer between the two levels. 

There are several sources for error during population transfer using adiabatic passage. First, there are errors inherent in the way the adiabatic transfer is performed:
\begin{equation}
    \mathcal{E}_{prep}(\Omega, \Delta_0) = \sin^2\theta_0,
\end{equation}
\begin{equation}
    \mathcal{E}_{LZ}(\Omega, \dot\Delta) = e^{-\pi^2\Omega^2/|\dot\Delta|},
\end{equation}
where \({\tan{2\theta_0} = \frac{\Omega}{\Delta_0}}\) and \({\alpha = \dot{\Delta}}\) is the linear sweep rate of the laser frequency. \(\mathcal{E}_{prep}\) is the error from imperfect adiabatic state preparation - ideally, for rectangular pulses, one would start and end the sweep at detuning \({\lvert\Delta_0\rvert = \infty}\) for the adiabatic states to correspond exactly to a diabatic state. This is not reasonable to do, so one must instead start at some finite detuning, resulting in this preparation error. \(\mathcal{E}_{LZ}\) is the Landau-Zener probability \cite{Zener-1932,Landau-1932}, accounting for how adiabatic the transfer is.

Considering the atomic energy structure of the shelving transition \(^{137}\mathrm{Ba}^+\), there are the additional error sources of off-resonant coupling to other transitions, and decay from the finite lifetime of the shelving state:
\begin{equation}
    \mathcal{E}_{OR}(\Omega, \Delta_i, C_i) = C_i^2\frac{\Omega^2}{2\tilde\Omega_i'^2},
\end{equation}
\begin{equation}
    \mathcal{E}_{dec}(t) = 1 - e^{-t/t_{dec}}
\end{equation}
where \(i\) is an unwanted transition outside of the laser frequency sweep, \(\Delta_i = \textrm{min}\lvert \Delta(t) - \Delta_{i0}\rvert\) is the minimum detuning of this unwanted transition from the laser frequency, \(\Delta_{i0}\) is the detuning of the unwanted transition from the desired transition, \({\tilde{\Omega}_i' = \sqrt{\Delta_i^2 + \Omega^2}}\) is the effective Rabi frequency of the laser coupling to this unwanted transition, and \(C_i\) is the overall strength of the \( i^{\mathrm{th}} \) transition compared to the desired transition. In the decay error \(\mathcal{E}_{dec}\), \(t_{dec}\) is decay time of the shelving state (\({t_{dec}\sim \SI{30}{s}}\) for \({\mathrm{Ba}^+}\)); this is an overestimation of the error as it assumes that the qudit was in the first state we shelved and that we leave it shelved the longest.

Another consideration is the dephasing error from the finite linewidth of the shelving laser \cite{Lacour-et-al-2007}. This error comes in as an exponential of the form \(e^{-2\pi^2\Delta\nu\Omega /\alpha}\), where \(\Delta\nu = \Gamma/2\pi\) is the FWHM laser linewidth. Putting all of these errors together, we have the fidelity of a single adiabatic transfer given by 
\begin{align}
    \label{Eq:Measurement-Fidelity1}
    &\mathcal{F}_{trans}(\Omega, \Delta(t), \{i\}, \alpha, t) = \left(1 - \mathcal{E}_{prep}(\Omega, \Delta_0)\right)^2
    \nonumber\\
    &\times\left(1/2 + e^{-2\pi^2\Delta\nu\Omega /\alpha}\big(\mathcal{E}_{LZ}(\Omega, \alpha) - 1/2\big)\right)\mathcal{E}_{dec}(t).
    \nonumber\\
    &\quad\quad\quad\quad\times\prod_{\{i\}}\big(1- \mathcal{E}_{OR}(\Omega, \Delta_i, C_i)\big) 
\end{align}
Here, we square the preparation factor, since this error occurs at the beginning and at the end of a transfer. The off-resonant error is a product over the set \(\{i\}\) of all off-resonant transitions.

Another error source comes from additional motional sidebands offset from all of the carrier frequencies by the secular trap frequency. In a standard ``blade" style trap, one would expect this frequency to be \({\sim\SI{2}{MHz}}\). In this case, for some shelving transitions, some motional sidebands are within the laser sweep range of a transition we wish to drive. This results in coherent adiabatic passage on these transitions. However, because the Lamb-Dicke parameter for this transition is \({\eta = 0.0243}\), these couplings are weakened: their strength scales as \(\eta^N\) compared to the carrier, where \(N\) is the order of the motional sideband. The probability of driving such a motional transition is given by
\begin{align}
    \label{Eq:Measurement-FidelityMot}
    \mathcal{F}_{mot}&(\Omega, \Delta(t), \alpha, t, \Delta_{j0}, N_j) = \nonumber\\
    &\mathcal{F}_{trans}(C_j\Omega\eta^{N_j}, \mathrm{max}\lvert \Delta(t) - \Delta_{j0}\rvert, \{\}, \alpha, t),
\end{align}
where \(C_j\) is the relative strength of this transition compared with the desired transition, and \(\Delta_{j0}\) is the detuning of the unwanted motional sideband transition from the desired transition. Here, we do not consider off-resonant coupling and leave the set \(\{i\}\) empty. The overall fidelity to successfully shelve a state is now given by
\begin{align}
    \label{Eq:Measurement-Fidelity}
    \mathcal{F}_{shelve}&(\Omega, \Delta(t), \{i\}, \alpha, t, j) = \nonumber\\
    &\mathcal{F}_{trans}(\Omega, \Delta(t), \{i\}, \alpha, t)
    \nonumber\\
    &\times\prod_{\{j\}}\left(1 - \mathcal{F}_{mot}(\Omega, \Delta(t), \alpha, t, \Delta_{j0}, N_j)\right)
\end{align}
where \(\{j\}\) is the set of all motional sidebands driven through during this transfer. Note that for most transfers, \(\{j\} = \{\}\).
\begin{figure*}[t!]%
    \centering
    \subfloat[\label{sfig:testaa1}]{\includegraphics[width=0.49\linewidth]{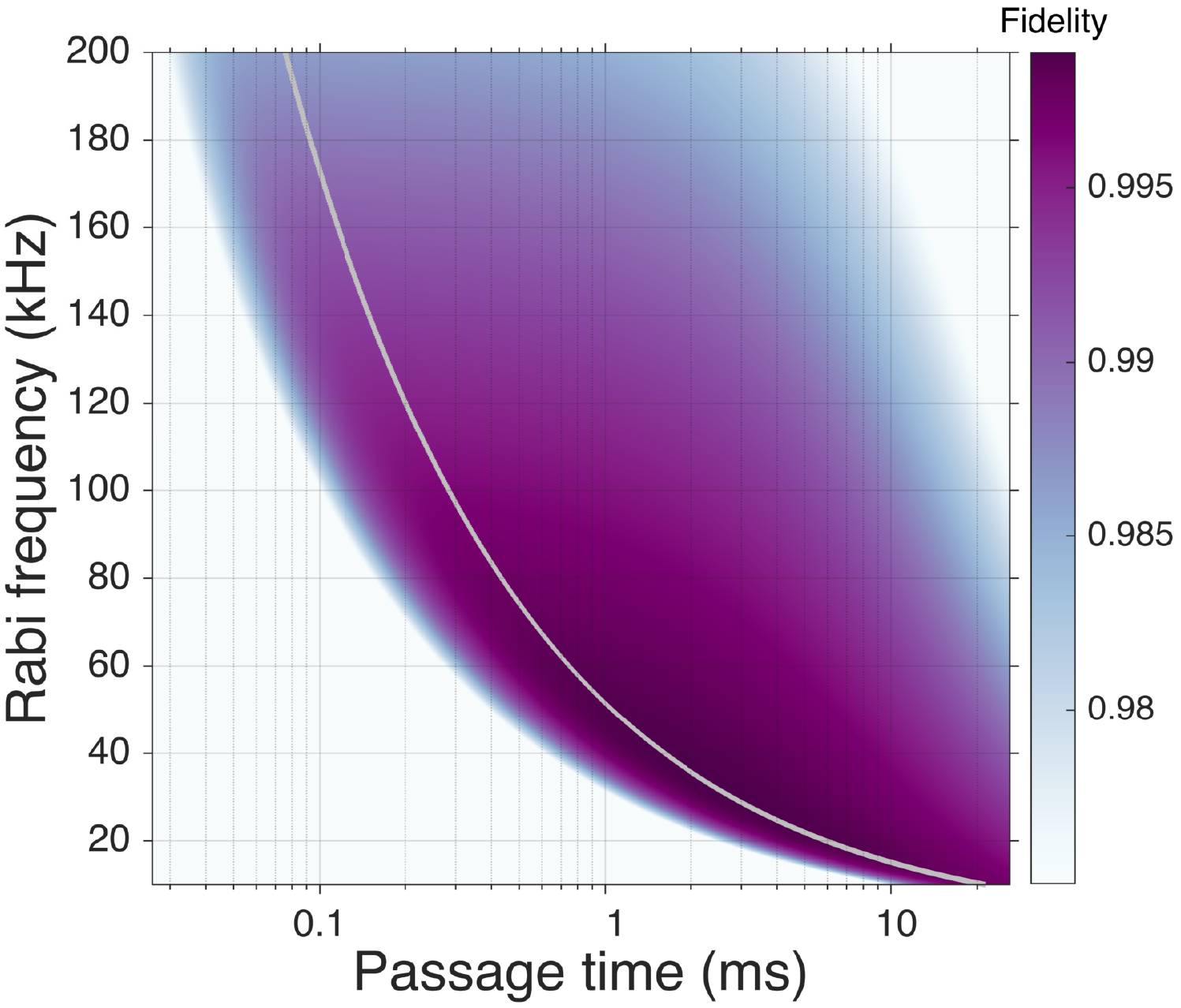}}\hfill%
    \subfloat[\label{sfig:testaa2}]{\includegraphics[width=0.49\linewidth]{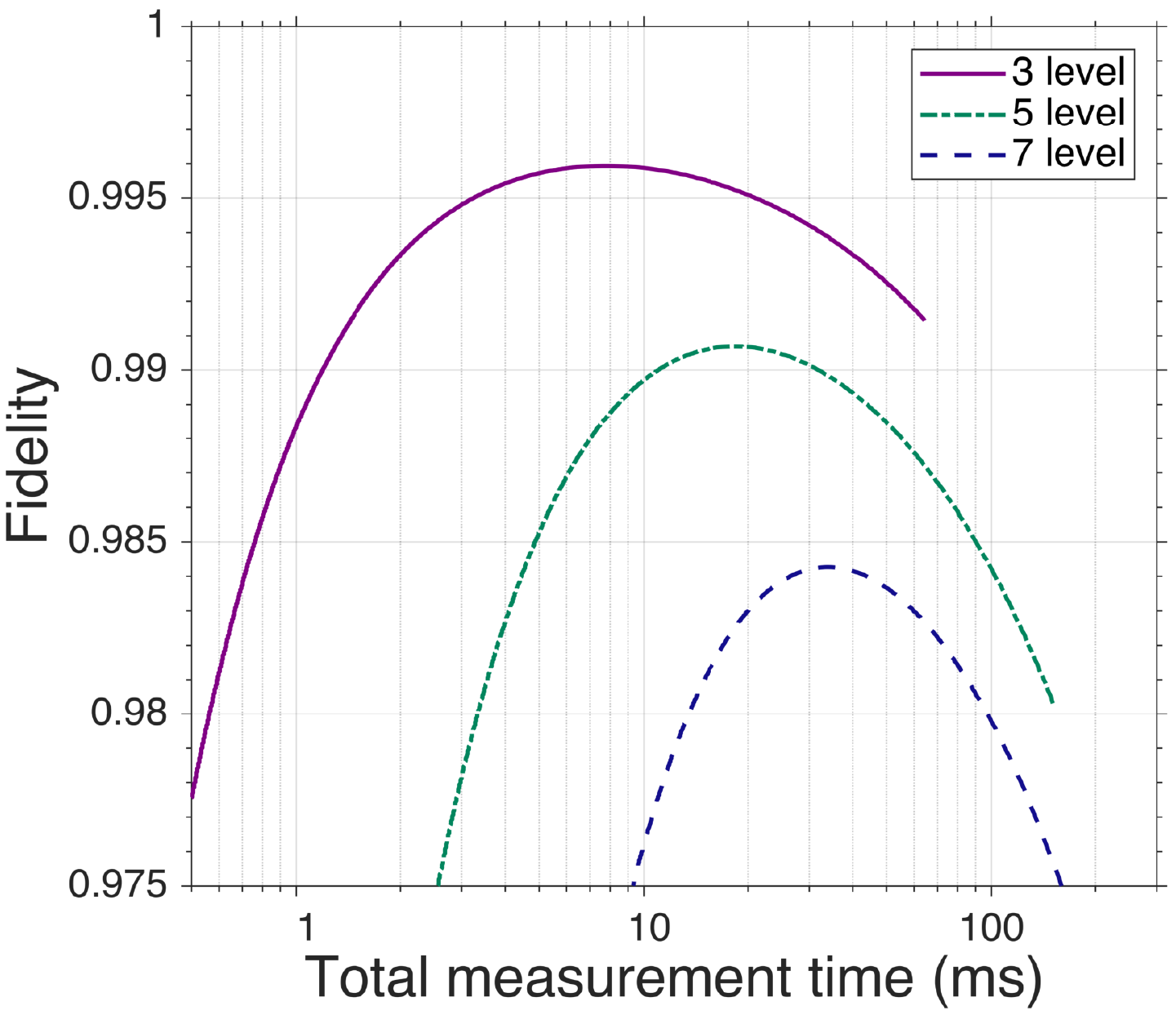}}\hfill%
    \caption{(a) The fidelity of population transfer (Equation \ref{Eq:Measurement-Fidelity}) plotted for various applied Rabi frequencies and passage times \({t = \frac{2\Delta}{\alpha}}\). The gray curve follows the optimal parameters. (b) The measurement time and fidelity for different prime-dimensional qudits. We use the optimal Rabi frequency for each measurement time. Fluorescence time is included in the measurement duration, and we assume that the amount of adiabatic passages needed is \({2d-3}\) (the maximum amount of transfers needed for an arbitrary measurement).}
    \label{fig:Adiabatic-Passage}
\end{figure*}
\setlength\tabcolsep{5pt}
\begin{table*}[htb!]
    \centering
    \begin{tabular}{ccccc}
        \toprule\toprule
        \textbf{Error Source} & \textbf{3-level} & \textbf{5-level} & \textbf{7-level}
        \\\midrule
        Preparation (\(\mathcal{E}_{prep}\))& \({9.3\times 10^{-4}}\) & \({2.2\times 10^{-3}}\) & \({3.4\times 10^{-3}}\)
        \\
        Off-resonant coupling (\(\mathcal{E}_{OR}\))& \({1.6\times 10^{-4}}\)& \({3.7\times 10^{-4}}\) & \({5.9\times 10^{-4}}\)
        \\
        Adiabadicity (\(\mathcal{E}_{LZ}\))& \({2.0\times 10^{-4}}\) & \({4.7\times 10^{-4}}\) & \({7.4\times 10^{-4}}\)
        \\
        Dephasing (linewidth) & \({1.8\times 10^{-3}}\) & \({4.2\times 10^{-3}}\) & \({6.5\times 10^{-3}}\)
        \\
        State decay (\(\mathcal{E}_{dec}\))& \({2.4\times 10^{-4}}\) & \({5.7\times 10^{-4}}\) & \({8.9\times 10^{-4}}\)
        \\
        Motional sideband driving & n/a & n/a & \({1.2\times 10^{-3}}\)
        \\
        Fluorescence (\(\mathcal{E}_{phot}\))& \({5.8\times 10^{-4}}\) & \({1.2\times 10^{-3}}\) & \({1.7\times 10^{-3}}\)
        \\
        \bottomrule\bottomrule
    \end{tabular}
    \caption{Error budget for the measurement sequences. Details for the 5- and 7-level sequences can be found in Appendix \ref{app:Measurement}. Coherent sideband driving error is where a motional sideband transition frequency lies within an adiabatic frequency sweep, given in Equation \ref{Eq:Measurement-FidelityMot}.}
    \label{tab:MeasurementError}
\end{table*}

With laser stabilization, one can achieve a laser linewidth of less than $\SI{2}{Hz}$ \cite{SLS}; we assume a linewidth of exactly $\SI{2}{Hz}$ for the following calculations. We choose a quantization magnetic field of \({\SI{470}{\mu T}}\). In \({^{137}\mathrm{Ba}^+}\), the \({F=3,4}\) levels are separated by less than \({\SI{1}{MHz}}\), so we choose to ignore them and shelve into the \({F=1,2}\) levels, which are separated by \({\sim \SI{70}{MHz}}\). For simplicity, we choose the shelving transitions to drive \({\ket{S_{1/2};F,m_F}}\) states to \({\ket{D_{5/2};F' = F, m_F' = m_F}}\) states in the shelving manifold. By orienting the \({\SI{1762}{nm}}\) laser wavevector and polarization in a useful geometry \cite{Roos-2000, James-1998}, \({q=\pm 1}\) transitions (\({q = -(m_F - m_F')}\)) are completely suppressed, and the strengths of \({q=\pm 2}\) transitions are reduced. 

With these properties, we calculate the fidelity of population transfer for different Rabi frequencies and overall passage times for the 3-level qudit \({\ket{F=2, m_F=0}\leftrightarrow \ket{F'=2,m_F'=0}}\) shelving transition using Equation \ref{Eq:Measurement-Fidelity} (see Figure \ref{fig:Adiabatic-Passage}(a)). We assume that the initial detuning \({\Delta_0 = \SI{1.3}{MHz}}\), which is \({\SI{500}{kHz}}\) detuned from the carrier's nearest motional frequency: the tilt mode at \({\SI{1.8}{MHz}}\). If we pick optimal parameters (along the gray line in Figure \ref{fig:Adiabatic-Passage}), we can get better than ${99.8\%}$ fidelity for individual transfers.

For the overall shelving procedure, there are important errors to avoid, and we have several tricks that allow us to do so. The desired \({q=0}\) transitions are, in the smallest case, \({\sim\SI{3.5}{MHz}}\) apart in frequency. However, there are some \({q=\pm2}\) transitions we wish to avoid driving; the \({F=2, m_F=2\leftrightarrow F'=2, m_F'=0}\) transition is just \({\sim\SI{800}{kHz}}\) detuned from the desired \({F=2, m_F=-2\leftrightarrow F'=2, m_F'=-2}\) transition. This transition also happens to have the highest motional sideband coupling. In the shelving procedure, we have the freedom to not shelve one of the encoded states; if we encode using this \(F=2, m_F = -2\) level, we make sure to measure it first, to avoid the need to shelve it and incur these errors. A useful trick to avoid coherent transfers of unwanted transitions is to hide an encoded state; if shelving one state drives a motional sideband which transfers population from another encoded state, we can move this other encoded state to a different hyperfine manifold, hiding it from the motional sideband transition. The error from this transfer is negligible in comparison to the other errors in the overall measurement, which will be discussed in Section \ref{Sec:Single-Qudit}. Finally, we also have the choice to not deshelve one of the shelved states, since we only need to fluoresce all but one of the encoded states. Using these principles, we can perform the 3- and 5-level measurements without driving any motional sideband transitions coherently; for the 7-level measurement, we must coherently drive four (second-order) motional sidebands. If we are not careful about constructing the sequence, undesired carrier transitions could reduce our measurement fidelity to under \(10\%\), and first-order motional transitions could reduce it to below \(95\%\). More details about the measurement sequences can be found in Appendix \ref{app:Measurement}. 

Next, we have to consider each fluorescence measurement. We assume that the imaging system has \({\text{NA} = 0.5}\) and a quantum efficiency of \({80\%}\). Assuming over-saturation for all fluorescing lasers, a good estimate for our fluorescence rate is \({(R_f/2) \approx 1/\tau_{P_{1/2}}\times 1/3\times B_{P_{1/2}-S_{1/2}}\approx \SI{30}{MHz}}\), where \(\tau_{P_{1/2}} = \SI{7.92}{ns}\) is the lifetime of the \(P_{1/2}\) state, \(B_{P_{1/2}-S_{1/2}} = 0.756\) is the branching ratio from the \(P_{1/2}\) state to the \(S_{1/2}\) state. Assuming around 10 bright-state photons are needed to discriminate between a bright or dark reading, each fluorescence step takes \({t_{fluor} \approx \SI{6}{\mu s}}\). Each fluorescence measurement also introduces a decay error of \(\mathcal{E}_{dec}(t_{fluor})\), and an error stemming from dark and background counts in the photon detector. In Reference \cite{Harty-2013}, this error is estimated to be \(\mathcal{E}_{phot}\sim 2.8\times 10^{-4}\); their setup has higher background counts and a lower collection efficiency than our assumption, so we use this number as an over-estimation.

Figure \ref{fig:Adiabatic-Passage}(b) considers the entire shelving measurement process for different qudits up to 7-levels. During a typical measurement, the procedure is complete once fluorescence has been seen, so in most cases, not all of the transfers in the shelving procedure need to be performed. Here we assume the worst case where we have to do all of the transfers (for \(d\) levels, this is \({2d-3}\) transfers). Again, we assume that the initial detuning for each transfer is \({\SI{1.3}{MHz}}\). The overall fidelity is calculated by 
\begin{align}
    \mathcal{F}_{meas} = &\big(\left(1 - \mathcal{E}_{dec}(t_{fluor})\right)\left(1 - \mathcal{E}_{phot}\right)\big)^{d-1}
    \nonumber\\
    &\quad\times\prod_{\{k\}}\mathcal{F}_{shelve},
\end{align}
where \(\{k\}\) is the set of all of the shelving and deshelving transitions for the measurement. Details about the sequences we chose can be found in Appendix \ref{app:Measurement}. We plot the best fidelity for different sweep rates in Figure \ref{fig:Adiabatic-Passage}; as shown, it is possible to get better than \(98\%\) overall measurement fidelity for up to 7-level qudits. Both the 3- and 5- level qudits can be measured with better than \({99\%}\) fidelity. Table \ref{tab:MeasurementError} summarizes all of the error contributions for the different qudits. The biggest contributors are the imperfect state preparation (from finite detuning), and the dephasing from the laser linewidth. Note that utilizing chirped pulses would eliminate the preparation error entirely. Only the 7-level qudit measurement involves coherently driving motional sidebands, suffering an appreciable amount of error. This highlights the importance of choosing a good measurement sequence, to avoid all first-order and as many second-order sidebands as possible. Note that over-estimations were made in this analysis, so the experimental fidelity is likely to be better than what is presented here. With the ability to measure qudit states with high fidelity, combined with universal single qudit gates, quantum state tomography can be carried out in a straightforward manner \cite{Thew-et-al-2002}. A protocol to implement single qudit gates for trapped ions is introduced in Section IV.

There are multiple avenues by which the error rates for our shelving measurement could be improved beyond the analysis in this paper. As mentioned earlier in this section, shaped pulses could be used rather than rectangular pulses. This would drastically reduce one of the major error sources: adiabatic state preparation. Alternatives to adiabatic passage could also be considered, such as composite pulses and optimal control theory. Additionally, when performing statistical measurements, one could use an adaptive algorithm to do state fluorescence on the state that the qudit is most likely in, based on the previous measurements. Such an adaptive measurement would make the number of adiabatic passages necessary approach \({d-1}\), dramatically decreasing the measurement error.

\section{Single Qudit Gates}
\label{Sec:Single-Qudit}
Single qudit gates can be performed using well-studied decompositions like those in \cite{O'Leary-Brennen-Bullock-2006, Schirmer-et-al-2001, Ivanov-et-al-2006}. Here we describe the essential features of such schemes, and present simulations on the resulting error rates when implemented with microwave fields of uncontrolled polarization or with stimulated Raman transitions.

To enable universal quantum computation, in terms of single-qubit operations, the Pauli gate set along with a non-Clifford gate is sufficient. Similarly, for qudits, generalized versions of the Pauli gates and a non-Clifford gate fulfill the requirement for universal quantum computing in terms of single-qudit operations \cite{Howard-Vala-2012}. A convenient non-Clifford gate is typically chosen to be the \(\pi/8\) gate. We refer the reader to Appendix \ref{app:Single-Qudit-Library} for the definitions of these gates.

To physically realize these gates, one way is to decompose them into a sequence of simpler physical operations. It is known that single qudit gates can be decomposed into sequences of two-level operations as outlined in \cite{O'Leary-Brennen-Bullock-2006,Schirmer-et-al-2001}. Physically, these operations are implemented using sequences of microwave or laser pulses, each implementing an evolution operator of the form 
\begin{equation}
    G(j,k;\theta,\phi) = \exp{\left(i\theta\left(e^{i\phi}\ketbra{j}{k} + e^{-i\phi}\ketbra{k}{j}\right)\right)}.
\end{equation}
Here \({\ket{j}}\) and \({\ket{k}}\) are two of the qudit basis states, \({\theta}\) represents a pulse angle (which physically depends on the Rabi frequency for the transition \({\ket{j}\leftrightarrow\ket{k}}\) and the pulse duration), and \({\phi}\) is a phase that can be controlled by manipulating the phase of the microwave or optical radiation. These constituent operations \({G(j,k;\theta,\phi)}\) are referred to as Givens rotations, in keeping with prior nomenclature.

Single-qudit gate decompositions can be made provided that the allowed transitions form a connected graph. For a qudit of dimension \(d\), at most \({d(d-1)/2}\) Givens rotations are required to synthesize an arbitrary single-qudit unitary, up to a set of phase factors on the qudit basis states. Essentially, the desired gate can be written as a sequence of population transfers between two levels from the qudit space:
\begin{equation}
\label{eq:Unitery-Decomp}
    \Hat{U} = \Hat{V}_K\Hat{V}_{K-1}\ldots\Hat{V}_1\Hat{\Theta},
\end{equation}
where \({\Hat{V}_i}\) are unitaries generated by individual pulses applied to a transition between states and \({\Hat{\Theta}}\) is a set of phase factors in a diagonal matrix. If necessary, these phase factors can further be eliminated by at most \({2(d-1)}\) additional rotations. Further details of this decomposition are given in Appendix \ref{app: single qudit gate decomposition}. There are other possible gate decomposition schemes that can be more efficient, such as the Householder reflection decomposition \cite{Ivanov-et-al-2006}. However, this decomposition requires an ancillary state which has direct transitions to all encoded states. This limits the generalizability to higher dimensional qudits for trapped ions. In comparison, the requirement for Givens rotation decomposition is more lenient, and can be easily generalized.

We use this decomposition method with Givens rotation to construct the aforementioned generalized Pauli gates and the \(\pi/8\) gate. In addition, we also decompose the quantum Fourier transform of a single qudit, which is equivalent to generalization of the Hadamard gate to higher dimensions and we denote it with \(\hat{H}_d\). These decompositions are written out explicitly in Appendix \ref{app:Single-Qudit-Library}.

\setlength\tabcolsep{5pt}
\begin{table*}[htb!]
    \centering
    \begin{tabular}{cccc}
        \toprule\toprule
        \textbf{Error Source} & \textbf{3-level} & \textbf{5-level}
        \\\midrule
        Magnetic field noise & \({(2.5 \pm 0.2)\times 10^{-10}}\) & \({(2.5 \pm 0.2)\times 10^{-9}}\)
        \\
        Off-resonant coupling* & \({(1.12 \pm 0.01)\times 10^{-4}}\)& \({(1.35 \pm 0.02)\times 10^{-3}}\)
        \\
        Scattering** & \({4.62\times 10^{-5}}\) & \({1.94\times 10^{-4}}\)
        \\
        \bottomrule\bottomrule
    \end{tabular}
    \caption{Error budget for \(\hat{H}_d\) gate with a \({\SI{10}{kHz}}\) Rabi frequency, and a magnetic field noise of \({\SI{2.7}{pT}}\). Off-resonant coupling simulations were run 500 times each and magnetic field noise simulations were run 300 times each, varying the initial state randomly; the average error from these are shown. *Only present for gates driven with microwaves. **Only present for gates driven with Raman transitions.}
    \label{tab:SingleQudit}
\end{table*}
 To drive a single qudit gate, we consider two possible methods. The first method uses direct transitions with a microwave source while the other is via Raman transitions with laser beams in the visible range. For microwaves, an unpolarized source is assumed for simplicity.

In order to assess the practicality of a trapped ion qudit system, the expected errors for the constructed gate set should be evaluated. From previous qubit experiments, sources of error for a trapped ion system are well-understood \cite{Gaebler-et-al-2016}. As discussed in the introduction, we focus on known errors that are fundamental to our trapped ion system and errors due to environmental factors. Errors due to experimental controls are excluded from our analysis.   For single qudit gates, the first error we consider is magnetic field noise. Fluctuations in the overall magnetic field result in magnetic sub-level energy fluctuations, decohering the qudits while transitions are being driven. Another source of error will be off-resonant coupling. This error is relevant for an unpolarized driving source as it is unable to make use of selection rules for state transitions to mitigate off-resonant coupling. Finally, when using Raman transitions to drive gates, off-resonant coupling to the \(P\) states induces photon scattering, which can lead to state decoherence.

To estimate the errors of single qudit gates in ion traps using microwaves, we simulate the full Hamiltonian with a fluctuating magnetic field and off-resonant coupling. We perform these simulations for all the aforementioned gates: generalized Pauli gates, the \(\hat{H}_d\) generalized Hadamard gate, and a generalized \(\pi/8\) gate \cite{Howard-Vala-2012}; additional details for how the simulations were performed and the full results can be found in Appendix \ref{app:Single-Qudit}.

It is found that for higher Rabi frequencies, the gate times decrease, and thus the magnetic field fluctuation errors are reduced; however off-resonant coupling error becomes worse. With a Rabi frequency of \({\SI{100}{kHz}}\), the \(\hat{H}_d\) error increases to around \({1\%}\) and \({10\%}\) for 3- and 5-level qudits respectively. Thus, for further analysis in this article, the Rabi frequency for microwave transitions is limited to \({\SI{10}{kHz}}\). Note that polarization control of microwaves is possible with a specialized ion trap \cite{Shappert-et-al-2013}, which would allow us to reduce this off-resonant coupling error almost entirely.

Using Raman transitions with polarized laser beams avoids errors from off-resonant coupling, so we are able to assume a Rabi frequency of \({\SI{100}{kHz}}\). However, there is another significant error stemming from photon scattering. To obtain a characteristic value for this error, we pick the error from transition that has the largest photon scattering error among all the \(\ket{l}\) to \(\ket{l+1}\) transitions. Only Raman scattering decoheres the qudit states for single qudit gates \cite{Ozeri-2007}. Thus, the state with the largest Raman scattering rate, \(R_{Raman}\), is chosen to characterize the error, which is obtained by computing the difference between the total and Rayleigh scattering rates. 
To compute the scattering probability during the gate time, the calculation is done using the Kramers-Heisenberg formula \cite{Loudon-1983}.
For \( d=3 \) and using the zig-zag encoding shown in Figure \ref{fig:Qudit_MS_frequencies}(a), the total and Rayleigh scattering rates that result in the largest Raman scattering rate are
\begin{align}
    \label{Eq: Single qudit total scattering d=3}
    R_{total}^{(3)} &= \frac{\sqrt{3} \Omega}{6} \left( \frac{\Delta_{1/2} \Delta_{3/2}}{\Delta_{1/2} - \Delta_{3/2}} \right) \times
    \nonumber\\
    &\Bigg[ \gamma_{P_{1/2} \rightarrow S_{1/2}} \frac{\omega_R^3}{\omega_{P_{1/2} \rightarrow S_{1/2}}^3} \left( 5\frac{1}{\Delta_{1/2}^2} + 7\frac{1}{\Delta_{3/2}^2} \right) 
    \nonumber\\
    &+ 5 \gamma_{P_{1/2} \rightarrow D_{3/2}} \frac{(\omega_R-\omega_{D_{3/2}})^3}{\omega_{P_{1/2} \rightarrow D_{3/2}}^3} \frac{1}{\Delta_{1/2}^2}
    \nonumber\\
    &+ 7 \gamma_{P_{3/2} \rightarrow D_{3/2}} \frac{(\omega_R-\omega_{D_{3/2}})^3}{\omega_{P_{3/2} \rightarrow D_{3/2}}^3} \frac{1}{\Delta_{3/2}^2}
    \nonumber\\
    &+ 7 \gamma_{P_{3/2} \rightarrow D_{5/2}} \frac{(\omega_R-\omega_{D_{5/2}})^3}{\omega_{P_{3/2} \rightarrow D_{5/2}}^3} \frac{1}{\Delta_{3/2}^2} \Bigg],
\end{align}
\begin{align}
    \label{Eq: Single qudit Rayleigh scattering d=3}
    R_{Rayleigh}^{(3)} &= \frac{\sqrt{3} \Omega}{36} \left( \frac{\Delta_{1/2} \Delta_{3/2}}{\Delta_{1/2} - \Delta_{3/2}} \right) \gamma_{P_{1/2} \rightarrow S_{1/2}} \times
    \nonumber\\
    &\!\!\!\frac{\omega_R^3}{\omega_{P_{1/2} \rightarrow S_{1/2}}^3} \left( \frac{13}{\Delta_{1/2}^2} + \frac{34}{\Delta_{1/2} \Delta_{3/2}} + \frac{25}{\Delta_{3/2}^2} \right),
\end{align}
where \( R_{total} \) is the total scattering rate, \( R_{Rayleigh} \) is the Rayleigh scattering rate, \( \Omega \) is the Rabi frequency, \({\gamma_{P_{1/2} \rightarrow S_{1/2}} = \SI{9.53e7}{s^{-1}} }\) is the decay rate from \({6P_{1/2}}\) to the \({6S_{1/2}}\) state of \({\mathrm{Ba}^+}\), \({\gamma_{P_{1/2} \rightarrow D_{3/2}} = \SI{3.10e7}{s^{-1}} }\) is the decay rate from \({6P_{1/2}}\) to the \({5D_{3/2}}\) state, \({\gamma_{P_{3/2} \rightarrow D_{3/2}} = \SI{6.00e6}{s^{-1}}} \) is the decay rate from \({6P_{3/2}}\) to the \({5D_{3/2}}\) state, \({\gamma_{P_{3/2} \rightarrow D_{5/2}} = \SI{4.12e7}{s^{-1}} }\) is the decay rate from \({6P_{3/2}}\) to \({5D_{5/2}}\) states \cite{NIST-ASD-Team-2018}, \( \omega_R \) is the laser frequency, \( \hbar \omega_i \), \( i \in \{ D_{3/2}, D_{5/2} \} \), corresponds to the energy level in the \( i^{\text{th}} \) state, with the energy of the \( 6S_{1/2} \) state set to be zero, \( \omega_{j \rightarrow k} \) corresponds to the transition frequency between states \( j \) and \( k \), \({\Delta_{1/2}}\) is the detuning of our laser frequency for a Raman transition from the \({6P_{1/2}}\) state, and \({\Delta_{3/2}}\) is the laser detuning from the \({6P_{3/2}}\) state. With a laser frequency of \SI{532}{nm} assumed, we have \( \Delta_{1/2} = \SI{-44.08}{THz} \) and \( \Delta_{3/2} = \SI{-94.78}{THz} \). For \( d=5 \) and the zig-zag encoding scheme as shown in Figure \ref{fig:Qudit_MS_frequencies}(b), the total and Rayleigh scattering rates that result in the largest Raman scattering rate are
\begin{align}
    \label{Eq: Single qudit total scattering d=5}
    R_{total}^{(5)} &= \frac{\Omega}{2} \left( \frac{\Delta_{1/2} \Delta_{3/2}}{\Delta_{1/2} - \Delta_{3/2}} \right) \times
    \nonumber\\
    &\Bigg[ \gamma_{P_{1/2} \rightarrow S_{1/2}} \frac{\omega_R^3}{\omega_{P_{1/2} \rightarrow S_{1/2}}^3} \left( 5\frac{1}{\Delta_{1/2}^2} + 7\frac{1}{\Delta_{3/2}^2} \right) 
    \nonumber\\
    &+ 5 \gamma_{P_{1/2} \rightarrow D_{3/2}} \frac{(\omega_R-\omega_{D_{3/2}})^3}{\omega_{P_{1/2} \rightarrow D_{3/2}}^3} \frac{1}{\Delta_{1/2}^2}
    \nonumber\\
    &+ 7 \gamma_{P_{3/2} \rightarrow D_{3/2}} \frac{(\omega_R-\omega_{D_{3/2}})^3}{\omega_{P_{3/2} \rightarrow D_{3/2}}^3} \frac{1}{\Delta_{3/2}^2}
    \nonumber\\
    &+ 7 \gamma_{P_{3/2} \rightarrow D_{5/2}} \frac{(\omega_R-\omega_{D_{5/2}})^3}{\omega_{P_{3/2} \rightarrow D_{5/2}}^3} \frac{1}{\Delta_{3/2}^2} \Bigg],
\end{align}
\begin{align}
    \label{Eq: Single qudit Rayleigh scattering d=5}
    R_{Rayleigh}^{(5)} &= \frac{\Omega}{12} \left( \frac{\Delta_{1/2} \Delta_{3/2}}{\Delta_{1/2} - \Delta_{3/2}} \right) \gamma_{P_{1/2} \rightarrow S_{1/2}} \times
    \nonumber\\
    &\!\!\!\frac{\omega_R^3}{\omega_{P_{1/2} \rightarrow S_{1/2}}^3} \left( \frac{13}{\Delta_{1/2}^2} + \frac{34}{\Delta_{1/2} \Delta_{3/2}} + \frac{25}{\Delta_{3/2}^2} \right).
\end{align}
Further details on the calculations on the scattering rates can be found in Appendix \ref{app:Photon_Scatt}. The Raman scattering probability, \( P_{Raman} \), which is treated to be equivalent to the errors, is then
\begin{equation} \label{Eq: Scattering probability RRamantg}
    P_{Raman} = R_{Raman} t_g,
\end{equation}
where \( t_g \) is the gate time. The \(\hat{H}_d\) gate, which has the longest duration, is used to obtain an upper bound for the error. The \(\hat{H}_d\) gate has a gate time \( t_g = \SI{28.04}{\mu s} \) for \( d = 3 \) and \( t_g = \SI{67.85}{\mu s} \) for \( d = 5 \) with a Rabi frequency of \SI{100}{kHz}. 

For a single \(\pi\)-pulse population transfer, assuming a \(\SI{10}{kHz}\) Rabi frequency for microwave transitions and \(\SI{100}{kHz}\) for Raman transitions, the error is on the order of \(10^{-6}\). For all of the gates of interest, the overall fidelities are better than \({99.8\%}\). To isolate the contribution of each effect, simulations are also run for each error by itself, and the average results for the \(\hat{H}_d\) gate with a Rabi frequency of \({\SI{10}{kHz}}\) are shown in Table \ref{tab:SingleQudit}. Note that the photon scattering error is independent of the Rabi frequency, so the error figure for this error in Table \ref{tab:SingleQudit} is representative for a Rabi frequency of \({\SI{100}{kHz}}\) as well. With such a well controlled magnetic field, off-resonant coupling and photon scattering errors are more than five orders of magnitude larger than the magnetic field noise error. For the magnetic field noise error to be of the same order of magnitude as the others, which we take to be \(10^{-4}\), the fluctuations could be relaxed. The extents of relaxation of the requirement on magnetic field noise for each transition technique and \( d=3 \) and \( d=5 \) are tabulated in Table \ref{tab: Single qudit magnetic field relaxation}. Overall the error estimates for known error sources are sufficiently small, and there is no limiting factor for achieving fidelities higher than typical fault tolerant thresholds for single qudit gates, whether they are driven by microwaves or Raman transitions. Further details of the calculations can be found in Appendices \ref{app:Single-Qudit-Mag} and \ref{app:Magnetic_Field_Threshold}.

\setlength\tabcolsep{5pt}
\begin{table}[htb!]
    \centering
    \begin{tabular}{cccc}
        \toprule\toprule
        {} & \multicolumn{2}{c}{\( \sqrt{\langle \Delta B^2 \rangle} \) \textbf{/ \SI{}{nT}}}
        \\
        \textbf{Transition Technique} & \textbf{3-level} & \textbf{5-level}
        \\\midrule
        Microwave & \(0.811\) & \(0.167\)
        \\
        Raman transition & \(8.111\)& \(1.676\)
        \\
        \bottomrule\bottomrule
    \end{tabular}
    \caption{Estimation of thresholds of the standard deviation of magnetic field, \( \sqrt{\langle \Delta B^2 \rangle} \), for error values from magnetic field noise to be below \(10^{-4}\) for a \(\hat{H}_d\) gate with \({\SI{10}{kHz}}\) and \({\SI{100}{kHz}}\) Rabi frequencies for microwave and Raman transitions respectively.}
    \label{tab: Single qudit magnetic field relaxation}
\end{table}

\section{Two Qudit(Entangling) Gates}
\label{Sec:Two-Qudit}
In addition to the single-qudit operations described in the previous section, to form a universal gate set for quantum computing, an entangling gate or a two-qudit gate is required. In this section, we derive such an entangling gate and simulate the errors associated with this scheme.

A two-qudit gate can be performed using generalizations of a common technique often referred to as the M\o{}lmer-S\o{}rensen (MS) gate \cite{Sorensen-Molmer-2000}. Lasers applying optical dipole forces to the ion crystal can be used to implement a state-dependent force; combined with the Coulomb repulsion between ions, this force can mediate an entangling interaction.

In this section, we give a detailed derivation of a MS-like entangling protocol for qudits (part A). In addition, we investigate the effects of a variety of possible error sources (part B).

\subsection{Ideal Generalized MS Gate}
We describe a generic approach to implementing two-qudit gates by addressing an appropriate combination of motional sideband transitions. We assume that the qudit levels are chosen such that there are dipole-allowed transitions between each pair of levels \({\ket{l}\leftrightarrow\ket{l+1}}\), and that the energies are chosen in a zig-zag configuration. Our entangling gate can be used to generate any arbitrary two-qudit unitary with the addition of the single qudit gates described in Section \ref{Sec:Single-Qudit} and forms a universal gate set \cite{Brylinski-2001}.

In analogy to the qubit case, we generalize the MS gate to a qudit system with the following Hamiltonian
\begin{equation} 
\label{Eq:Qudit-MS}
    \mathcal{H}_{ideal} = \hbar\eta_M\bar{\Omega}(\hat{a}^{\dagger}e^{i(\omega_M-\mu)t}+\hat{a}e^{-i(\omega_M-\mu)t})\sum_{n=1}^{N}\hat{S}_{x,n},
\end{equation}
where 
\begin{align}
    \hat{S}_{x} &= \frac{\hat{S}_{+}+\hat{S}_{-}}{2}
    \nonumber\\
    \hat{S}_{+} &= \sum_{l'=-s}^{s-1}\sqrt{s(s+1)-l'(l'+1)}\ket{l'+s+1}\bra{l'+s}
    \nonumber\\
    \hat{S}_{-} &= \sum_{l'=-s}^{s-1}\sqrt{s(s+1)-l'(l'+1)}\ket{l'+s}\bra{l'+s+1},
\end{align}
\( \eta_M \) is the Lamb-Dicke parameter of the motional normal mode used for MS gate entanglement, \( \bar{\Omega} \) is the qudit MS gate Rabi frequency, \( \hat{a} \) and \( \hat{a}^{\dagger} \) are the lowering and raising operators of the motional normal mode used for entanglement, \( \omega_M \) is the motional frequency used for entanglement, \( \mu \) is the frequency detuned from resonant transitions, \({d=2s+1}\) is the total qudit levels. This is analogous to generalizing a spin half system in the qubit case to an arbitrary spin system in the qudit case. This can be done by applying red- and blue-detuned transition frequencies between each \( \ket{l} \) and \( \ket{l+1} \) states on the 2 qudits to be entangled (see Figure \ref{fig:Qudit_MS_frequencies}), which is known to be straightforward \cite{Senko-et-al-2014}. Further details of the derivations can be found in Appendix \ref{app:Qudit-Ent}.

The desired time evolution unitary operator generated by the Hamiltonian in Equation \ref{Eq:Qudit-MS} is
\begin{equation}
    \hat{U}\left( t \right) = \exp{\left( i \theta_0 \left( \sum_{n=1}^N \hat{S}_x,n \right)^2 \right)},
\end{equation}
where \( \theta_0 \) is the qudit MS gate phase.

\begin{figure}[htb!]
    \centering
    \subfloat[\label{sfig:Yo001}]{\includegraphics[width=0.48\textwidth]{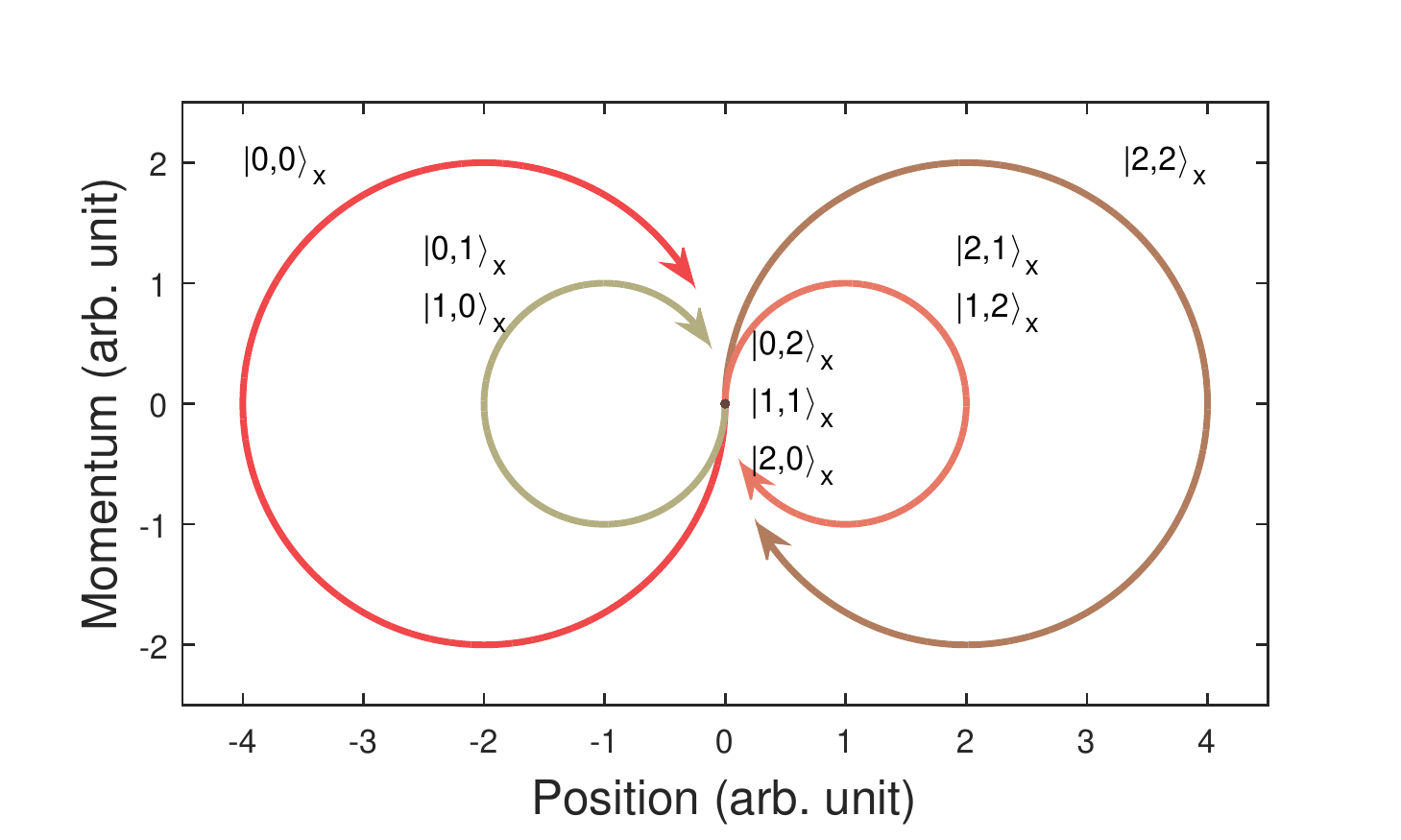}}
    \\\vspace{-12px}
    \subfloat[\label{sfig:Yo002}]{\includegraphics[width=0.48\textwidth]{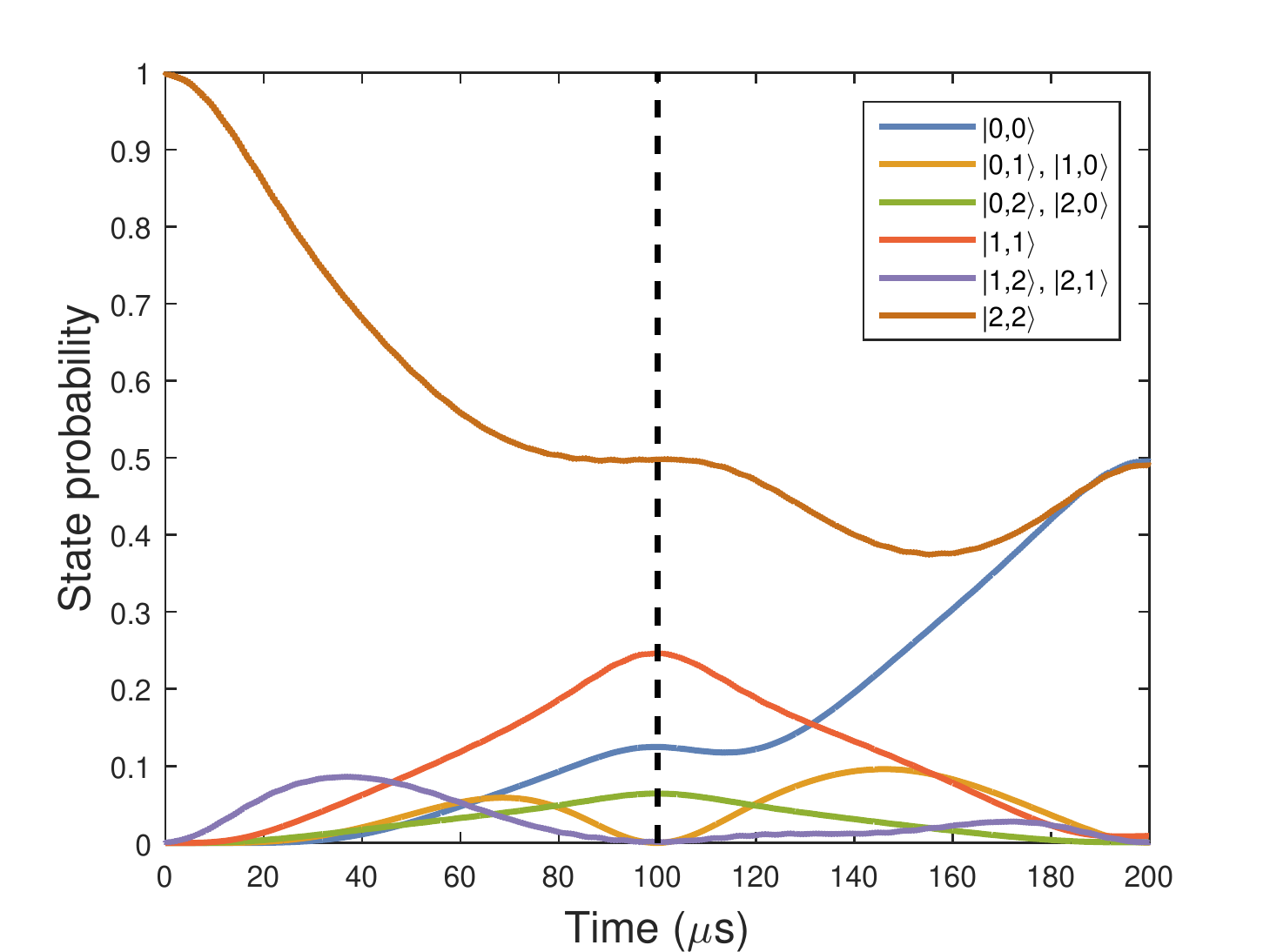}}
    \caption{(a) Illustration of evolution of 3-level qudits in position-momentum phase space during the qudit MS gate. The displacement of all states return to the original position at the end of the gate. The phase gained by each basis state is proportional to the area enclosed by the trajectory in phase space. (b) State probability evolution of the 3-level qudit MS gate. The gate is set to complete a loop in the position-momentum phase space every \(\SI{100}{\mu s}\). \( \theta_0 \) is set to \({\theta_0=-\frac{\pi}{4}}\). After one loop, at \({t=\frac{2\pi}{\omega_C-\mu}=\SI{100}{\mu s}}\) (dashed line), multilevel entanglement is achieved.}
    \label{fig:Entangling_gate_evolution}
\end{figure}

\begin{figure*}[t]
    \centering
    \subfloat[\label{sfig:Yooo1}]{\includegraphics[width=0.5\textwidth]{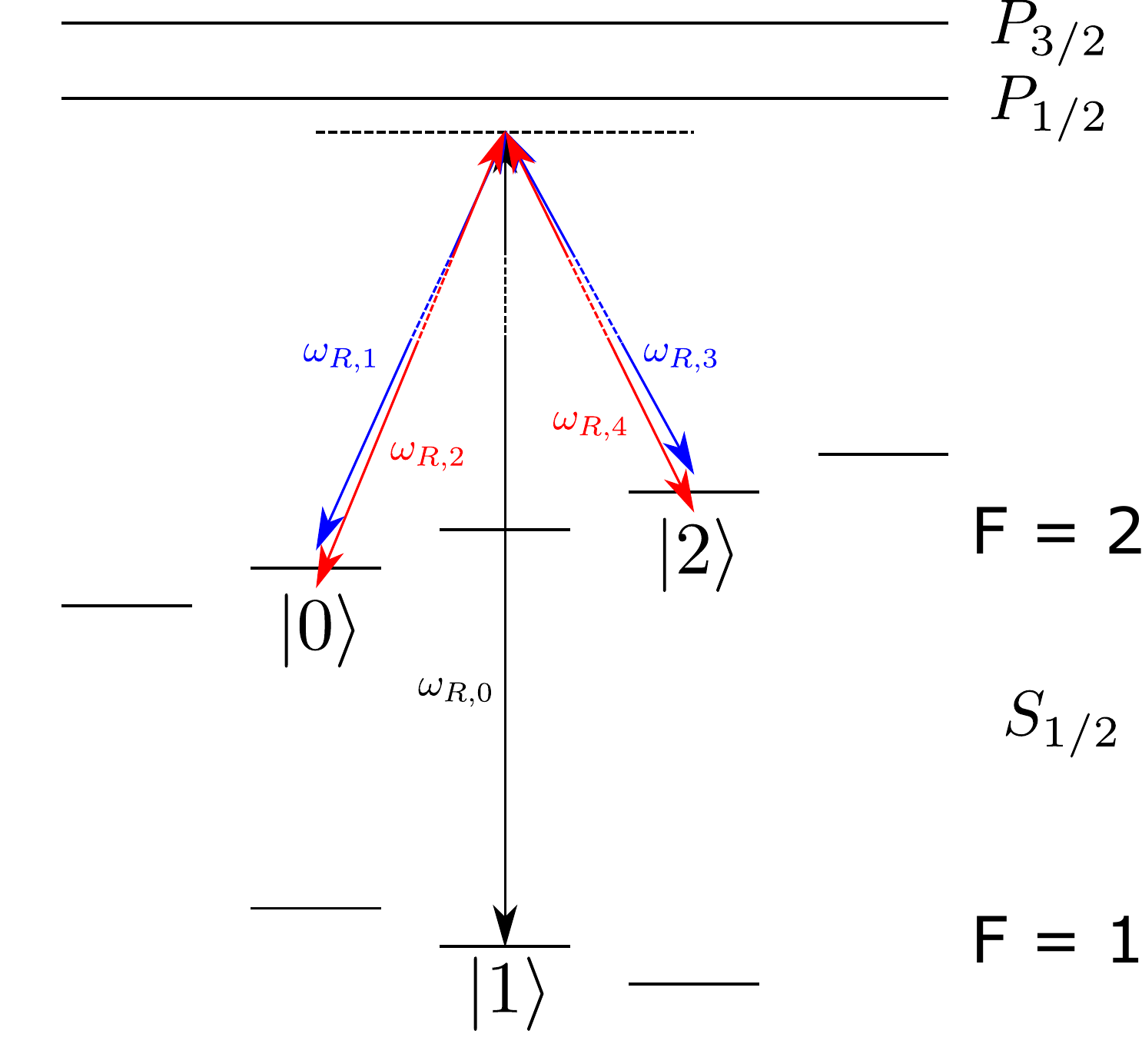}}
    \subfloat[\label{sfig:Yoo2}]{\includegraphics[width=0.5\textwidth]{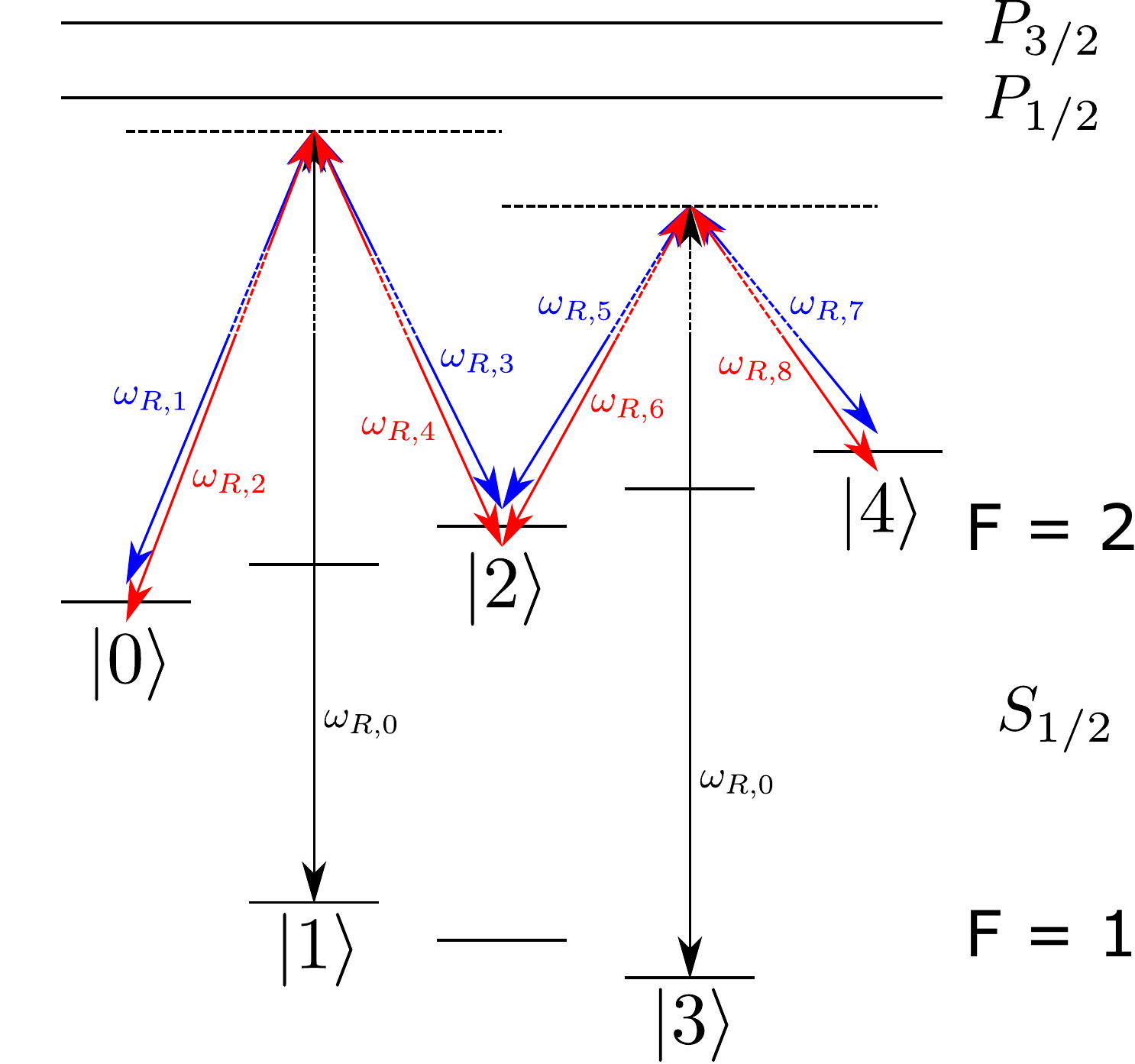}}
    \\
    \subfloat[\label{sfig:Yoo3}]{\includegraphics[width=0.3\textwidth]{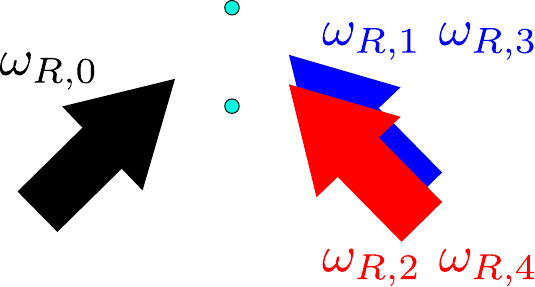}}
    \caption{Schematics of laser perturbations applied to \({^{137}\mathrm{Ba}^+}\) for (a) 3-level qudit and (b) 5-level qudit for qudit MS gate. \({\omega_{R,n}}\) denotes the \({n^{th}}\) laser frequency applied for the gate. (c) Illustration of physical implementation of laser beams and frequencies for the qudit MS gate. Cyan circles represent the ions.}
    \label{fig:Qudit_MS_frequencies}
\end{figure*}

In the phase space picture as shown in Figure \ref{fig:Entangling_gate_evolution}(a), this operation corresponds to displacing the system in the phase space with a radius proportional to \({{S}_{x,1}+{S}_{x,2}}\). The geometric phase gained after closing the loop is proportional to the area enclosed by the trajectory, which is proportional to \({({S}_{x,1}+{S}_{x,2})^2}\).

\subsection{Error Estimates}
In order to estimate the expected error of the qudit MS gate, we consider sources of error that are intrinsic to the formulation as well as errors from imperfect environment. The intrinsic sources of error are:
\begin{enumerate}
    \item Inaccuracy from Lamb-Dicke approximation (LDA).
    \item Inaccuracy from rotating wave approximation (RWA).
    \item Presence of spectator phonon modes.
    \item Photon scattering.
\end{enumerate}
The errors from an imperfect environment are:
\begin{enumerate}[resume]
    \item Imperfect cooling of ions.
    \item Motional heating of ions.
    \item Magnetic field noise.
\end{enumerate}

Choosing the center-of-mass mode as the coupled mode for qudit MS entanglement, we have \( \eta_M = \eta_C \) and \( \omega_M = \omega_C \), where the subscript \( C \) denotes center-of-mass mode. To model a realistic ion trap, the parameters used are \({\eta_C = 5.07 \times 10^{-2}}\), \({\omega_C = 2\pi \times \SI{2}{MHz}}\), \({\omega_{T} = 2\pi\times\SI{1.8}{MHz}}\) (the frequency of spectator mode, subscript T denotes tilt mode), \({\mu = 2\pi\times \SI{2.01}{MHz}}\), and a gate time of \({t_{g}=\frac{2\pi}{|\omega_{C}-\mu|}=\SI{100}{\mu s}}\). We set \({\theta_0 = -\frac{\pi}{4}}\) as an example. This value of \({\theta_0}\) is chosen as it results in a non-trivial entanglement result that is not replicable by a single qubit MS gate for a 3-level qudit system. For example, \({\theta_0 = -\frac{\pi}{2}}\) acting on the state \({\ket{2,2}}\) of a 3-level qudit system can be shown to give the same output as a qubit MS gate acting on the appropriate transition levels (see Figure \ref{fig:Entangling_gate_evolution}(b) at \({t=\SI{200}{\mu s}}\)). We kept \({\theta_0 = -\frac{\pi}{4}}\) for the 5-level qudit for simplicity. The motional heating rate is assumed to be \({\SI{100}{\; quanta/s}}\), which is a realistic estimate \cite{Senko-2014}. The initial two-qudit state is chosen to be \( \ket{d-1}\ket{d-1} \). Variations of the parameters and Hamiltonian were used to pinpoint the magnitude of error contributed by a certain error source.

In the case of a qubit MS gate, we desire an interaction Hamiltonian of the form \cite{Sorensen-Molmer-2000}
\begin{equation}
    \mathcal{H} = \hbar\eta\Omega(\hat{a}^{\dagger}e^{i(\omega_M-\mu)t}+\hat{a}e^{-i(\omega_M-\mu)t})\sum_{n=1}^{N} \frac{\hat{\sigma}_{x,n}}{2},
\end{equation}
where \({\eta}\) is the Lamb-Dicke parameter, \({\Omega}\) is the resonant Rabi frequency between the two levels, \({\hat{a}}\) and \({\hat{a}^\dagger}\) are the lowering and raising operators of a vibrational mode in an ion chain respectively, \({\omega_M}\) is the target vibrational mode frequency, \({\mu}\) is the detuning of laser frequencies from the resonant frequency, \({t}\) is time, and \({\hat{\sigma}_x}\) is the Pauli x operator. \({n}\) is the index for each ion in a chain of \({N}\) ions.

In order to arrive at the desired qudit MS Hamiltonian (Equation \ref{Eq:Qudit-MS}), we start with a chain of \({N}\) ion qudits. The static Hamiltonian is
\begin{align}
    \mathcal{H}_0&=\mathcal{H}_{0,M}+\mathcal{H}_{0,S}
    \nonumber\\
    \mathcal{H}_{0,M}&=\sum_{m=1}^{N}\hbar\omega_m\left(\hat{a}_m^{\dagger}\hat{a}_m+\frac{1}{2}\right)
    \nonumber\\
    \mathcal{H}_{0,S}&=\sum_{n=1}^{N}\sum_{l=0}^{d-1}E_{l,n}\ket{l}\bra{l}_n,
\end{align}
where \({\mathcal{H}_{0,M}}\) describes the Hamiltonian of the motional state, \({\mathcal{H}_{0,S}}\) describes the Hamiltonian of the internal energy states, the subscript \( m \) denote the \( m^{\text{th}} \) vibrational normal mode, \({E_{l}}\) is the energy of state \({\ket{l}}\), and \({l=l'+s}\). We assume that for each transition level between \({\ket{l}}\) and \({\ket{l+1}}\), we apply a laser perturbation with frequency \({\omega_{L,l}}\), close to the transition energy between the two target levels and far off-resonant to (or forbidden by selection rules for) transitions to the other levels. The interaction Hamiltonian is then approximately
\begin{align}
    \mathcal{H}_{int}=\sum_{n=1}^{N}\sum_{l=0}^{d-1}\hbar\Omega_{l,n}\cos(k\hat{x}_n-\omega_{L,l}t+\phi_l) 
    \nonumber\\
    \times\left(\ket{l+1}\bra{l}_n+\ket{l}\bra{l+1}_n\right),
\end{align}
where \( \Omega_{l,n} \) is the Rabi frequency for the target transition from \(\ket{l}\) to \(\ket{l + 1}\) for the \( n^{\text{th}} \) ion, \({\hat{x}}\) is the position of an ion along the motion of the phonon mode being used for entanglement, \({k}\) is the wavevector of the laser perturbation along \({\hat{x}}\), and \({\phi}\) is the initial laser phase. The total Hamiltonian is then
\begin{equation}
    \mathcal{H}=\mathcal{H}_{0,M}+\mathcal{H}_{0,S}+\mathcal{H}_{int}.
\end{equation}

Assigning odd qudit levels to lower energy levels and even qudit levels to higher ones in a zig-zag pattern, we define
\begin{equation}
    E_{l+1}-E_{l}=-(-1)^l\hbar\omega_l.
\end{equation}
In the interaction picture with respect to \({\mathcal{H}_0}\), the effective Hamiltonian is then
\begin{align}
    \mathcal{H}_I=\sum_{n=1}^{N}\sum_{l=0}^{d-1}\hbar\Omega_{l,n}\cos(k\hat{x}'_{n}-\omega_{L,l}t+\phi_{l})
    \nonumber\\
    \times\left(e^{-i(-1)^{l}\omega_l t}\ketbra{l+1}{l}_{n}+e^{i(-1)^{l}\omega_l t}\ketbra{l}{l+1}_{n}\right),
\end{align}
where \({\hat{x}' = e^{\frac{i\mathcal{H}_{0,M}t}{\hbar}}\hat{x}e^{-\frac{i\mathcal{H}_{0,M}t}{\hbar}}}\) is the position operator in the interaction picture.

The above description assumes one laser frequency per transition. In analogy to the MS scheme, let us have two laser perturbations tuned close to resonant for each \({\ket{l}}\) to \({\ket{l+1}}\) transition. One set of laser fields is blue-detuned while the other is red-detuned, with frequencies
\begin{align}
    \omega_{L,l}&=\omega_l+\mu \;\; \text{(blue-detuned)}
    \nonumber\\
    \omega_{L,l}&=\omega_l-\mu \;\; \text{(red-detuned)}.
\end{align}

For small \({\mu}\), we can apply a RWA to obtain the effective Hamiltonians for blue and red-detuned laser perturbations respectively:
\begin{align}
    \label{Eq:Red_Blue_sideband_Hamiltonians}
    \mathcal{H}_b \approx \sum_{n=1}^{N}\sum_{l=0}^{d-1}\frac{\hbar\Omega_l}{2}\left[e^{-i(-1)^{l}(k\hat{x}'_{n}-\mu t+\phi_{b,l})}\ketbra{l+1}{l}_{n}\right.
    \nonumber\\
    \left.+e^{i(-1)^{l}(k\hat{x}'_n-\mu t+\phi_{b,l})}\ketbra{l}{l+1}_{n}\right]
    \nonumber\\
    \mathcal{H}_r \approx \sum_{n=1}^{N}\sum_{l=0}^{d-1}\frac{\hbar\Omega_l}{2}\left[e^{-i(-1)^{l}(k\hat{x}'_{n}+\mu t+\phi_{r,l})}\ketbra{l+1}{l}_{n}\right.
    \nonumber\\
    \left.+e^{i(-1)^{l}(k\hat{x}'_n+\mu t+\phi_{r,l})}\ketbra{l}{l+1}_{n}\right].
\end{align}
Defining
\begin{align}
    \phi_{M,l}&=\frac{\phi_{r,l}-\phi_{b,l}}{2}
    \nonumber\\
    \phi_{S,l}'&=\frac{\phi_{r,l}+\phi_{b,l}}{2},
\end{align}
and adding \({\mathcal{H}_b}\) and \({\mathcal{H}_r}\) gives the resulting effective Hamiltonian of the form
\begin{align}
    \label{Eq:MS-gate no approx}
    \mathcal{H}_{total} &=\mathcal{H}_b+\mathcal{H}_r
    \nonumber\\   
    &=\sum_{n=1}^{N}\sum_{l=0}^{d-1}\hbar\Omega_{l}\cos(\mu t+\phi_{M,l})
    \nonumber\\
    &\times\left[e^{-i(-1)^{l}\phi_{S,l}'}e^{-i(-1)^{l}k\hat{x}'_{n}}\ketbra{l+1}{l}_{n}\right.
    \nonumber\\
    &\left.+e^{i(-1)^{l}\phi_{S,l}'}e^{i(-1)^{l}k\hat{x}'_{n}}\ketbra{l}{l+1}_{n}\right].
\end{align}

For small \({k\hat{x}'}\), we can apply the Lamb-Dicke approximation (LDA), which gives
\begin{equation}
    e^{\pm i(-1)^{l}k\hat{x}'} \approx 1 \pm i(-1)^{l}k\hat{x}'.
\end{equation}
Expressing \({k\hat{x}'}\) in terms of raising and lowering operators of ion chain vibrational modes,
\begin{equation}
    k\hat{x}'_n = \sum_{m=1}^{N}\eta_{m,n}\left(e^{i\omega_{m}t}\hat{a}^{\dagger}_{m}+e^{-i\omega_{m}t}\hat{a}_{m}\right),
\end{equation}
where \( \eta_{m,n} \) is the Lamb-Dicke parameter for the \( m^{\text{th}} \) motional mode and the \( n^{\text{th}} \) ion, \( \omega_m \) is the motional frequency of the \( m^{\text{th}} \) motional mode. We arrive at
%\vskip10px
\begin{widetext}
\begin{align}
\label{Eq:MS-Hamiltonian-withoutRWA}
    \mathcal{H}_{total}=&\sum_{n=1}^{N}\sum_{l=0}^{d-1}\hbar\Omega_{l}\cos(\mu t+\phi_{M,l})
    \left[e^{-i(-1)^{l}\phi_{S,l}'}\ketbra{l+1}{l}_{n}+e^{i(-1)^{l}\phi_{S,l}'}\ketbra{l}{l+1}_{n}\right]+\sum_{m=1}^{N}\sum_{n=1}^{N}\sum_{l=0}^{d-1}\eta_{m,n}\Omega_{l}
    \nonumber\\
    &\times\cos(\mu t+\phi_{M,l})\left(e^{i\omega_{m}t}\hat{a}^{\dagger}_{m}+e^{-i\omega_{m}t}\hat{a}_{m}\right)
    \left[e^{-i(-1)^{l}(\phi_{S,l}'+\frac{\pi}{2})}\ketbra{l+1}{l}_{n}
    +e^{i(-1)^{l}(\phi_{S,l}'+\frac{\pi}{2})}\ketbra{l}{l+1}_{n}\right].
\end{align}
Without loss of generality, we let the mode \({m=M}\) to be close to the laser frequency detuning \({\mu \approx \omega_{M}}\), and far-detuned from the other vibrational mode frequencies. With the condition \({\Omega_{l} \ll \mu}\), another RWA can be applied, which gives the resultant effective Hamiltonian of
\begin{align}
    \label{Eq:MS-Hamiltonian-pre}
    \mathcal{H}_{total}=\sum_{n=1}^{N}\sum_{l=0}^{d-1}&\hbar\frac{\eta_{M,n}\Omega_{l}}{2}
    \left(e^{i(\omega_{M}-\mu)t-i\phi_{M,l}}\hat{a}^{\dagger}_{M}+e^{-i(\omega_{M}-\mu)t+i\phi_{M,l}}\hat{a}_{M}\right)
    \nonumber\\
    &\times\left[(e^{-i(-1)^{l}(\phi_{S,l}'+\frac{\pi}{2})}\ketbra{l+1}{l}_{n}
    +e^{i(-1)^{l}(\phi_{S,l}'+\frac{\pi}{2})}\ketbra{l}{l+1}_{n})\right].
\end{align}
\end{widetext}

For simplicity, we specialize to the case where the near-resonant motional mode is the center-of-mass mode, \({\eta_{M,n}=\eta_{C}}\) and we rewrite \({\omega_M=\omega_C}\). To arrive at Equation \ref{Eq:Qudit-MS}, we let
\begin{align}
    \label{Eq:para_req1}
    \phi_{S}&=-(-1)^{l}\left(\phi_{S,l}'+\frac{\pi}{2}\right)
    \nonumber\\
    \phi_{M}&=\phi_{M,l}
    \nonumber\\
    \Omega_{l}&=\bar{\Omega} \sqrt{s(s+1)-l'(l'+1)}.
\end{align}
For the case where the spin phase \({\phi_{S} = 0}\), and the motional phase \({\phi_{M} = 0}\), with Equations \ref{Eq:MS-Hamiltonian-pre} and \ref{Eq:para_req1}, we get the exact form in Equation \ref{Eq:Qudit-MS}. This dictates the phases of the blue and red-detuned laser perturbations \footnote{For the case where the even numbered states are lower in energies while the odd number states energies are higher, i.e. \( E_{l+1}-E_{l}=(-1)^l\hbar\omega_l \), the spin phase is transformed to \( \phi_{S}=(-1)^{l}\left(\phi_{S,l}'+\frac{\pi}{2}\right) \).}
\begin{align}
    \phi_{b,l}+\phi_{r,l}&=-\pi
    \nonumber\\
    \phi_{b,l}&=\phi_{r,l}.
\end{align}

The time-evolution operator generated by the Hamiltonian of Equation \ref{Eq:Qudit-MS} is obtained by solving Schr\"{o}dinger's equation
\begin{equation} \label{Eq:Schrodinger time evolution}
    \frac{d\hat{U}}{dt}=-\frac{i}{\hbar}\mathcal{H}\hat{U}.
\end{equation}
We evaluate the time-evolution operator with the Magnus expansion
\begin{equation}
    \hat{U}(t)=e^{\sum_{k=1}^{\infty}M_{k}(t)},
\end{equation}
where \({M_{k}(t)}\) is the \({k^{th}}\) term in the Magnus expansion. For the Hamiltonian at hand, the generated Magnus expansions are
\begin{align}
\label{Eq:Magnus_expansions}
    M_{1}(t)&=-\frac{i}{\hbar} \int_0^t \mathcal{H}(t_1) \; dt_1 
    \nonumber\\
    &= \left(\alpha(t)\hat{a}^{\dagger}-\alpha^{*}(t)\hat{a}\right)\sum_{n=1}^{N}\hat{S}_{x,n}
    \nonumber\\
    M_{2}(t)&=-\frac{1}{2\hbar^2} \int_0^t \; dt_1 \int_0^{t_1} [\mathcal{H}(t_1),\mathcal{H}(t_2)] \; dt_2
    \nonumber\\
    &\hspace{-10px}= i \frac{\eta_C^2 \Omega^2}{\omega_C - \mu} \left( t - \frac{ \sin{\left( \left( \omega_C - \mu \right) t \right)} }{\omega_C - \mu} \right) \left( \sum_{n = 1}^N \hat{S}_{x,n} \right)^2
    \nonumber\\
    &\approx i\frac{\eta_C^2\Omega^2}{\omega_C-\mu}t\left(\sum_{n=1}^{N}\hat{S}_{x,n}\right)^2
    \nonumber\\
    M_{k}(t) &= 0 \; \text{for} \; k>2,
\end{align}
where \({\alpha(t)=\frac{\eta_{C}\Omega}{\omega_{C}-\mu}\left[1-e^{i(\omega_M-\mu)t}\right]}\). Here, we have neglected terms in \( M_2(t) \) which are bounded with \( t\). To minimize coupling to the phonon states (which is equivalent to minimizing \({M_{1}(t)}\) and closing the loop in the phase space picture in Figure \ref{fig:Entangling_gate_evolution}(a)) and obtain the desired entangling gate, we require
\begin{equation} \label{Eq:gate_time}
    t=K\frac{2\pi}{|\omega_{C}-\mu|},
\end{equation}
where \({K}\) is a positive integer.
The resultant unitary of the qudit entangling gate is then
\begin{equation} \label{Eq:MS-Unitary}
    \hat{U}=\exp{\left(\frac{2i\eta_C^2\Omega^2\pi}{(\omega_C-\mu)|\omega_{C}-\mu|}K\left(\sum_{n=1}^{N}\hat{S}_{x,n}\right)^2\right)}.
\end{equation}
The ion qudits in eigenstates of \({\hat{S}_{x}}\) after the gate in Equation \ref{Eq:MS-Unitary} gain phases of
\begin{align}
    \label{Eq:Geometric phase}
    \theta &= \theta_0\left(\sum_{n=1}^{N}\hat{\lambda}_{n}\right)^2
    \nonumber\\
    \theta_0 &= \frac{2 K \eta_C^2\Omega^2\pi}{(\omega_C-\mu)|\omega_{C}-\mu|},
\end{align}
where \({\lambda_{n}}\) is the eigenvalue of the \({n^{th}}\) ion with respect to \({\hat{S}_{x}}\). For a two-qudit gate, \({N=2}\), and the output is an entangled 2-qudit state in general.

In order to obtain the output state and thus the fidelity without making the second RWA and LDA approximations, the time evolution of an input state is simulated by numerical integration according to the differential equation in Equation \ref{Eq:Schrodinger time evolution} using the Hamiltonian in Equation \ref{Eq:MS-gate no approx}. For estimating error from spectator modes, simulations are done with and without coupling to other phonon modes, and the results are compared. 

In order to obtain the error due to imperfect cooling of ions, the input phonon state is modeled to be in a thermal state with phonon Fock state population distribution of \cite{Fox-2006}
\begin{equation}
    P_{n}=\frac{\Bar{n}^n}{(\Bar{n}+1)^{n+1}},
\end{equation}
where \({\Bar{n}}\) is the average phonon number. In order to obtain a crude (over)estimation of the error due to motional heating of ions during the gate, the phonon state of the motional mode is increased by one when the phase space displacement is maximal, from which we compute \({F_{heat}}\). The overall fidelity is then
\begin{equation}
    \mathcal{F}=(1-P_{heat})\mathcal{F}_{0}+P_{heat}\mathcal{F}_{heat},
\end{equation}
where \({P_{heat}}\) is the probability that a phonon hop happens due to motional heating from the environment, \({F_{0}}\) is the fidelity when no phonon hop happens, and \({F_{heat}}\) is the fidelity when a phonon hop has happened.

Since it is more computationally intensive to implement a time-varying magnetic field noise for the simulation of an entangling gate, we obtain an estimate of the error introduced by magnetic field noise by setting a constant magnetic field offset error of \(\SI{2.7}{pT}\). This modifies the original Hamiltonian to
\begin{equation}
    \mathcal{H}=\mathcal{H}_{orig}+\sum_{l=0}^{d-1} \Delta E_{l} \ketbra{l}{l},
\end{equation}
where \({\mathcal{H}_{orig}}\) is the original Hamiltonian and \({\Delta E_{l}}\) is the energy shift of state \({\ket{l}}\) due to magnetic field shift.

Photon scattering is another significant source of error for the qudit MS gate. A Raman scattering event immediately decoheres a qudit state. Although a Rayleigh scattering event does not directly decohere the quantum states, momentum transfer due to photon recoil still introduces some error during MS gate \cite{Ozeri-2007}. Accounting for the scattering probability for two ions, the gate fidelity with photon scattering error is
\begin{equation} \label{Eq: F_scatter expression}
    \mathcal{F}_{scatter} = \left( 1 - P_{Raman} - P_{Rayleigh} \varepsilon_{recoil}\right)^2
\end{equation}
where \( P_{Rayleigh} \) is the probability of a Rayleigh scattering event and \( \varepsilon_{recoil} \) is the error introduced due to photon recoil when a Rayleigh scattering event has occurred. The Rayleigh scattering photon recoil error generalized to qudits from Ref. \cite{Ozeri-2007} is
\begin{equation}
    \varepsilon_{recoil} = \frac{5 \left( d-1 \right)^2}{6 \pi} \eta_C^2 \theta_0
\end{equation} 
\( P_{Raman} \) is computed using Equation \ref{Eq: Scattering probability RRamantg}. \( P_{Rayleigh} \) is obtained from
\begin{equation}
    P_{Rayleigh}=R_{Rayleigh}t_g
\end{equation}
For the case of a 3-level qudit system, the photon scattering probability is calculated with the states \({\ket{0}}\), \({\ket{1}}\) and \({\ket{2}}\) in Figure \ref{fig:Qudit-Encoding}(a) being the 3 computational qudit states. The total photon scattering rate (for any encoded state) in this case is derived to be
\begin{align}
    \label{Eq:3-level MS total Scatter}
    R_{total}&=\frac{10}{\sqrt{6}} \bar{\Omega} \left( \frac{\Delta_{1/2} \Delta_{3/2}}{\Delta_{1/2} - \Delta_{3/2}} \right) \times
    \nonumber\\
    &\Bigg[ \gamma_{P_{1/2} \rightarrow S_{1/2}} \frac{\omega_R^3}{\omega_{P_{1/2} \rightarrow S_{1/2}}^3} \left( \frac{1}{\Delta_{1/2}^2} + \frac{2}{\Delta_{3/2}^2} \right) 
    \nonumber\\
    &+ \gamma_{P_{1/2} \rightarrow D_{3/2}} \frac{(\omega_R-\omega_{D_{3/2}})^3}{\omega_{P_{1/2} \rightarrow D_{3/2}}^3} \frac{1}{\Delta_{1/2}^2} 
    \nonumber\\
    &+ \gamma_{P_{3/2} \rightarrow D_{3/2}} \frac{(\omega_R-\omega_{D_{3/2}})^3}{\omega_{P_{3/2} \rightarrow D_{3/2}}^3} \frac{2}{\Delta_{3/2}^2} 
    \nonumber\\
    &+ \gamma_{P_{3/2} \rightarrow D_{5/2}} \frac{(\omega_R-\omega_{D_{5/2}})^3}{\omega_{P_{3/2} \rightarrow D_{5/2}}^3} \frac{2}{\Delta_{3/2}^2} \Bigg],
\end{align}
For the Rayleigh scattering rate, the state with the smallest rate is chosen to maximize Raman scattering probability to obtain a conservative error estimate. This Rayleigh scattering rate for 3-level qudits is derived to be
\begin{align}
    R_{Rayleigh} &= 5 \sqrt{6} \bar{\Omega} \left( \frac{\Delta_{1/2} \Delta_{3/2}}{\Delta_{1/2} - \Delta_{3/2}} \right) \gamma_{P_{1/2} \rightarrow S_{1/2}} \times
    \nonumber\\
    &\frac{\omega_R^3}{\omega_{P_{1/2} \rightarrow S_{1/2}}^3} \left( \frac{1}{3}\frac{1}{\Delta_{1/2}} + \frac{2}{3} \frac{1}{\Delta_{3/2}} \right)^2.
\end{align}

With the 5-level qudit in the zig-zag encoding, the total and Rayleigh photon scattering rates are derived to be
\begin{align}
    \label{Eq:5-level MS total Scatter}
    R_{total}&=\frac{49 \sqrt{6}}{9} \bar{\Omega} \left( \frac{\Delta_{1/2} \Delta_{3/2}}{\Delta_{1/2} - \Delta_{3/2}} \right) \times
    \nonumber\\
    &\Bigg[ \gamma_{P_{1/2} \rightarrow S_{1/2}} \frac{\omega_R^3}{\omega_{P_{1/2} \rightarrow S_{1/2}}^3} \left( \frac{1}{\Delta_{1/2}^2} + \frac{2}{\Delta_{3/2}^2} \right) 
    \nonumber\\
    &+ \gamma_{P_{1/2} \rightarrow D_{3/2}} \frac{(\omega_R-\omega_{D_{3/2}})^3}{\omega_{P_{1/2} \rightarrow D_{3/2}}^3} \frac{1}{\Delta_{1/2}^2} 
    \nonumber\\
    &+ \gamma_{P_{3/2} \rightarrow D_{3/2}} \frac{(\omega_R-\omega_{D_{3/2}})^3}{\omega_{P_{3/2} \rightarrow D_{3/2}}^3} \frac{2}{\Delta_{3/2}^2} 
    \nonumber\\
    &+ \gamma_{P_{3/2} \rightarrow D_{5/2}} \frac{(\omega_R-\omega_{D_{5/2}})^3}{\omega_{P_{3/2} \rightarrow D_{5/2}}^3} \frac{2}{\Delta_{3/2}^2} \Bigg],
\end{align}
\begin{align}
    R_{Rayleigh} &= \frac{49 \sqrt{6}}{3} \bar{\Omega} \left( \frac{\Delta_{1/2} \Delta_{3/2}}{\Delta_{1/2} - \Delta_{3/2}} \right) \gamma_{P_{1/2} \rightarrow S_{1/2}} \times
    \nonumber\\
    &\frac{\omega_R^3}{\omega_{P_{1/2} \rightarrow S_{1/2}}^3} \left( \frac{1}{3}\frac{1}{\Delta_{1/2}} + \frac{2}{3} \frac{1}{\Delta_{3/2}} \right)^2,
\end{align}
With photon scattering error, the fidelity is then transformed as
\begin{equation}
    \mathcal{F} \rightarrow \mathcal{F} \left( 1 - P_{Raman} - P_{Rayleigh} \varepsilon_{recoil}\right)^2
\end{equation}
The details for these derivations are available in the Appendix \ref{app:Photon_Scatt}.
\setlength\tabcolsep{5pt}
\begin{table}
\centering
\begin{tabular}{ccc}
\toprule\toprule
\textbf{Error Source} & \textbf{3-level} & \textbf{5-level}\\
\midrule
LDA & \({3 \times 10^{-4}}\) & \({3.0 \times 10^{-3}}\)\\
RWA & \({4 \times 10^{-4}}\) & \({2.6 \times 10^{-3}}\)\\
Spectator mode & \({2.7 \times 10^{-3}}\) & \({1.10 \times 10^{-2}}\)\\
Photon scattering* & \({7 \times 10^{-4}}\) & \({2.4 \times 10^{-3}}\)\\
Imperfect cooling & \({< 10^{-4}}\) & \({< \times 10^{-4}}\)\\
Motional heating & \({3.3 \times 10^{-3}}\) & \({4.6 \times 10^{-3}}\)\\
Magnetic field noise & \({<10^{-4}}\) & \({<10^{-4}}\) \\ 
\bottomrule\bottomrule
\end{tabular}
\caption{Error estimate from error sources for the qudit entangling gate. Each error estimate except for photon scattering is obtained by the increase in fidelity when the error source is removed from the simulation. *Error for photon scattering listed here is computed from \(1-\mathcal{F}_{scatter}\), where \(\mathcal{F}_{scatter}\) is fidelity with photon scattering and is defined in Equation \ref{Eq: F_scatter expression}. **The error estimates for \({d=5}\) listed here are obtained for the case without the large error from off-resonant frequencies (see text).}
\label{tab:SimFidelities}
\end{table}

 The fidelity obtained with all the error sources taken into consideration for \({d=3}\) is \({\mathcal{F}=0.9932}\). For \({d=5}\), off-resonant transition frequencies distorts the Hamiltonian significantly from the encoding scheme in Figure \ref{fig:Qudit-Encoding}(c), and results in a fidelity much smaller than 1, which is \( \mathcal{F} = 0.0296 \). We note that this error is present due to symmetry of the zig-zag encoding scheme that we have chosen, and may be overcome with other encoding schemes. Neglecting this error, an overall fidelity of \({\mathcal{F}=0.9789}\) is obtained for \({d=5}\) with these parameters. From Table \ref{tab:SimFidelities}, the spectator phonon mode and motional heating of ions are the major sources of error. 
 To reduce the error due to a spectator phonon mode, a direct way is to tune the trap parameters such that the spectator mode is detuned farther from the desired phonon mode frequency. This would reduce the contribution to the state evolution from the spectator modes. To eliminate the spectator mode contribution without the need to tune the trap parameters, clever pulse shaping could be performed which removes spin-phonon coupling of spectator modes, which is shown for the qubit case \cite{Choi-et-al-2014}. Assuming that spectator mode error can be eliminated by clever pulse shaping techniques, the fidelity for this 3-level qudit entangling gate can be increased to \({\mathcal{F}=0.9959}\). Neglecting the error due to off-resonant frequencies again, the fidelity for the 5-level qudit entangling gate is \( \mathcal{F}=0.9899 \) if the error from spectator mode can be overcome. 
 
 It is also experimentally relevant to estimate the thresholds of magnetic field noise such that the error from this source starts to become significant. From Table \ref{tab:SimFidelities}, error values from most other error sources are in the order of at least \(10^{-4}\). To be below this benchmark, the requirements on the standard deviations of the magnetic field noise can be relaxed to \( \sqrt{\langle \Delta B^2 \rangle} = \SI{1.608}{nT} \) and \( \sqrt{\langle \Delta B^2 \rangle} = \SI{0.804}{nT} \) for \( d=3 \) and \( d=5 \) respectively. 
 
 \begin{comment}Reducing the error from LDA can be achieved by reducing the resonant Rabi frequency and thus increasing the number of cycles in the phase space i.e. increasing \({K}\) in Equation \ref{Eq:gate_time}.\end{comment}
 
 Overall, it is possible to achieve more than \({99\%}\) for 3-level qudits with this generalized entangling gate. For \({d \geq 5}\), this gate is not applicable for our specific encoding scheme using \({^{137}\mathrm{Ba}^+}\) due to error from off-resonant frequencies.

\section{Summary}
\label{Sec:Summary}
\setlength\tabcolsep{3pt}
\begin{table}
\centering
\begin{tabular}{ccc}
\toprule\toprule
\textbf{Error Source} & \textbf{3-level} & \textbf{5-level}\\
\midrule
State preparation & \multicolumn{2}{c}{\({\geq 99.9\%}\)} \\
Measurement &${99.59\%}$ & $99.07\%$\\
Single-Qudit gates & \({>99.99\%}\)& \({>99.9\%}\)\\
Two-Qudit gates &\({99.32\%}\) & \({2.96\%}\) \\
\bottomrule\bottomrule
\end{tabular}
\caption{\label{tab:ErrorSummary}Overall error budget for qudits using \({^{137}\mathrm{Ba}^+}\). By controlling the polarization for optical pumping and using additional microwave pulses, \cite{Harty-et-al-2014} was able to achieve \({99.93\%}\) fidelity state preparation. Other numbers are from simulations described in this paper.}
\end{table}
In this paper, we have described a suitable operation set to implement qudit-based quantum computing. We discussed how to satisfy all of DiVencenzo's requirements for quantum information processing with qudits. Using the hyperfine sub-levels of a trapped ion, we are able to encode the different qudit states such that we can arbitrarily create any superposition state for the system. Standard optical pumping can be used to initialize our qudits reliably. An optical shelving method using a metastable state was discussed in detail, which allows us to measure the state of the qudit with low error. Finally, we presented a M\o{}lmer-S\o{}rensen-like entangling gate which, along with single qudit gates, allows us to create any arbitrary entangled quantum state in our qudits \cite{Brylinski-2001}. With these conditions satisfied, our proposed trapped ion system can be considered a universal quantum computer. 

As a comparison for our operations' fidelities, we use the \({99.25\%}\) fault-tolerant error threshold found for qubits \cite{Fowler-et-al-2012} on the grounds that there is significant evidence that qudit-based codes will have more relaxed thresholds \cite{Campbell-2014}. Table \ref{tab:ErrorSummary} lays out all of the error sources considered for our qudit platform. The simulation codes for the error estimations presented in this paper can be found in the repository \cite{GIT}. We acknowledge as well that this is not an exhaustive study; more details could be included such as noise from Rabi rate fluctuations. However, our main goal is to show that there are no fundamental roadblocks towards qudit implementations, and we have taken measures to ensure the errors considered are upper bounds for this study. For qutrits, we find no fundamental obstacles to achieving this error threshold. For 5-level qudits, more work needs to be done to improve the entangling gate, but if we succeed in overcoming the parasitic coupling, these gates could be done with a fidelity of at least \({97.89\%}\).

\begin{acknowledgments}
This research was supported in part by the Natural Sciences and Engineering Research Council of Canada (NSERC) and the Canada First Research Excellence Fund (CFREF).
\end{acknowledgments}

\newpage
\appendix
\section{Measurement}
\label{app:Measurement}
\begin{figure*}[t!]
    \centering
    \subfloat[\label{fig:XZ-1130MHz}]{\includegraphics[width=\linewidth]{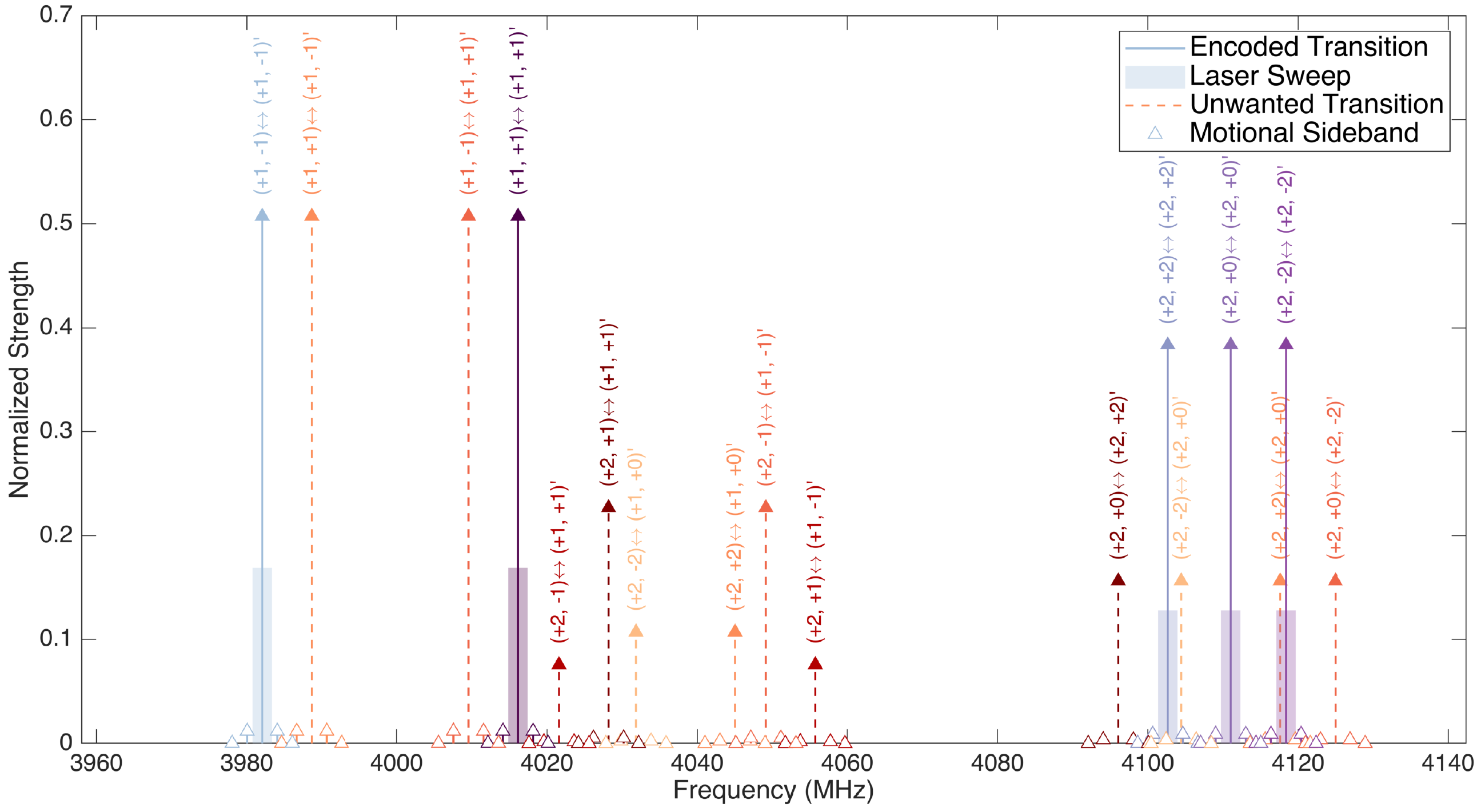}}
    \caption{Absolute frequency spectrum for shelving 5-level qudits. Shaded areas show the laser sweep with \(\SI{1.3}{MHz}\) detuning. We use a -1130 MHz laser carrier frequency. Text on top of transitions are denoted as \((F, m_F)\leftrightarrow (F, m_F)'\), where the prime denotes the shelving state.}
\end{figure*}
\begin{figure}
    \centering\label{fig:5-Level_Shelving-Sequence_All}
    \includegraphics[width=\linewidth]{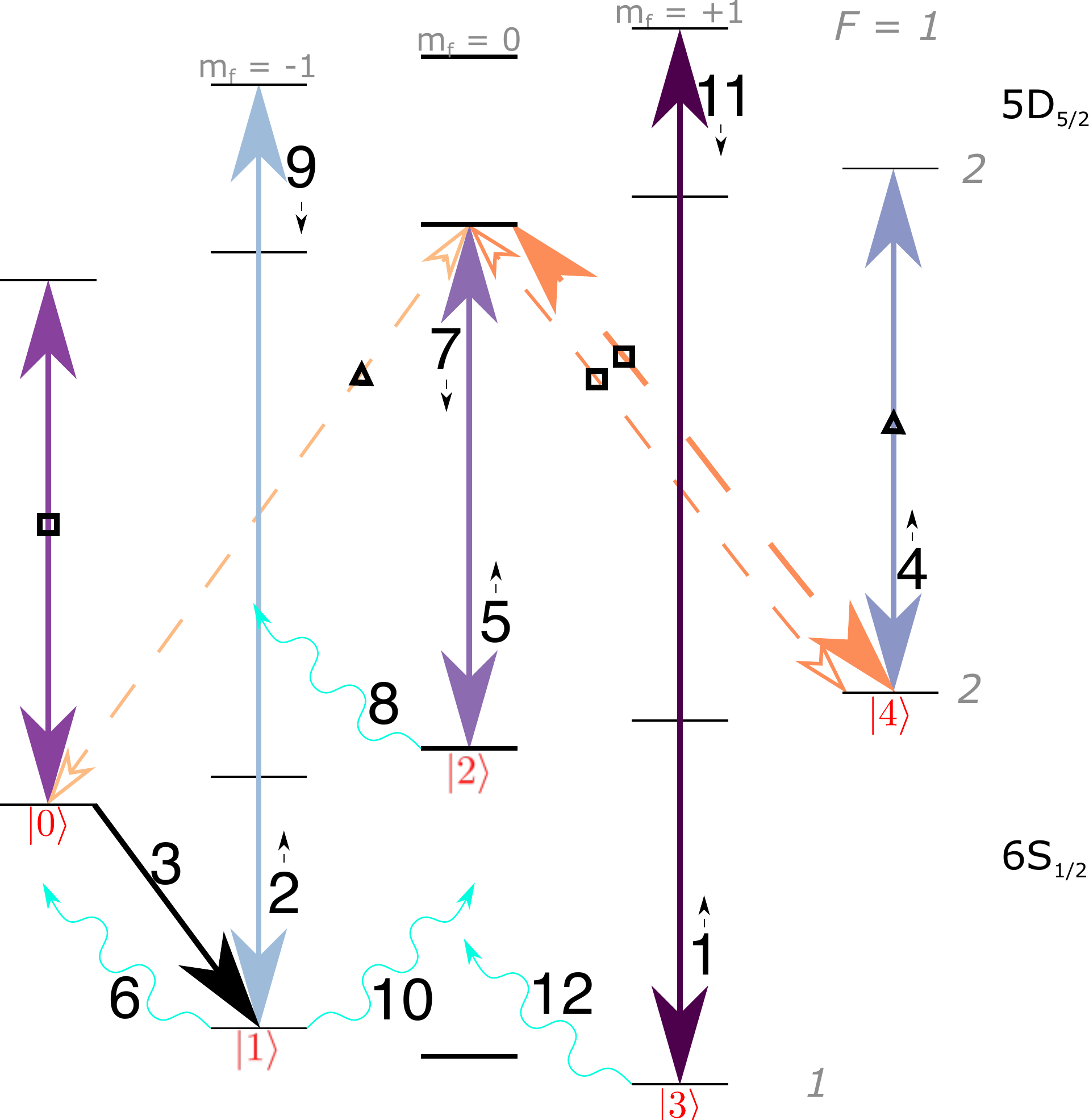}
    \caption{Measurement sequence for a 5-level qudit. Solid blue/purple straight arrows are desired transitions and dashed red/orange arrows are unwanted transitions. Black straight arrows are microwave or Raman transitions used to hide population. Line sizes denote the order of the transition: carrier transitions are thickest (left-most purple arrow) and first order are thinner (orange arrow driving from \(\ket{0}\)). Non-carrier transitions also have white arrows. Symbols on lines denote that a transition's frequency is within an encoded transition's laser frequency sweep (with the same symbol) and it will be driven. Large black numbers denote the step number and the small arrows pointing up (down) denote shelving (Deshelving) and hiding. Red vector states denote the encoded states. Wiggly teal arrows denote fluorescence measurement of an encoded state.}
    \label{fig:my_label}
\end{figure}
The shelving transition from \(S_{1/2}\) to \(D_{5/2}\) is a quadrupole transition with a wavelength of \(\SI{1762}{nm}\). As shown in reference \cite{Roos-2000}, Section 3.4.3., we can suppress some transitions while retaining others by choosing different directions and polarizations for this laser. There are two useful orientations: one orientation suppresses all \(\Delta m_F = \pm 1\) transitions completely, while reducing the strength of \(\Delta m_F = \pm 2\) transitions, and retaining the \(\Delta m_F = 0\) transitions; we call this the XZ orientation. Another orientation suppresses everything but \(\Delta m_F = \pm 2\) transitions; we call this the orthogonal orientation. While using the orthogonal orientation cleans up the frequency spectrum, we chose to use the XZ orientation for the simpler mapping to the shelving state: \(F, m_F \rightarrow F' = F, m_{F'} = m_F\).
\begin{table}[htb!]
    \centering
    \begin{tabular}{cl}
    \toprule\toprule
        Step & Desired Transition\\\midrule
        1 & Shelve \((1, +1)\rightarrow(1, +1)'\)\\
        2 & Shelve \((1, -1)\rightarrow(1, -1)'\)\\
        3 & Hide \((2, -2)\rightarrow(1, -1)\)\\
        4 & Shelve\((2, +2)\rightarrow(2, +2)'\)\\
        5 & Shelve \((2, 0)\rightarrow(2, 0)'\)\\
        6 & Fluoresce \(\ket{0}\) state \\
        7 & Deshelve \((2, 0)'\rightarrow(2, 0)\)\\
        8 & Fluoresce \(\ket{2}\) state \\
        9 & Deshelve \((1, -1)'\rightarrow(1, -1)\)\\
        10 & Fluoresce \(\ket{1}\) state \\
        11 & Deshelve \((1, +1)'\rightarrow(1, +1)\)\\
        12 & Fluoresce \(\ket{3}\) state \\\bottomrule\bottomrule
    \end{tabular}
    \caption{The steps for measuring a 5-level qudit. Transitions are denoted by \((F, m_F)\), and apostrophes denote states within the \(D_{5/2}\) shelving manifold. Shelving/deshelving is done by adiabadic passage on the \(\SI{1762}{nm}\) transition, hiding is done by a microwave or Raman transition, and fluorescence is driven on the \(\SI{493}{nm}\) \(S_{1/2}\leftrightarrow P_{1/2}\) transition.}
    \label{tab:5-Level-Sequence}
\end{table}
\begin{figure*}
    \centering
    \begin{minipage}[b]{.49\textwidth}
    \subfloat[\label{fig:7-Level_Shelving-Sequence_Freqs}]{\includegraphics[width=\linewidth]{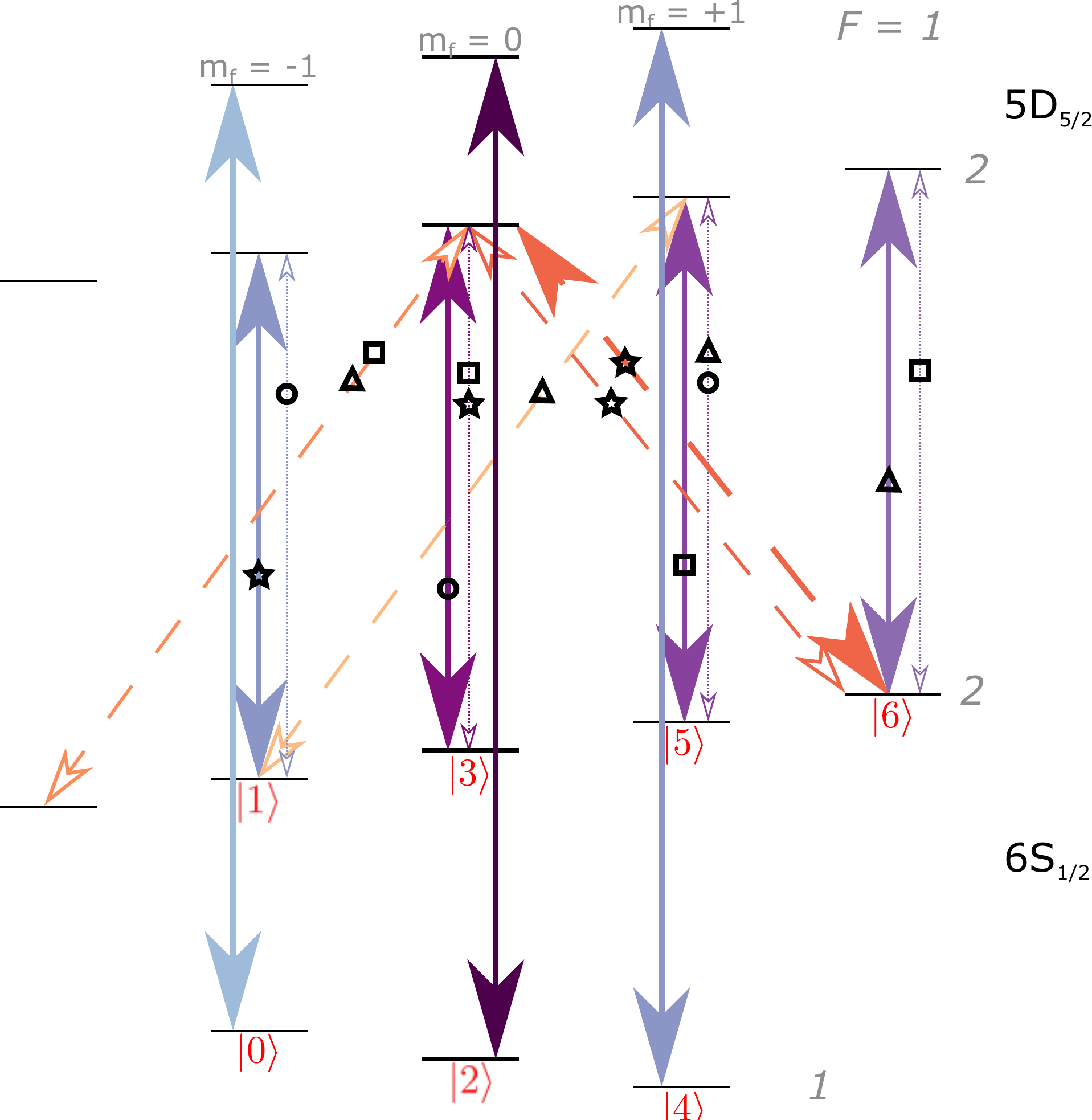}}
    \end{minipage}
    \begin{minipage}[b]{.49\textwidth}
    \subfloat[\label{fig:7-Level_Shelving-Sequence_Shelving}]{\includegraphics[width=\linewidth]{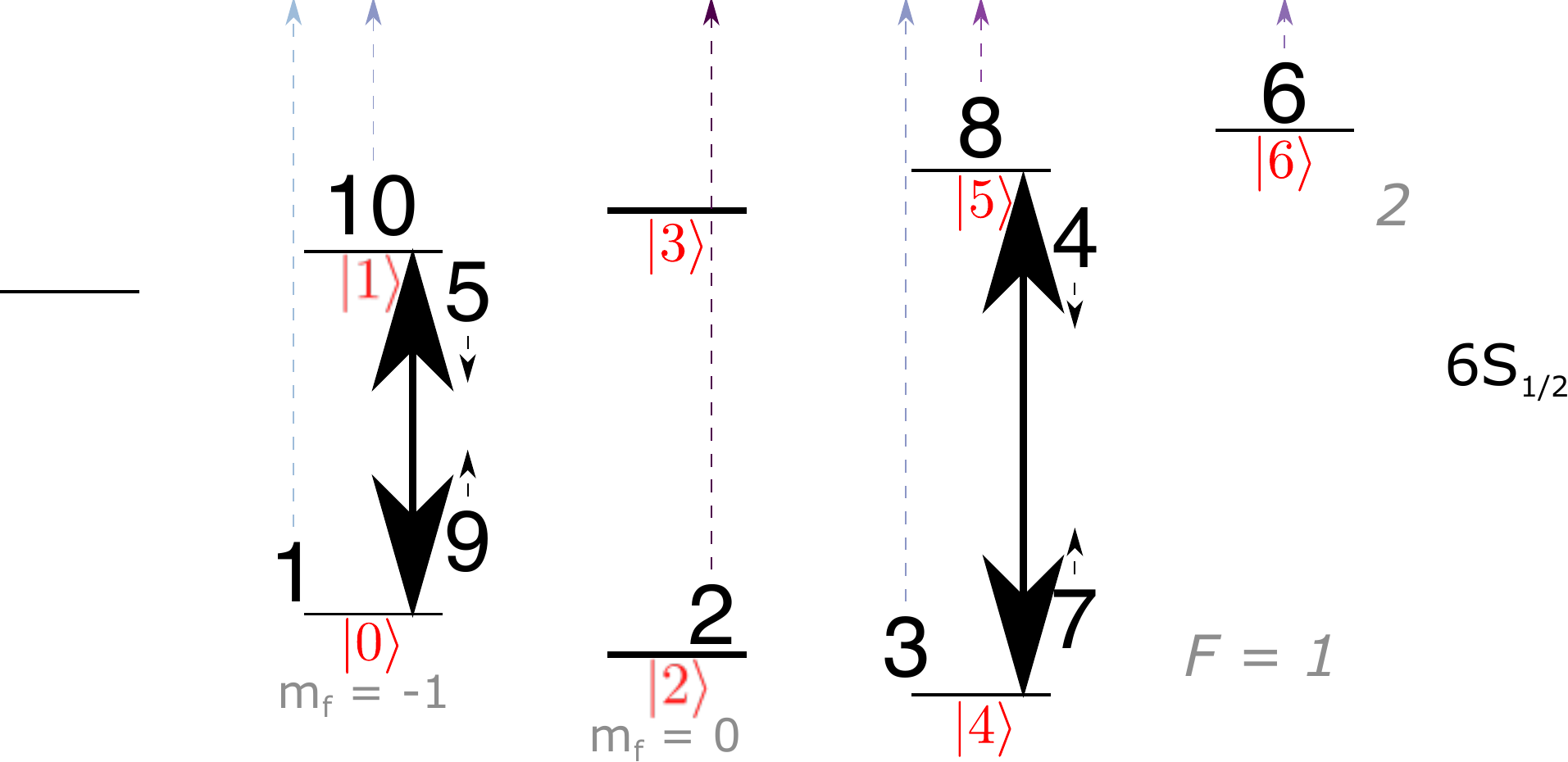}}
    \\
    \subfloat[\label{fig:7-Level_Shelving-Sequence_Deshelving}]{\includegraphics[width=\linewidth]{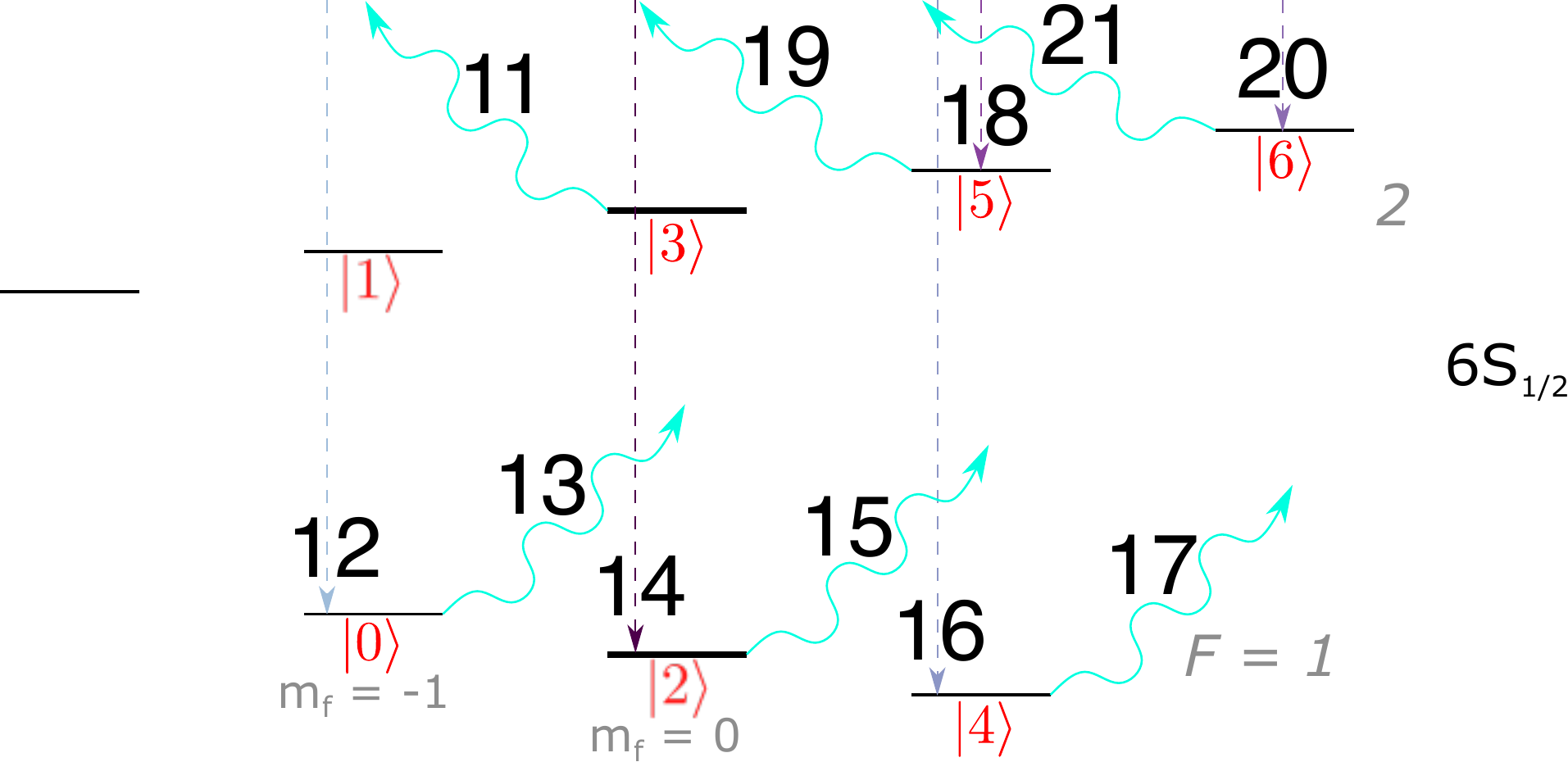}}
    \end{minipage}
    \caption{\label{fig:7-Level_Shelving-Sequence}Measurement sequence for a 7-level qudit. Red vector states denote the encoded states. (a) Solid blue/purple arrows are desired transitions and dashed red/orange arrows are unwanted transitions. The line sizes denote the order of the transition: carrier transitions are thickest (left most purple arrow) and first order are thinner (orange arrow driving from \(\ket{0}\)). Non-carrier transitions also have white arrows. Symbols on lines denote that a transition's frequency is within an encoded transition's laser frequency sweep (with the same symbol) and it will be driven. (b, c) Large black numbers denote the step number and small arrows pointing up/down denote moving population, either shelving, deshelving, or hiding. (b) Black transitions are microwave or Raman transitions used to hide population. (c) Wiggly teal arrows denote fluorescence measurement of an encoded state.}
\end{figure*}
\begin{table}[htb!]
    \centering
    \begin{tabular}{clc}
    \toprule\toprule
        Step & Desired Transition & Other Transitions\\\midrule
        1 & Shelve \((1, -1)\rightarrow(1, -1)'\) & \\
        2 & Shelve \((1, 0)\rightarrow(1, 0)'\) & \\
        3 & Shelve \((1, +1)\rightarrow(1, +1)'\) & \\
        4 & Hide \((2, +1)\rightarrow(1, +1)\) & \\
        5 & Hide \((2,-1)\rightarrow(1, -1)\) & \\
        6 & Shelve \((2, +2)\rightarrow(2, +2)'\) & \\
        7 & Return \((1,+1)\rightarrow(2, +1)\) & \\
        \multirow{2}{*}{8} & \multirow{2}{*}{Shelve \((2, +1)\rightarrow(2, +1)'\)} & \((2, 0)\leftrightarrow(2, 0)'\)\\ 
        & & \((2, +2)\leftrightarrow(2, +2)'\)\\
        9 & Return \((1,-1)\rightarrow(2, -1)\) & \\
        10 & Shelve \((2, -1)\rightarrow(2, -1)'\) & \((2, 0)\leftrightarrow(2, 0)'\)\\
        11 & Fluoresce \(\ket{3}\) state & \\
        12 & Deshelve \((1, -1)'\rightarrow(1, -1)\) & \\
        13 & Fluoresce \(\ket{0}\) state & \\
        14 & Deshelve \((1, 0)'\rightarrow(1, 0)\) & \\
        15 & Fluoresce \(\ket{2}\) state & \\
        16 & Deshelve \((1, +1)'\rightarrow(1, +1)\) & \\
        17 & Fluoresce \(\ket{4}\) state & \\
        18 & Deshelve \((2, +1)'\rightarrow(2, +1)\) & \((2, +2)\leftrightarrow(2, +2)'\)\\
        19 & Fluoresce \(\ket{5}\) state & \\
        20 & Deshelve \((2, +2)'\rightarrow(2, +2)\) & \\
        21 & Fluoresce \(\ket{6}\) state & \\
        \\\bottomrule\bottomrule
    \end{tabular}
    \caption{The steps for measuring a 7-level qudit. Transitions are denoted by \((F, mF)\), and apostrophes denote states within the \(D_{5/2}\) shelving manifold. Shelving/deshelving is done by adiabadic passage on the \(\SI{1762}{nm}\) transition, hiding is done by a microwave or Raman transition, and fluorescence is driven on the \(\SI{493}{nm}\) \(S_{1/2}\leftrightarrow P_{1/2}\) transition. The third column lists any undesired motional sideband transitions that are also driven by this transition.}
    \label{tab:7-Level-Sequence}
\end{table}

In order to successfully measure out a qudit state using our shelving scheme described in Section \ref{Sec:Measurement}, we must first come up with a way to modulate the laser frequency to drive all of the desired transitions. The range of transition frequencies is \(\sim\SI{8}{GHz}\) wide, the ground state hyperfine splitting in \(^{137}\mathrm{Ba}^+\). However, we can set the laser frequency near halfway between the frequencies and use an electro-optic modulator with \(\sim\SI{4}{GHz}\) modulation to create sidebands both above and below this laser carrier frequency, allowing us to drive all of the transitions.

\begin{comment}The next challenge is to pick the carrier frequency for the laser. We define the carrier frequency of this transition as the offset frequency from the fine-structure transition. \(F = 1\rightarrow F'=1\) transitions are grouped together, and \(F=2\rightarrow F'=2\) transitions are grouped together (as we change the laser frequency, the grouped transitions do not change frequency relative to one another). If we pick \(f_{carrier} = \SI{-1092}{MHz}\), we get the frequency spectrum shown in Figure \ref{fig:XZ-1092MHz}; the two groups are just barely separated from each other. However, some other \(\Delta F=1\) transitions lie near the desired transitions. By separating them further, as in Figure \ref{fig:XZ-1130MHz}, which uses a carrier frequency of \(f_{carrier} = \SI{-1130}{MHz}\), these unwanted transitions are no longer near desired transitions. In this case, the range of frequencies is \(\sim\SI{150}{MHz}\), which can easily be generated by an arbitrary waveform generator.\end{comment}

Figure \ref{fig:XZ-1130MHz} shows all of the relevant frequencies involved in the shelving transition for a carrier frequency of \(f_{carrier} = \SI{-1130}{MHz}\). Some transitions lying near our desired transitions are unavoidable. In addition, many of the motional sideband frequencies of these unwanted transitions lie within our adiabatic laser sweep. In fact, some of the unwanted carrier transitions themselves lie within the laser sweep. If we drove such an adiabatic transfer on the desired transition, the unwanted transition would also be adiabatically driven, resulting in additional error. For second-order motional sidebands, which are very weak, this error is quite small. However, we must avoid sweeping through carrier and first-order sidebands. The carrier frequency was chosen to minimize the number of undesired frequencies near encoded transitions and to thereby maximize the overall fidelity of a measurement. 

These transitions are only driven if our qudit actually contains population in one of the states involved in the transition. So by picking a particular order of shelving, we can reduce the number of unwanted transitions coherently driven. In addition, we can use microwaves or Raman lasers to transfer population around within the ground state. This can essentially hide the population, so it cannot be driven by unwanted transitions. The 3-level qudit with zig-zag encoding has no unwanted transitions lying within the laser sweep, but the 5-level zig-zag encoded qudit, as seen in Figure \ref{fig:XZ-1130MHz}, has several. Using the shelving sequence shown in Figure \ref{fig:5-Level_Shelving-Sequence_All}, we can avoid all of the unwanted transitions.

For the 7-level qudit, the frequency space is even more cluttered. Using the measurement sequence depicted in Figure \ref{fig:7-Level_Shelving-Sequence}, we can avoid most of these, driving only a total of 4 second-order motional sidebands.

\section{Single-Qudit gates}
\label{app:Single-Qudit}
\begin{figure}[htb]
    \centering
    \includegraphics[width=\linewidth]{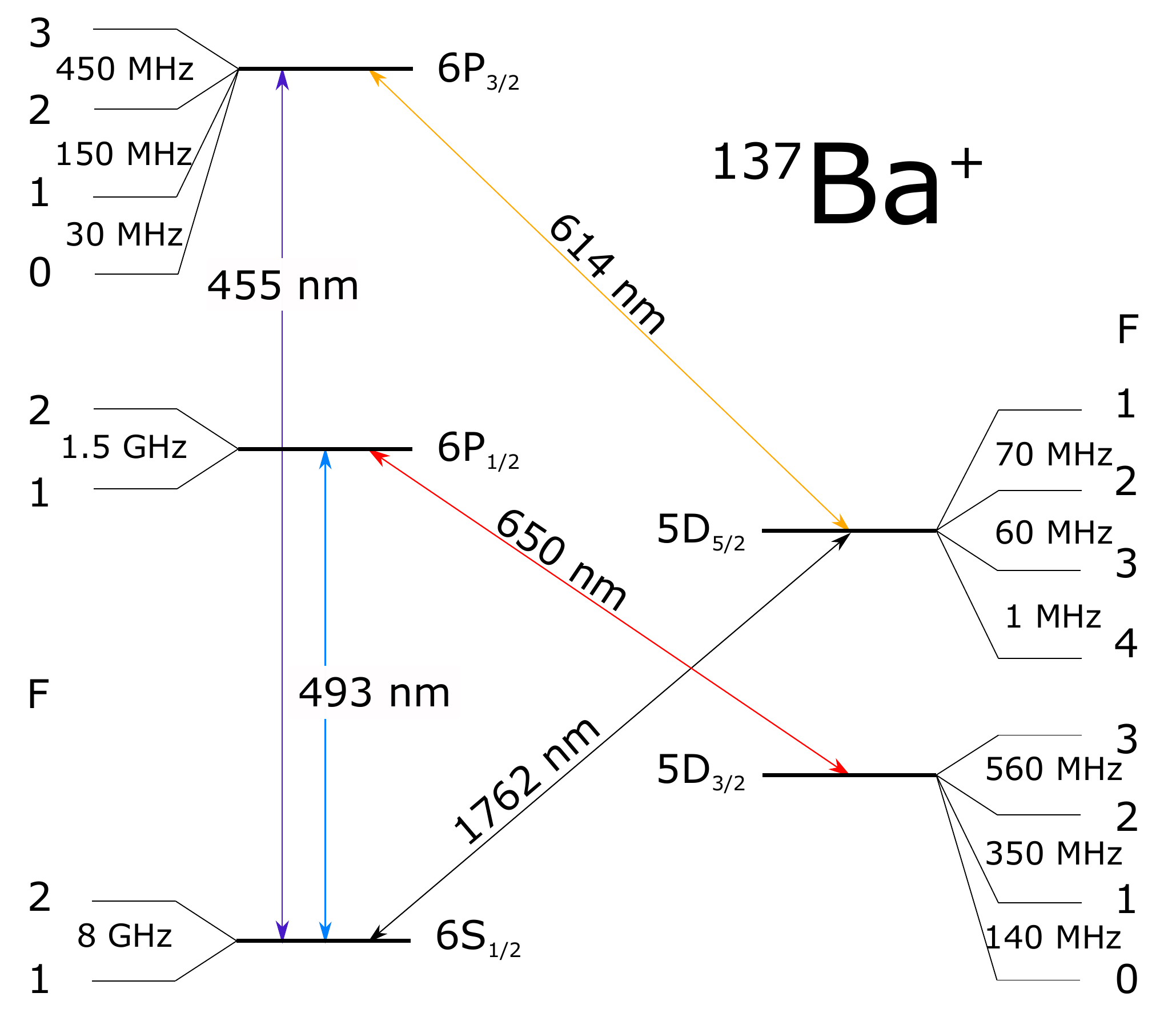}
    \caption{\label{fig:Ba137}\({^{137}\mathrm{Ba}^+}\) energy structure \cite{Curry-2004}. The \({6S_{1/2}\leftrightarrow 6P_{1/2}}\) optical transition is used for optical pumping, Doppler cooling, and fluorescence measurement. The \({6S_{1/2}\leftrightarrow5D_{5/2}}\) transition is used to shelve qudit states. The \({5D_{3/2}\leftrightarrow6P_{1/2}}\) transition is used to repump dark states back into the cooling/fluorescence cycle. The \({5D_{5/2}\leftrightarrow6P_{3/2}}\) transition is used to empty the \({5D_{5/2}}\) state. Because of its nuclear spin \({3/2}\), each level is split into hyperfine levels: the frequency differences between these levels are shown \cite{Blatt-Werth-1982,Silverans-et-al-1986,Villemoes-et-al-1993}}
\end{figure}
\subsection{Gate Decomposition} \label{app: single qudit gate decomposition}

The overall goal is to decompose an arbitrary d-dimensional unitary like
\begin{equation}
    \Hat{U} = \Hat{V}_K\Hat{V}_{K-1}\ldots\Hat{V}_1\hat{\Theta}\in SU(d),
    \label{eq:Unitery-DecompApp}
\end{equation}
where each operator \({\hat{V}_K}\) is a two-level pulse between adjacent energy levels \({\ket{k}}\) and \({\ket{k+1}}\) of the form
\begin{equation}
    \hat{V}_K= e^{-iC_K(e^{i\phi}\ket{k}\bra{k+1} + e^{-i\phi}\ket{k+1}\ket{k})},
    \label{Eq:Pulse}
\end{equation} 
for some phase \({\phi\in \mathbb{R}}\), and \({\hat{\Theta}}\) is a diagonal matrix of some arbitrary phases. For an arbitrary d-dimensional unitary \({\hat{U}\in SU(d)}\), define the unitary \({\hat{U}^{(0)} = e^{-i\Gamma/d}\hat{U}}\). Denote \({U^{(0)}_{i,j}}\) as the \(i^{\text{th}}\) row and \(j^{\text{th}}\) column entry of \({\hat{U}^{(0)}}\). The algorithm starts by finding an operator of the form in Equation \ref{Eq:Pulse} such that

\begin{equation}
    \hat{V}^\dagger_1\begin{pmatrix}
    U^{(0)}_{1, d}\\
    U^{(0)}_{2, d}\\
    \vdots
    \end{pmatrix} = 
    \begin{pmatrix}
    0\\
    C\\
    \vdots
    \end{pmatrix},
\end{equation}
for some number \({C}\). Defining the new operator \({\hat{U}^{(1)} = \hat{V}^\dagger_1\hat{U}^{(0)}}\), we continue finding pulses \({\hat{V}^\dagger_n}\) in this way until we end up with the last column decomposed as
\begin{equation}
    \hat{V}^\dagger_{d-1}\ldots\hat{V}^\dagger_1\begin{pmatrix}
    U^{(0)}_{1, d}\\
    \vdots\\
    U^{(0)}_{d, d}
    \end{pmatrix} = 
    \begin{pmatrix}
    0\\
    \vdots\\
    e^{i\theta_d}
    \end{pmatrix}.
\end{equation}

We repeat this process for each column, resulting in the diagonal matrix given by
\begin{equation}
    \hat{V}^\dagger_{d(d-1)/2}\ldots\hat{V}^\dagger_n\ldots\hat{V}^\dagger_1\hat{U}^{(0)} = \text{diag}(e^{i\theta_1},\ldots,e^{i\theta_d}),
\end{equation}
or
\begin{equation}
    \hat{U}^{(0)} = \hat{V}_{d(d-1)/2}\ldots\hat{V}_n\ldots\hat{V}_1\text{diag}(e^{i\theta_1},\ldots,e^{i\theta_d}).
\end{equation}
We define mappings \({\sigma: \mathbb{Z} \rightarrow \mathbb{Z}_d}\) and \({\tau:\mathbb{Z} \rightarrow \mathbb{Z}_d}\) such that at step \({K}\), \({\sigma(K)}\) gives the row to decompose and \({\tau(K)}\) gives the column to decompose. Forcing all pulses to be unitary with positive \({C_K}\), we have
\begin{equation}
    \hat{V}_K= e^{-iC_K(e^{i\phi_K}\ket{\sigma(K)}\bra{\sigma(K)+1} + e^{-i\phi_K}\ket{\sigma(K)+1}\ket{\sigma(K)})},
    \label{Eq:Pulse-Dec}
\end{equation}
where the pulse angles and phases are
\begin{align}
    C_K = \begin{cases}
    \text{cot}^{-1}\left(\frac{\left|U^{(n-1)}_{\sigma(K),\tau(K)}\right|}{\left|U^{(n-1)}_{\sigma(K)+1,\tau(K)}\right|}\right) & \left|U^{(n-1)}_{\sigma(K) + 1,\tau(K)}\right| \neq 0\nonumber\\
    0 & \left|U^{(n-1)}_{\sigma(K) + 1,\tau(K)}\right| = 0
    \end{cases}
    \nonumber\\
    \phi_K = \frac{\pi}{2} + arg\left(U^{(n-1)}_{\sigma(K),\tau(K)}\right) - arg\left(U^{(n-1)}_{\sigma(K) + 1,\tau(K)}\right).
\end{align}

To eliminate the phases \({e^{i\theta_j}}\) in each column j, we simply perform two pulse of the form of Equation \ref{Eq:Pulse-Dec} with pulse angle \({C_K = \pi/2}\) and phases \({\phi_K = -\pi/2 -\theta_j}\), \({\phi_K = \pi/2}\).
\begin{figure*}[t!]
    \centering
    \subfloat[\label{sfig:Yo1}]{\includegraphics[width=\textwidth]{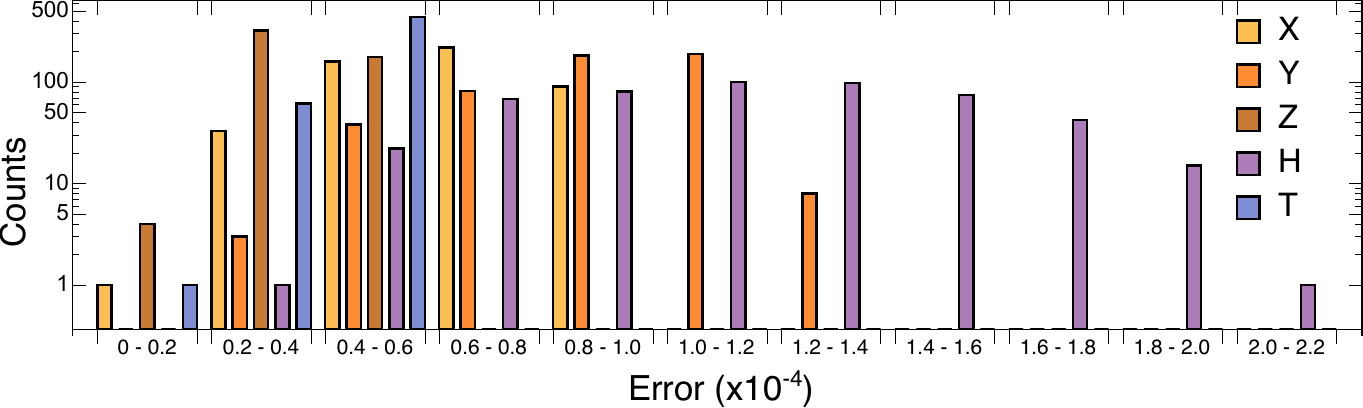}}
    \\
    \subfloat[\label{sfig:Yo2}]{\includegraphics[width=\textwidth]{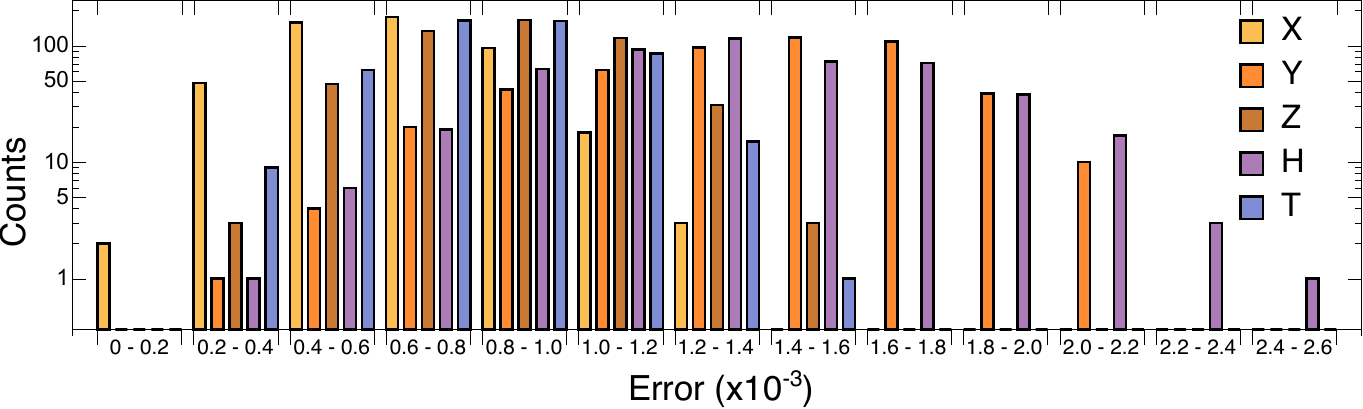}}
    \caption{Simulation histograms of gate errors for all Pauli gates (\(\hat{X}_d\), \(\hat{Y}_d\), \(\hat{Z}_d\)), the generalized Hadamard gate \(\mathcal{H}_d\), and \({\pi/8}\) gates (\(\hat{T}_d\)) \cite{Howard-Vala-2012} for (a) 3- and (b) 5-level qudits. Each gate has a sample size of 500 where the initial qudit state is a randomized superposition of the encoded states. A magnetic field offset is set at the standard deviation of \( \sqrt{ \vev{\Delta B (t)^2} } \approx \SI{2.7}{pT} \) (see supplementary material) and off-resonant coupling to other states are incorporated into the simulations. The Rabi frequency is set at \({\SI{10}{kHz}}\) for the data sets in this figure.}
    \label{fig:MagSimFidelities}
\end{figure*}
The overall maximum number of pulses needed to generate an arbitrary \({d\times d}\) unitary \({\hat{U}}\) with phases eliminated is \({(d-1)(d+4)/2}\).

Next, we show how to physically implement these pulses. If we apply a magnetic field \({B}\), our atomic Hamiltonian in the ground manifold becomes
\begin{align}
    \label{eq:Atomic-Hamiltonian}
    \mathcal{H}_A = \sum_{m_g = -F_g}^{F_g}&(E_g - g_{F}\mu_Bm_gB\ketbra{F_g, m_g}{F_g, m_g} 
    \nonumber\\
    +\sum_{m_e = -F_e}^{F_e}&(E_e + g_{F}\mu_Bm_eB\ketbra{F_e, m_e}{F_e, m_e},
\end{align}
where \({g}\) and \({e}\) correspond to the \({F=1, 2}\) hyperfine levels respectively, and \({E_i}\) is the energy of hyperfine level \({i}\). Driving a transition between connected qudit states \({m_g}\) and \({m_e}\) with a resonant microwave \({E(t) = E_0\cos{\omega t + \phi}}\) (\({\omega = \omega_e - \omega_g}\)), the perturbing Hamiltonian in the static Hamiltonian interaction picture looks like
\begin{align}
    \label{eq:Microwave-Interaction-Hamiltonian}
    \mathcal{H}_I = \sum_{q=-1}^1\frac{1}{2}&\left(e^{i\phi}\tilde{\Omega}_{m_g - m_e}^*\ketbra{F_g, m_g}{F_e, m_e} \right.
    \nonumber\\
    &\left.+ e^{-i\phi}\tilde{\Omega}_{m_g - m_e}\ketbra{F_e, m_e}{F_g, m_g}\right),
\end{align}
where
\begin{align}
    \label{eq:EffectiveRabi}
    \tilde{\Omega}_q &= 2(-1)^{F_e + J_g + I}\sqrt{(2F_e + 1)(2J_g + 1)}\bra{J_g}|\mathbf{\hat{d}}|\ket{J_e}    \nonumber\\
    &\times\sixj{J_e}{J_g}{1}{F_g}{f_e}{1}E_{0, q}\sum_{m_g, m_e}\bracket{F_g, m_g}{F_e, m_e;1, q}.
\end{align}
Note that some transitions have the same frequency; we could align our electromagnetic polarization to the magnetic field to select a single transition. Each decomposed pulse \({R[\theta,\rho]_{ij}}\) corresponds to the physical parameters \({\theta \leftrightarrow |\tilde{\Omega}_{m_i - m_j}|(t - t_0)/2}\), \({\rho \leftrightarrow \phi - arg(\tilde{\Omega}_{m_i - m_j})}\).

\subsection{Gate Library}
\label{app:Single-Qudit-Library}
\begin{table}[htb!]
    \centering
    \begin{tabular}{ccccc}
    \toprule\toprule
        \multicolumn{1}{c|}{Unitary} & \multicolumn{1}{c|}{Pulse} & \multicolumn{1}{c|}{Transition} & \multicolumn{1}{c|}{Pulse Angle, \(C\)} & Phase, \(\phi\)  \\ \midrule
        \multirow{3}{*}{\(\hat{X}_3\)} 
        & 1 & \(\ket{0}\leftrightarrow\ket{1}\) & \(\pi\) & \(0\) \\
        & 2 & \(\ket{1}\leftrightarrow\ket{2}\) & \(\pi/2\) & \(\pi/2\) \\
        & 3 & \(\ket{0}\leftrightarrow\ket{1}\) & \(\pi/2\) & \(\pi/2\) \\
        \midrule
        \multirow{4}{*}{\(\hat{Y}_3\)} 
        & 1 & \(\ket{0}\leftrightarrow\ket{1}\) & \(\pi/2\) & \(\pi/2\) \\
        & 2 & \(\ket{0}\leftrightarrow\ket{1}\) & \(\pi/2\) & \(7\pi/6\) \\
        & 3 & \(\ket{1}\leftrightarrow\ket{2}\) & \(\pi/2\) & \(\pi/2\) \\
        & 4 & \(\ket{0}\leftrightarrow\ket{1}\) & \(\pi/2\) & \(\pi/2\) \\
         \midrule
        \multirow{2}{*}{\(\hat{Z}_3\)} 
        & 1 & \(\ket{1}\leftrightarrow\ket{2}\) & \(\pi/2\) & \(\pi/2\) \\
        & 2 & \(\ket{1}\leftrightarrow\ket{2}\) & \(\pi/2\) & \(\pi/6\) \\
        \midrule
        \multirow{7}{*}{\(\hat{H}_3\)} 
        & 1 & \(\ket{0}\leftrightarrow\ket{1}\) & \(\pi/2\) & \(\pi/2\) \\
        & 2 & \(\ket{0}\leftrightarrow\ket{1}\) & \(\pi/2\) & \(3\pi/2\) \\
        & 3 & \(\ket{0}\leftrightarrow\ket{1}\) & \(\pi/4\) & \(5\pi/6\) \\
        & 4 & \(\ket{1}\leftrightarrow\ket{2}\) & \(\pi/2\) & \(\pi/2\) \\
        & 5 & \(\ket{1}\leftrightarrow\ket{2}\) & \(\pi/2\) & \(2\pi/3\) \\
        & 6 & \(\ket{1}\leftrightarrow\ket{2}\) & \(\arctan{\sqrt{2}}\) & \(7\pi/6\) \\
        & 7 & \(\ket{0}\leftrightarrow\ket{1}\) & \(\pi/4\) & \(7\pi/6\) \\
        \midrule
        \multirow{2}{*}{\(\hat{T}_3\)} 
        & 1 & \(\ket{1}\leftrightarrow\ket{2}\) & \(\pi/2\) & \(\pi/2\) \\ 
        & 2 & \(\ket{1}\leftrightarrow\ket{2}\) & \(\pi/2\) & \(31\pi/18\) \\
        \bottomrule\bottomrule
    \end{tabular}
    \caption[Three dimensional unitary decompositions]{Unitary decomposition for various three dimensional unitary gates of interest.}
    \label{tab:decomp_three}
\end{table}
\begin{table}[htb!]
    \centering
    \begin{tabular}{ccccc}
    \toprule\toprule
        \multicolumn{1}{c|}{Unitary} & \multicolumn{1}{c|}{Pulse} & \multicolumn{1}{c|}{Transition} & \multicolumn{1}{c|}{Pulse Angle, \(C\)} & Phase, \(\phi\)  \\ \midrule
        \multirow{6}{*}{\(\hat{X}_5\)} 
        & 1 & \(\ket{0}\leftrightarrow\ket{1}\) & \(\pi\) & 0 \\ 
        & 2 & \(\ket{2}\leftrightarrow\ket{3}\) & \(\pi\) & 0 \\ 
        & 3 & \(\ket{3}\leftrightarrow\ket{4}\) & \(\pi/2\) & \(\pi/2\) \\ 
        & 4 & \(\ket{2}\leftrightarrow\ket{3}\) & \(\pi/2\) & \(\pi/2\) \\ 
        & 5 & \(\ket{1}\leftrightarrow\ket{2}\) & \(\pi/2\) & \(\pi/2\) \\ 
        & 6 & \(\ket{0}\leftrightarrow\ket{1}\) & \(\pi/2\) & \(\pi/2\) \\
        \midrule
        \multirow{10}{*}{\(\hat{Y}_5\)} 
        & 1 & \(\ket{0}\leftrightarrow\ket{1}\) & \(\pi/2\) & \(\pi/2\) \\
        & 2 & \(\ket{0}\leftrightarrow\ket{1}\) & \(\pi/2\) & \(9\pi/10\) \\ 
        & 3 & \(\ket{1}\leftrightarrow\ket{2}\) & \(\pi/2\) & \(\pi/2\) \\ 
        & 4 & \(\ket{1}\leftrightarrow\ket{2}\) & \(\pi/2\) & \(7\pi/10\) \\
        & 5 & \(\ket{2}\leftrightarrow\ket{3}\) & \(\pi/2\) & \(\pi/2\) \\
        & 6 & \(\ket{2}\leftrightarrow\ket{3}\) & \(\pi/2\) & \(9\pi/10\) \\ 
        & 7 & \(\ket{3}\leftrightarrow\ket{4}\) & \(\pi/2\) & \(\pi/2\) \\ 
        & 8 & \(\ket{2}\leftrightarrow\ket{3}\) & \(\pi/2\) & \(\pi/2\) \\ 
        & 9 & \(\ket{1}\leftrightarrow\ket{2}\) & \(\pi/2\) & \(\pi/2\) \\ 
        & 10 & \(\ket{0}\leftrightarrow\ket{1}\) & \(\pi/2\) & \(\pi/2\) \\
        \midrule
        \multirow{6}{*}{\(\hat{Z}_5\)} 
        & 1 & \(\ket{1}\leftrightarrow\ket{2}\) & \(\pi/2\) & \(\pi/2\) \\
        & 2 & \(\ket{1}\leftrightarrow\ket{2}\) & \(\pi/2\) & \(19\pi/10\) \\
        & 3 & \(\ket{2}\leftrightarrow\ket{3}\) & \(\pi/2\) & \(\pi/2\) \\
        & 4 & \(\ket{2}\leftrightarrow\ket{3}\) & \(\pi/2\) & \(7\pi/10\) \\
        & 5 & \(\ket{3}\leftrightarrow\ket{4}\) & \(\pi/2\) & \(\pi/2\) \\
        & 6 & \(\ket{3}\leftrightarrow\ket{4}\) & \(\pi/2\) & \(19\pi/10\) \\
        \midrule
        \multirow{18}{*}{\(\hat{H}_5\)} 
        & 1 & \(\ket{0}\leftrightarrow\ket{1}\) & \(\pi/2\) & \(\pi/2\) \\
        & 2 & \(\ket{0}\leftrightarrow\ket{1}\) & \(\pi/2\) & \(3.30265\) \\
        & 3 & \(\ket{0}\leftrightarrow\ket{1}\) & \(\pi/4\) & \(0.63627\) \\
        & 4 & \(\ket{1}\leftrightarrow\ket{2}\) & \(\pi/2\) & \(\pi/2\) \\
        & 5 & \(\ket{1}\leftrightarrow\ket{2}\) & \(\pi/2\) & \(6.18626\) \\
        & 6 & \(\ket{1}\leftrightarrow\ket{2}\) & \(0.95532\) & \(1.53005\) \\
        & 7 & \(\ket{0}\leftrightarrow\ket{1}\) & \(0.60641\) & \(4.57966\) \\
        & 8 & \(\ket{2}\leftrightarrow\ket{3}\) & \(\pi/2\) & \(\pi/2\) \\
        & 9 & \(\ket{2}\leftrightarrow\ket{3}\) & \(\pi/2\) & \(\pi/2\) \\
        & 10 & \(\ket{2}\leftrightarrow\ket{3}\) & \(\pi/3\) & \(1.981884\) \\
        & 11 & \(\ket{1}\leftrightarrow\ket{2}\) & \(0.85289\) & \(3.74954\) \\
        & 12 & \(\ket{0}\leftrightarrow\ket{1}\) & \(0.60641\) & \(3.69336\) \\
        & 13 & \(\ket{3}\leftrightarrow\ket{4}\) & \(\pi/2\) & \(\pi/2\) \\
        & 14 & \(\ket{3}\leftrightarrow\ket{4}\) & \(\pi/2\) & \(9\pi/10\) \\
        & 15 & \(\ket{3}\leftrightarrow\ket{4}\) & \(1.10714\) & \(9\pi/10\) \\
        & 16 & \(\ket{2}\leftrightarrow\ket{3}\) & \(\pi/3\) & \(9\pi/10\) \\
        & 17 & \(\ket{1}\leftrightarrow\ket{2}\) & \(0.95532\) & \(9\pi/10\) \\
        & 18 & \(\ket{0}\leftrightarrow\ket{1}\) & \(\pi/4\) & \(9\pi/10\) \\
        \midrule
        \multirow{6}{*}{\(\hat{T}_5\)} 
        & 1 & \(\ket{1}\leftrightarrow\ket{2}\) & \(\pi/2\) & \(\pi/2\) \\
        & 2 & \(\ket{1}\leftrightarrow\ket{2}\) & \(\pi/2\) & \(7\pi/10\) \\
        & 3 & \(\ket{2}\leftrightarrow\ket{3}\) & \(\pi/2\) & \(\pi/2\) \\
        & 4 & \(\ket{2}\leftrightarrow\ket{3}\) & \(\pi/2\) & \(3\pi/10\) \\
        & 5 & \(\ket{3}\leftrightarrow\ket{4}\) & \(\pi/2\) & \(\pi/2\) \\
        & 6 & \(\ket{3}\leftrightarrow\ket{4}\) & \(\pi/2\) & \(11\pi/10\) \\
        \bottomrule\bottomrule
    \end{tabular}
    \caption[Five dimensional unitary decompositions]{Unitary decomposition for various five dimensional unitary gates of interest.}
    \label{tab:decomp_five}
\end{table}
Here, we give a list of useful qudit gates decomposed into these pulses. We assume that all qudits are in the zig-zag configuration, so that \({\ket{l}}\) is connected to \({\ket{l\pm 1}}\) and we only perform pulses on these transitions. Different pulses are notated \({R[\theta, \rho]_{ij}}\), where \({\theta}\) is the angle from the z-axis(pulse angle), \({\rho}\) is the angle from the y-axis(phase) in the bloch sphere picture, and \({ij}\) denotes the transition \({\ket{i}\leftrightarrow\ket{j}}\).

The first set of useful gates are the generalized Pauli gates. The simplest way to generalize the Pauli gates to \(d\)-dimensions is by the following prescription:\({\hat{X}\ket{j} = \ket{j + 1 \text{ mod } d}}\), \({\hat{Z}\ket{j} = \omega^j\ket{j}}\), and \({\hat{Y}\ket{j} = i\hat{X}\hat{Z}}\), where \({\omega = e^{2\pi i/d}}\) \cite{Howard-Vala-2012}. The matrix forms for the \(\hat{X}_d\), \(\hat{Y}_d\) and \(\hat{Z}_d\) operators for \(d=3\) are given as follows:

\begin{equation}
    \hat{X}_3 = 
    \begin{pmatrix}
    0 & 0 & 1 \\
    1 & 0 & 0 \\
    0 & 1 & 0
    \end{pmatrix},
\end{equation}
\begin{equation}
    \hat{Y}_3 = 
    \begin{pmatrix}
    0 & 0 & ie^{-i\frac{2\pi}{3}} \\
    i & 0 & 0 \\
    0 & ie^{i\frac{2\pi}{3}} & 0
    \end{pmatrix},
\end{equation}
\begin{equation}
    \hat{Z}_3 = 
    \begin{pmatrix}
    1 & 0 & 0 \\
    0 & e^{i\frac{2\pi}{3}} & 0 \\
    0 & 0 & e^{-i\frac{2\pi}{3}}
    \end{pmatrix}.
\end{equation}

For \(d=5\), they are

\begin{equation}
    \hat{X}_5 = 
    \begin{pmatrix}
    0 & 0 & 0 & 0 & 1 \\
    1 & 0 & 0 & 0 & 0 \\
    0 & 1 & 0 & 0 & 0 \\
    0 & 0 & 1 & 0 & 0 \\
    0 & 0 & 0 & 1 & 0 \\
    \end{pmatrix},
\end{equation}
\begin{equation}
    \hat{Y}_5 = 
    \begin{pmatrix}
    0 & 0 & 0 & 0 & ie^{-i\frac{2\pi}{5}} \\
    i & 0 & 0 & 0 & 0 \\
    0 & ie^{i\frac{2\pi}{5}} & 0 & 0 & 0 \\
    0 & 0 & ie^{i\frac{4\pi}{5}} & 0 & 0 \\
    0 & 0 & 0 & ie^{-i\frac{4\pi}{5}} & 0 \\
    \end{pmatrix},
\end{equation}
\begin{equation}
    \hat{Z}_5 = 
    \begin{pmatrix}
    1 & 0 & 0 & 0 & 0 \\
    0 & e^{i\frac{2\pi}{5}} & 0 & 0 & 0 \\
    0 & 0 & e^{i\frac{4\pi}{5}} & 0 & 0 \\
    0 & 0 & 0 & e^{-i\frac{4\pi}{5}} & 0 \\
    0 & 0 & 0 & 0 & e^{-i\frac{2\pi}{5}}
    \end{pmatrix}.
\end{equation}

The next set of gates are the single-qudit quantum Fourier transform, defined as \({\hat{H}_d\ket{j} = \frac{1}{\sqrt{d}}\sum\limits_{l = 0}^{p}e^{2\pi ijl/d}\ket{l}}\). In matrix forms, they are

\begin{equation}
    \hat{H}_3 = \frac{1}{\sqrt{3}}
    \begin{pmatrix}
    1 & 1 & 1 \\
    1 & e^{i\frac{2\pi}{3}} & e^{-i\frac{2\pi}{3}} \\
    1 & e^{-i\frac{2\pi}{3}} & e^{i\frac{2\pi}{3}}
    \end{pmatrix}.
\end{equation}
\begin{equation}
    \hat{H}_5 = \frac{1}{\sqrt{5}}
    \begin{pmatrix}
    1 & 1 & 1 & 1 & 1 \\
    1 & e^{i\frac{2\pi}{5}} & e^{i\frac{4\pi}{5}} & e^{-i\frac{4\pi}{5}} & e^{-i\frac{2\pi}{5}} \\
    1 & e^{i\frac{4\pi}{5}} & e^{-i\frac{2\pi}{5}} & e^{i\frac{2\pi}{5}} & e^{-i\frac{4\pi}{5}} \\
    1 & e^{-i\frac{4\pi}{5}} & e^{i\frac{2\pi}{5}} & e^{-i\frac{2\pi}{5}} & e^{i\frac{4\pi}{5}} \\
    1 & e^{-i\frac{2\pi}{5}} & e^{-i\frac{4\pi}{5}} & e^{i\frac{4\pi}{5}} & e^{i\frac{2\pi}{5}}
    \end{pmatrix}.
\end{equation}

The final set of gates we present are called \({\pi/8}\)(pi-over-eight) gates; in qubit form, they look like \({\hat{T}_d = 
\begin{pmatrix}
e^{-i\frac{\pi}{8}} & 0 \\
0 & e^{i\frac{\pi}{8}}
\end{pmatrix}}\). These gates are useful in quantum information theory as supplements to the Pauli and Clifford gates \cite{Howard-Vala-2012}. We used the generalized 3- and 5-level \({\pi/8}\) gates presented in reference \cite{Howard-Vala-2012}. In matrix form, for \(d=3\), it is 

\begin{equation}
    \hat{T}_3 = 
    \begin{pmatrix}
    1 & 0 & 0 \\
    0 & e^{i\frac{2\pi}{9}} & 0 \\
    0 & 0 & e^{-i\frac{2\pi}{9}}
    \end{pmatrix}.
\end{equation}

For \(d=5\), it is

\begin{equation}
    \hat{T}_5 = 
    \begin{pmatrix}
    1 & 0 & 0 & 0 & 0 \\
    0 & e^{-i\frac{4\pi}{5}} & 0 & 0 & 0 \\
    0 & 0 & e^{-i\frac{2\pi}{5}} & 0 & 0 \\
    0 & 0 & 0 & e^{i\frac{4\pi}{5}} & 0 \\
    0 & 0 & 0 & 0 & e^{i\frac{2\pi}{5}}
    \end{pmatrix}.
\end{equation}

The Givens decomposition of all of these gates are listed out for 3- and 5-level qudits in tables \ref{tab:decomp_three} and \ref{tab:decomp_five}.

\subsection{Single Qudit Gate Error Simulations}
\label{app:Single-Qudit-Mag}
Magnetic field noise can be modeled by a perturbative Hamiltonian
\begin{align}
    \mathcal{H}_{noise} &= \sum_{m_e = -F_e}^{F_e}g_F\mu_Bm_e\Delta B\ketbra{F_e, m_e}{F_e, m_e} \text{ }
    \nonumber\\
    &- \sum_{m_g = -F_g}^{F_g}g_F\mu_Bm_g\Delta B\ketbra{F_g, m_g}{F_g, m_g},
\end{align}
where \({g_F}\) is the hyperfine g-factor, \({\mu_B}\) is the Bohr Magneton, and \({\Delta B(t)}\) is the random fluctuation of the magnetic field from the set magnetic field. The subscripts \( g \) and \( e \) denote the lower and higher energy state in the hyperfine splitting respectively. The resultant Hamiltonian is then
\begin{equation} \label{eq:noisy-gate}
    \mathcal{H} = \mathcal{H}_{ideal} + \mathcal{H}_{noise},
\end{equation}
where \( \mathcal{H}_{ideal} \) is the ideal Hamiltonian for a single qudit gate. The output state under this Hamiltonian is obtained by numerically solving Schr\"{o}dinger's equation.

To account for off-resonant coupling the Hamiltonian has to be modified to
\begin{equation}
    \mathcal{H} = \mathcal{H}_{ideal} + \mathcal{H}_{noise} + \mathcal{H}_{OR},
\end{equation}
where \({\mathcal{H}_{OR}}\) is the component of the Hamiltonian due to off-resonant coupling. It has the form
\begin{align}
    \mathcal{H}_{OR} &= -\frac{\Omega_l}{2} \left[ \ket{l} \bra{l+1} + \ket{l+1} \bra{l} \right] + \sum_k \sum_{k' \ne k} \frac{\Omega_{k,k'}}{2}
    \nonumber\\
    &\hspace{-12px}\times\exp{( i \left( \omega_k -\omega_{k'} \right) t - i \; \text{sgn} \left( \omega_k -\omega_{k'} \right) \omega_l t )} \ket{k}\bra{k'}
    \nonumber\\
    &= -\mathcal{H}_{ideal} + \sum_k \sum_{k' \ne k} \frac{\Omega_{k,k'}}{2} 
    \nonumber\\
    &\hspace{-12px}\times\exp{( i \left( \omega_k -\omega_{k'} \right) t - i \; \text{sgn} \left( \omega_k -\omega_{k'} \right) \omega_l t )} \ket{k}\bra{k'},
\end{align}
where \( \ket{l} \) and \( \ket{l+1} \) are the states where resonant transition is desired, \( \Omega_l \) is the Rabi frequency for the desired transition, \( \omega_l \) is the transition frequency between \( \ket{l} \) and \( \ket{l+1} \), \( \Omega_{k,k'} \) is the Rabi frequency for the transition between the states \( \ket{k} \) and \( \ket{k'} \), \( \hbar \omega_k \) is the energy for the \( \ket{k} \) state, \( \text{sgn} \left( x \right) \) is the sign function
\begin{equation}
    \text{sgn} \left( x \right) = \left\{ \begin{array}{cc} -1 & \text{if } x<0 \\ 0 & \text{if } x=0 \\ 1 & \text{if } x>0 \end{array} \right..
\end{equation}

For the simulation, we find that it is too computationally intensive to simulate both off-resonant and magnetic field noise error simultaneously with a time-varying noise. Thus, the deviation in magnetic field is set at a constant offset at the standard deviation of \( \SI{2.7}{pT} \) as an estimation for simulations with both errors taken into account. We found no discernible difference in the average fidelity obtained whether a magnetic field offset is present as the error is dominated by off-resonant coupling. For the simulations with only magnetic field noise present, we generate random magnetic field noise using a Ornstein–Uhlenbeck function with a mean of 0, inverse correlation time \({\gamma = \SI{0.5}{ms^{-1}}}\), and volatility \({\sigma = \sqrt{2\gamma \langle\Delta B^2\rangle}}\); we assume that the magnetic field noise is Gaussian and stationary \cite{Monz-2011}. For simulation of an entire gate, pulses are applied immediately after one another; we calculate the fidelity by comparing the final state from the evolution of Equation \ref{eq:noisy-gate} to the desired state from applying the gate.

\section{Qudit Entangling Gate Simulations}
\label{app:Qudit-Ent}
\subsection{General Approach}
For numerical simulations of the qudit MS gate, we have the capability to simulate the time evolution without making the Lamb-Dicke approximation (LDA). This gives us a more realistic fidelity that we expect to get when we carry out the experiment.

First, we choose some convenient motional and spin phases
\begin{align}
    \phi_s=-(-1)^l\left(\phi'_{s,l}+\frac{\pi}{2}\right)=0
    \nonumber\\
    \phi_m=\phi_{m,l}=0.
\end{align}
Then, the Hamiltonian is
\begin{align}
    \mathcal{H}=\sum_{n=1}^{N}\sum_{l=0}^{d-1}&\hbar\Omega\sqrt{s(s+1)-l'(l'+1)}\cos({\mu}t)
    \nonumber\\
    &\times\left[i(-1)^le^{-i(-1)^l{\Delta}k\hat{x}'_n}\ket{l+1}\bra{l}_n\right.
    \nonumber\\
    &\left.-i(-1)^le^{i(-1)^l{\Delta}k\hat{x}'_n}\ket{l}\bra{l+1}_n\right].
\end{align}

Due to computational limitations, we can only simulate 2 phonon modes (center-of-mass and tilt modes). Thus, the position operator for each ion is
\begin{align}
    {\Delta}k\hat{x}'_n&=\eta_{C}\left(e^{i{\omega_{C}}t}\hat{a}_{C}^{\dagger}+e^{-i{\omega_{C}}t}\hat{a}_{C}\right)
    \nonumber\\
    &-(-1)^n\eta_{T}\left(e^{i{\omega_{T}}t}\hat{a}_{T}^{\dagger}+e^{-i{\omega_{T}}t}\hat{a}_{T}\right),
\end{align}
where the subscripts \({C}\) and \({T}\) denote centre-of-mass mode and tilt mode respectively. The Hamiltonian is then
\begin{widetext}
\begin{align}
    \label{Eq:MS-gate for sim alpha}
    \mathcal{H} =\sum_{n=1}^{2}\sum_{l=0}^{d-1}&\hbar\Omega\sqrt{s(s+1)-l'(l'+1)}\cos({\mu}t)
    \nonumber\\
    &\left[i(-1)^le^{-i(-1)^l(\eta_{C}(e^{i{\omega_{C}}t}\hat{a}_{C}^{\dagger}+e^{-i{\omega_{C}}t}\hat{a}_{C})-(-1)^n\eta_{T}(e^{i{\omega_{T}}t}\hat{a}_{T}^{\dagger}+e^{-i{\omega_{T}}t}\hat{a}_{T}))}\ket{l+1}\bra{l}_n\right.
    \nonumber\\
    &\left.-i(-1)^le^{i(-1)^l(\eta_{C}(e^{i{\omega_{C}}t}\hat{a}_{C}^{\dagger}+e^{-i{\omega_{C}}t}\hat{a}_{C})-(-1)^n\eta_{T}(e^{i{\omega_{T}}t}\hat{a}_{T}^{\dagger}+e^{-i{\omega_{T}}t}\hat{a}_{T}))}\ket{l})\bra{l+1}_n\right].
\end{align}

Since the phonon operators for different modes commute,
\begin{align}
    \label{Eq:MS-gate for sim}
    \mathcal{H}=\sum_{n=1}^{2}\sum_{l=0}^{d-1}&\hbar\Omega\sqrt{s(s+1)-l'(l'+1)}\cos({\mu}t)
    \nonumber\\
    &\left[i(-1)^le^{-i(-1)^l\eta_{C}(e^{i{\omega_{C}}t}\hat{a}_{C}^{\dagger}+e^{-i{\omega_{C}}t}\hat{a}_{C})}e^{i(-1)^l(-1)^n\eta_{T}(e^{i{\omega_{T}}t}\hat{a}_{T}^{\dagger}+e^{-i{\omega_{T}}t}\hat{a}_{T})}\ket{l+1}\bra{l}_n\right.
    \nonumber\\
    &\left.-i(-1)^le^{i(-1)^l\eta_{C}(e^{i{\omega_{C}}t}\hat{a}_{C}^{\dagger}+e^{-i{\omega_{C}}t}\hat{a}_{C})}e^{-i(-1)^l(-1)^n\eta_{T}(e^{i{\omega_{T}}t}\widehat{a}_{T}^{\dagger}+e^{-i{\omega_{T}}t}\widehat{a}_{T})}\ket{l}\bra{l+1}_n\right].
\end{align}
\end{widetext}

In the presence of a magnetic field mismatch, the Hamiltonian is modified to
\begin{equation} \label{eq:Realistic H with mag shift}
    \mathcal{H} \rightarrow \mathcal{H} + \sum_{l=0}^{d-1} \Delta E_{l} \ketbra{l}{l}.
\end{equation}
With Equations \ref{Eq:MS-gate for sim} and \ref{eq:Realistic H with mag shift}, the time evolution operator can be solved numerically using Schr\"{o}dinger's equation.

We are only concerned about the output spin state and not the phonon states at the output. Thus, \({\ket{\psi_{ideal}}}\) is an ideal spin state and fidelity is computed after tracing out the phonon states of the output density operator. The fidelity at the end of the gate is
\begin{equation} \label{Eq:Numerical fidelity}
    \mathcal{F}=\langle{\psi_{ideal}}{\rvert}Tr_{phonon}(U(t){\rho}_0U^\dagger(t))\lvert{\psi_{ideal}}\rangle,
\end{equation}
where \({\rho_0}\) is the initial density operator before applying MS gate.

\subsection{Simulating Mixed Initial State}
We are taking into account error due to having an initial state that is not absolutely in the ground state, which is a realistic assumption. The initial motional state is assumed to be in a mixed state with the density operator
\begin{equation}
    \rho_0 = \sum_{m}\sum_{n}P_{C}(m)P_{T}(n)\lvert{\psi_0,m,n}\rangle\langle{ \psi_0,m,n}\rvert,
\end{equation}
where \({P}\) is the phonon Fock state population. The \({C}\) and \({T}\) subscripts again refer to the centre-of-mass and tilt modes respectively, and \({\ket{\psi_0}}\) is the initial qudit state. For a thermal state, we have \cite{Fox-2006}
\begin{equation}
    P(n)=\frac{\bar{n}^n}{(\bar{n}+1)^{n+1}},
\end{equation}
where \({\bar{n}}\) is the average phonon number.

The best strategy for a faster simulation is to evaluate the time evolution operator, then apply it to the initial density operator to get the output and compute the fidelity. However, this can be too memory-intensive, which is the case for us.

As an alternative approach, we compute the evolution of pure phonon Fock states, \(\lvert{\psi_0,m,n}\rangle\), then weigh each fidelity by the phonon populations, \({P_{C}(m)P_{T}(n)}\):
\begin{align}
    \mathcal{F}_{m,n} =& P_{C}(m)P_{T}(n)
    \nonumber\\
    &\times \sum_{m'}\sum_{n'} \langle{\psi_{ideal}}\rvert \langle{m',n'}\rvert U(t)\lvert{\psi_0,m,n}\rangle
    \nonumber\\
    &\times\langle{\psi_0,m,n}\rvert U^\dagger(t) \lvert{m',n'}\rangle \lvert{\psi_{ideal}}\rangle.
\end{align}
The total fidelity is then
\begin{align}
    \mathcal{F}_{total} =& \sum_{m}\sum_{n}P_{C}(m)P_{T}(n)
    \nonumber\\
    &\times \sum_{m'}\sum_{n'} \langle{\psi_{ideal}}\rvert \langle{m',n'}\rvert U(t)\lvert{\psi_0,m,n}\rangle
    \nonumber\\
    &\times\langle{\psi_0,m,n}\rvert U^\dagger(t) \lvert{m',n'}\rangle \lvert{\psi_{ideal}}\rangle,
\end{align}
which is equivalent to Equation \ref{Eq:Numerical fidelity}. Since we are numerically solving the problem, the summation over the Fock state population \({\sum_n P(n)}\) cannot be an infinite series. Thus, the number of allowed Fock states for the centre-of-mass, \({m_{max}}\) and tilt modes, \({n_{max}}\) have to be chosen such that they are large enough for accurate results. For our simulations, where \({\bar{n}_C=0.1}\) and \({\bar{n}_T=0}\), we have determined that \({m_{max}=20}\) and \({n_{max}=2}\) is accurate enough such that further increment of allowed Fock states do not increase the accuracy of the fidelity at the fourth decimal place.

To further speed up the process, phonon states where \({P_{C}(m)P_{T}(n)<10^{-5}}\) are ignored.

\subsection{Optimum Rabi Frequency}
The Rabi frequency-geometric phase relation in Equation 39 in the main text is derived with LDA and RWA. Without the RWA applied to Equation 30 (main text) to arrive at Equation 31 (main text), the Hamiltonian we arrive at is
\begin{equation}
    \mathcal{H}=2\hbar\eta\Omega \cos(\mu t) \left(\hat{a}^{\dagger}e^{i\omega_C t}+\hat{a}e^{-i\omega_C t}\right)\sum_{n=1}^{N}\hat{S}_{x,n},
\end{equation}
and \({\theta_0}\) is modified to
\begin{equation}
    \theta_0 = \frac{4{\omega_{C}}\Omega^2\eta_{C}^2\pi}{(\omega_{C}^2-{\mu}^2)|\omega_{C}-\mu|}.
\end{equation}
Without LDA, the geometric phase from the entangling gate for a certain phonon Fock state is \cite{Sorensen-Molmer-2000}
\begin{equation}
    \theta_n = \left(1-G(n,\eta)\right)\frac{4{\omega_{C}}\Omega^2\eta_{C}^2\pi}{(\omega_{C}^2-{\mu}^2)|\omega_{C}-\mu|},
\end{equation}
where
\begin{equation}
    G(n,\eta) = \left[\hat{a}g_1(\hat{n}-1,\eta),\hat{a}^{\dagger}g_1(\hat{n},\eta)\right]
\end{equation}
\begin{equation}
    g_1(\hat{n},\eta)= \frac{e^{-{\eta}^2/2}}{\hat{n}+1}L_{\hat{n}}^1({\eta}^2),
\end{equation}
where \({L_n^\alpha(x)}\) are the generalized Laguerre polynomials
\begin{equation}
    L_n^\alpha(x) = \sum_{m=0}^n(-1)^m\frac{(n+\alpha)!}{(n-m)!(m+\alpha)!}\frac{x^m}{m!}.
\end{equation}

Thus, to obtain the desired geometric phase with optimal fidelity for input qudits in phonon Fock state \({n}\), the laser amplitude should be tuned to the corresponding Rabi frequency of
\begin{align}
    \Omega_n &= \frac{1}{\sqrt{1-G(n,\eta)}}\sqrt{\frac{\theta(\omega_{C}^2-\mu^2)|\omega_{C}-\mu|}{4\omega_{C}\eta_{C}^2\pi}} 
    \nonumber\\
    &= \frac{1}{\sqrt{1-G(n,\eta)}}\Omega_{LDA},
\end{align}
where \({\Omega_{LDA}}\) is the optimal Rabi frequency with LDA.

For input states with a superposition of or mixed phonon states, the fidelity with errors only from the shifted geometric phase from the LDA case can be written as
\begin{equation}
    \mathcal{F} = \sum_{n=0}^{\infty}P_n \lvert \bra{\psi_0}e^{i(\theta_n-\theta_{ideal}) \left(\sum_{n=1}^{N}\hat{S}_{x,n}\right)^2} \ket{\psi_0} \rvert^2.
\end{equation}
We define
\begin{equation}
    f(\Delta \theta_n)=\lvert \bra{\psi_0}e^{i \Delta \theta_n \left(\sum_{n=1}^{N}\hat{S}_{x,n}\right)^2} \ket{\psi_0} \rvert^2,
\end{equation}
where \({ \Delta \theta_n = \theta_n-\theta_{ideal}}\). For small \({ \Delta \theta_n }\), \({ f(\Delta \theta_n) }\) can be approximated with Taylor series expansion
\begin{equation}
    f(\Delta \theta_n) = \sum_{l=0}^\infty \frac{d^l f(0)}{d\Delta \theta_n^l}\frac{\Delta \theta_n^l}{l!}.
\end{equation}
Since \({ f(0)=1 }\) is a maximum point, \({ \frac{d f(0)}{d\Delta \theta_n} = 0 }\). Keeping the largest non-zero term,
\begin{equation}
    f(\Delta \theta_n) \approx 1 + \frac{d^2 f(0)}{d\Delta \theta_n^2} \frac{\Delta \theta_n^2}{2}.
\end{equation}

To maximize the fidelity,
\begin{align}
    \frac{dF}{d\Omega} &= \sum_{n=0}^{\infty}P_n \frac{d f(\Delta \theta_n)}{d \Omega} 
    \nonumber\\
    &\approx \sum_{n=0}^{\infty}P_n \Delta \theta_n \frac{d^2 f}{d\Delta \theta_n^2}(0) \frac{d \Delta \theta_n}{d \Omega} = 0,
\end{align}
which implies
\begin{equation}
    \sum_{n=0}^\infty P_n ( (1-G(n,\eta))\Omega^2-\Omega_{LDA}^2 )(1-G(n,\eta))\Omega = 0.
\end{equation}
The solution where \({ \Omega = 0 }\) does not satisfy the condition \({ \Delta \theta_n \approx 0 }\). Thus, the optimum value of the Rabi frequency is approximately
\begin{equation} \label{Eq:Rabi_frequency_accurate}
    \Omega \approx \Omega_{LDA}\sqrt{1+\frac{\sum_{n=0}^{\infty}P_nG(n,\eta)(1-G(n,\eta))}{\sum_{n=0}^{\infty}P_n(1-G(n,\eta))^2}}.
\end{equation}
Since the objective is to obtain the error due to experimental imperfections and not inaccurate parameters, the Rabi frequency in Equation \ref{Eq:Rabi_frequency_accurate} is used for the simulations.

\subsection{Pin-pointing Error Sources}
To pin-point the contribution from each error source, modifications to the simulations are done accordingly.
\begin{enumerate}
    \item LDA
    
    To simulate the entangling gate with LDA, the matrix exponential in Equation \ref{Eq:MS-gate for sim alpha} is replaced by
\begin{widetext}
\begin{align}
    &e^{\pm i(-1)^l\left(\eta_{C}(e^{i{\omega_{C}}t}\hat{a}_{C}^{\dagger}+e^{-i{\omega_{C}}t}\hat{a}_{C})-(-1)^n\eta_{T}(e^{i{\omega_{T}}t}\hat{a}_{T}^{\dagger}+e^{-i{\omega_{T}}t}\hat{a}_{T})\right)} 
    \nonumber\\
    &\rightarrow 1\pm i(-1)^l\left(\eta_{C}\left(e^{i{\omega_{C}}t}\hat{a}_{C}^{\dagger}+e^{-i{\omega_{C}}t}\hat{a}_{C}\right)-(-1)^n\eta_{T}\left(e^{i{\omega_{T}}t}\hat{a}_{T}^{\dagger}+e^{-i{\omega_{T}}t}\hat{a}_{T}\right)\right).
\end{align}
\end{widetext}
    \item RWA:
    To minimize error from RWA, the frequency values \({\omega_T }\), \({\omega_C }\), and \({\mu}\) are increased from \({2\pi\times \SI{1.8}{MHz}}\), \({2\pi\times \SI{2}{MHz}}\), and \({2 \pi\times \SI{2.01}{MHz}}\) to \({2\pi \times \SI{49.8}{MHz}}\), \({2\pi \times \SI{50}{MHz}}\), and \({2\pi \times \SI{50.01}{MHz}}\) respectively for the simulations. The frequency values are then further increased to \({2\pi \times \SI{59.8}{MHz}}\), \({2\pi \times \SI{60}{MHz}}\), and \({2\pi \times \SI{60.01}{MHz}}\) to verify that the fidelity obtained the same up to the up to the fourth significant figure remain unchanged. These fidelity values are then taken to be the fidelity without RWA, up to the fourth significant figure.
    \item Spectator phonon mode:
    To eliminate tilt mode in the simulation, the Hamiltonian is simulated according to Equation \ref{Eq:MS-gate for sim} with \({\eta_T=0}\).
    \item Imperfect cooling:
    To obtain the fidelity with perfect ions cooling, the average phonon number in the mixed state, \({\bar{n}}\) is set to zero.
    \item Magnetic field noise:
    To obtain the fidelity without magnetic field noise, \({\Delta E_l}\) is set to zero.
\end{enumerate}

\subsection{Off-Resonant Error for 5-level Qudit Entangling Gate} \label{app:2qudit-Off-resonant}
\begin{figure*}[!htb]
    \centering
    \includegraphics[width=0.47\textwidth]{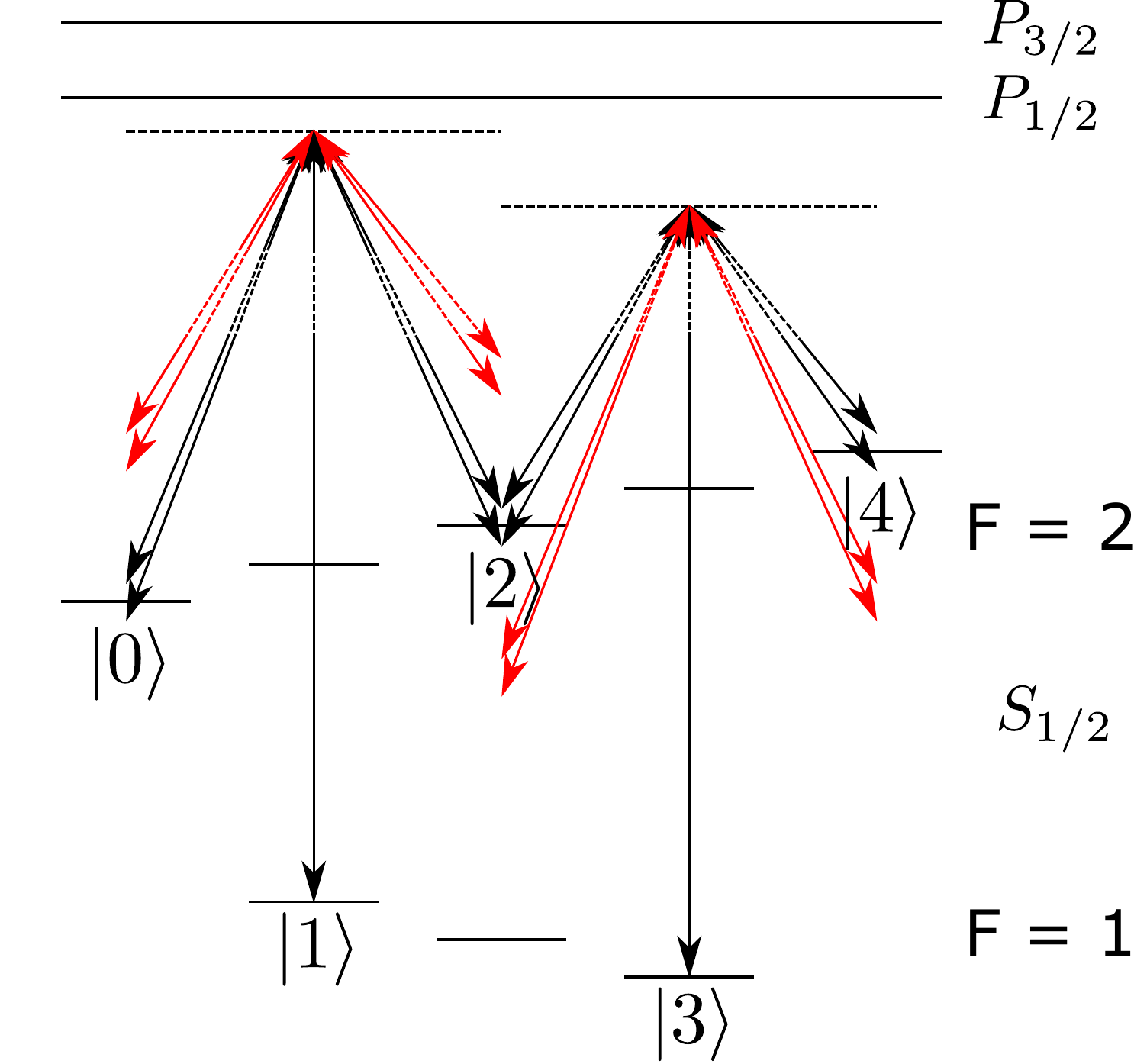}
    \caption{Schematic on laser frequencies applied to implement a 5-level qudit entangling gate for \({^{137}\mathrm{Ba}^+}\). Black arrows indicate the desired frequencies to be applied to the energy levels. Red arrows are the (unwanted) off-resonant frequencies.}
\label{fig:Entangling_Off_Resonant}
\end{figure*}

With the encoding scheme as shown in Figure \ref{fig:Entangling_Off_Resonant} for the 5-level qudits, we apply laser perturbations with frequencies as shown in the Figure to implement the entangling gate. However, there are some (unwanted) frequencies in each transition that are allowed by selection rules. For example, the required right and left-circularly polarized light acting on state \({\ket{3}}\) for the entangling gate acts on state \({\ket{1}}\) too, but at unwanted frequencies for \({\ket{1}}\) state. For the transitions \({\ket{0} \rightarrow \ket{1}}\) and \({\ket{1} \rightarrow \ket{2}}\), two additional blue-detuned off-resonant frequencies are introduced to each of the transition, whereas two additional red-detuned off-resonant frequencies are introduced to each of the transitions \({\ket{2} \rightarrow \ket{3}}\) and \({\ket{3} \rightarrow \ket{4}}\).

From Equation 25 (main text), the off resonant frequencies modify the ideal Hamiltonian in Equation 4 (main text) to
\begin{widetext}
\begin{align}
    \label{eq: H + H_OR}
    \mathcal{H}&=\mathcal{H}_{ideal}+\mathcal{H}_{OR}
    \nonumber\\
    \mathcal{H}_{OR}&=\frac{\hbar \Omega}{2}\sum_{n=1}^{N} \sum_{j=1}^{2}\left(\sum_{l=0}^{1} C_{l}\left[e^{-i(-1)^{l}(k\hat{x}'_{n}-\mu_j t-\frac{\pi}{2})}\ketbra{l+1}{l}_{n}+e^{i(-1)^{l}(k\hat{x}'-\mu_j t-\frac{\pi}{2})}\ketbra{l}{l+1}_{n}\right]\right.
    \nonumber\\
    & \quad\quad\quad\quad\quad\quad +\sum_{l=2}^{3} C_{l}\left[e^{-i(-1)^{l}(k\hat{x}'_{n}+\mu_j t-\frac{\pi}{2})}\ketbra{l+1}{l}_{n}\right.
    \left.+e^{i(-1)^{l}(k\hat{x}'+\mu_j t-\frac{\pi}{2})}\ketbra{l}{l+1}_{n}\right]\Bigg),
\end{align}
where \({\mu_1=4\Delta_{z}-\mu}\), \({\mu_2=4\Delta_{z}+\mu}\), \({C_{0}=C_{3}=6}\), \({C_{1}=C_{2}=\frac{\sqrt{6}}{3}}\), and the quantity \({\Delta_{z}}\) is the energy of the Zeeman splitting in frequency. We further simplify the problem by letting \({\Delta k \hat{x}' \rightarrow 0}\) in the off-resonant component of the Hamiltonian. The fidelity of \(0.0296\) in the main text is obtained from simulations with the Hamiltonian in Equation \ref{eq: H + H_OR}, which only has the error from off-resonant frequencies to verify that this error alone causes failure for the 5-level entangling gate. We employ the Magnus expansion again to evaluate the time-evolution operator generated by this Hamiltonian.

The first term in the Magnus expansion is
\begin{align}
    M_1(t) &= -\frac{i}{\hbar}\int_0^t \mathcal{H}_{ideal}(t_1) + \mathcal{H}_{OR}(t_1) \; dt_1
    \nonumber\\
    &= \left(\alpha(t)\hat{a}^{\dagger}-\alpha^{*}(t)\hat{a}\right)\sum_{n=1}^{N}\hat{S}_{x,n} 
    \nonumber\\
    &\quad + \frac{\hbar \Omega}{2}\sum_{n=1}^{N} \sum_{j=1}^{2} \left( \sum_{l=0}^{1} \frac{C_{l}}{\mu_j}\left[(e^{i(-1)^{l}\mu_j t}-1)\ketbra{l+1}{l}_{n}
    +(e^{-i(-1)^{l}\mu_j t}-1)\ketbra{l}{l+1}_{n}\right]\right.
    \nonumber\\
    &\quad\quad\quad\quad\quad\quad\:\:\:\left.+\sum_{l=2}^{3} \frac{C_{l}}{\mu_j}\left[(1-e^{-i(-1)^{l}\mu_j t})\ketbra{l+1}{l}_{n}
    +(1-e^{i(-1)^{l}\mu_j t})\ketbra{l}{l+1}_{n}\right]\right).
\end{align}

By changing the laser frequencies or Zeeman splitting, such that \({(1-e^{\pm i(-1)^{l}\mu_j t})=0}\) when \({t=K\frac{2\pi}{|\omega_{M}-\mu|}}\), it is still possible to minimize the contribution of the off-resonant in the first Magnus expansion. The second order Magnus expansion is
\begin{align}
    M_{2}(t)&=-\frac{1}{2\hbar^2} \int_0^t \; dt_1 \int_0^{t_1} \left[\mathcal{H}_{ideal}(t_1)+\mathcal{H}_{OR}(t_1),\mathcal{H}_{ideal}(t_2)+\mathcal{H}_{OR}(t_2)\right] \; dt_2
    \nonumber\\
    &=-\frac{1}{2\hbar^2} \int_0^t \; dt_1 \int_0^{t_1} \left[\mathcal{H}_{ideal}(t_1),\mathcal{H}_{ideal}(t_2)\right] + \left[\mathcal{H}_{ideal}(t_1),\mathcal{H}_{OR}(t_2)\right] 
    \nonumber\\
    &\quad\quad\quad\quad\quad\quad\quad\quad + \left[\mathcal{H}_{OR}(t_1),\mathcal{H}_{ideal}(t_2)\right] + \left[\mathcal{H}_{OR}(t_1),\mathcal{H}_{OR}(t_2)\right] \; dt_2.
\end{align}
The first term in the integral is the desired term, which is found in Equation 36 (main text). The rest of the terms can be approximated to be
\begin{align}
    &-\frac{1}{2\hbar^2} \int_0^t \; dt_1 \int_0^{t_1} \left[\mathcal{H}_{ideal}(t_1),\mathcal{H}_{OR}(t_2)\right] + \left[\mathcal{H}_{OR}(t_1),\mathcal{H}_{ideal}(t_2)\right] + \left[\mathcal{H}_{OR}(t_1),\mathcal{H}_{OR}(t_2)\right] \; dt_2
    \nonumber\\
    &\approx -\frac{i \Omega^2}{4}t \sum_{n=1}^N \sum_{j=1}^2 \left( \sum_{l=0}^1 \frac{C_l^2}{\mu_j}\left[-(-1)^l\ketbra{l}{l}_n+(-1)^l\ketbra{l+1}{l+1}_n\right]\right.
    \nonumber\\
    &\left.+ \sum_{l=2}^3 \frac{C_l^2}{\mu_j}\left[(-1)^l\ketbra{l}{l}_n-(-1)^l\ketbra{l+1}{l+1}_n\right]
    - \frac{C_0 C_1}{\mu_j} \left(\ketbra{0}{2}_n+\ketbra{2}{0}_n\right) + \frac{C_2 C_3}{\mu_j} \left(\ketbra{2}{4}_n+\ketbra{4}{2}_n\right) \right),
\end{align}
\end{widetext}
which consists of error terms due to Stark shifts and internal transitions of each qudit between \({\ket{0}}\) and \({\ket{2}}\) states and between \({\ket{2}}\) and \({\ket{4}}\) states. This term is comparable in magnitude to the desired term in the second Magnus expansion in Equation 36 (main text) and thus introduces a significant error.

\section{Photon Scattering Probability}
\label{app:Photon_Scatt}
From Kramers-Heisenberg formula, the scattering rate to some final state $\lvert f \rangle$ is \cite{Loudon-1983}
\begin{align}
    R_f = &\frac{(\omega_R-\omega_f)^3}{3 \pi \epsilon_0 \hbar c^3}\frac{\xi^2}{4 \hbar^2} \bigg\lvert 
    \nonumber\\
    &\times\sum_p \sum_q \frac{\langle f \lvert \hat{\sigma}_q \cdot \hat{d} \rvert p \rangle \langle p \lvert \hat{\sigma} \cdot \hat{d} \rvert g \rangle}{\omega_R - \omega_p} \bigg\rvert^2,
\end{align}
where \(\omega_R\) is the laser frequency, \(\hbar \omega_f\) is the energy of the final state, with the energy of the initial state state as zero, \( \hbar \omega_p \) is the energy of some intermediate state \( \lvert p \rangle \) that is coupled to during scattering, \( \epsilon_0 \) is the vacuum dielectric constant, \( c \) is the speed of light, \( \xi \) is the electric field amplitude, \( \hat{\sigma}_q \) is the polarization of the scattered photon, the subscript \( q \) denotes one of the spherical polarizations (sigma- and pi-polarization), \( \hat{\sigma} \) is the polarization of the incident laser beam, \( \hat{d} \) is the electric dipole operator, and \( \lvert g \rangle \) is the initial or ground state.

In our case, the initial/ground state is in the \( 6S \) state, \( \lvert g \rangle = \lvert 6S, J=1/2, F, m_F \rangle \), where \( J \) is the angular momentum number of the orbital angular momentum and electron spin, \( F \) is the angular momentum number of the hyperfine state and \( m_F \) is its \(z\)-projection. The intermediate states are in the \( 6P \) level, \( \lvert p \rangle = \lvert 6S, J, F, m_F \rangle \). The final state can be either be in the \( 6S \) or \( 5D \) state, \( \lvert f \rangle = \lvert 6S, J=1/2, F, m_F \rangle \) or \( \lvert f \rangle = \lvert 5D, J, F, m_F \rangle \).

To simplify the scattering rate equation, the transition matrix elements are reduced to the following forms with Wigner-Eckart theorem
\begin{equation}
    \langle p \lvert \hat{\sigma}_{Q} \cdot \hat{d} \rvert g \rangle = C_{gp,Q} \langle 6P, J \lvert \lvert \hat{d} \rvert \rvert 6S, J = 1/2 \rangle
\end{equation}
\begin{equation}
    \sum_q \langle f \lvert \hat{\sigma}_{q} \cdot \hat{d} \rvert p \rangle = C_{fp} \langle 6P, J \lvert \lvert \hat{d} \rvert \rvert 6S \text{ or } 6D, J\rangle,
\end{equation}
where the subscript \(Q\) denote the polarization of the incident laser beam in the spherical basis, \( C_{gp,Q} \) and \( C_{fp} \) are the coefficients that result from reduction of the transition matrix elements. Assuming that the laser polarization is purely in one direction in the spherical basis, the scattering rate can be written as
\begin{widetext}
\begin{align}
    \label{Eq: General scatter rate}
    R_{f,Q} &= \frac{(\omega-\omega_f)^3}{3 \pi \epsilon_0 \hbar c^3}\frac{\xi_{Q}^2}{4 \hbar^2} \Bigg| \frac{\langle 6P, J = 1/2 \lvert \lvert \hat{d} \rvert \rvert 6S \text{ or } 5D, J\rangle \langle 6P, J = 1/2 \lvert \lvert \hat{d} \rvert \rvert 6S, J = 1/2 \rangle}{\Delta_{1/2}} \sum_{p \in J = 1/2} \left( C_{fp}C_{gp,Q} \right) 
    \nonumber\\
    &+ \frac{\langle 6P, J = 3/2 \lvert \lvert \hat{d} \rvert \rvert 6S \text{ or } 5D, J\rangle \langle 6P, J = 3/2 \lvert \lvert \hat{d} \rvert \rvert 6S, J = 1/2 \rangle}{\Delta_{3/2}} \sum_{p \in J = 3/2} \left( C_{fp}C_{gp,Q} \right) \Bigg|^2.
\end{align}

The spontaneous decay rate from some state \( \lvert i \rangle \) to state \( \lvert j \rangle \) is given by \cite{Loudon-1983}
\begin{equation} \label{Eq: Decay rate expression}
    \gamma_{i \rightarrow j} = \frac{\lvert \langle i \lvert \hat{d} \rvert j \rangle \rvert^2 \omega_{i \rightarrow j}^3}{3 \pi \epsilon_0 \hbar c^3},
\end{equation}
where \( \omega_{i \rightarrow j} \) is the transition frequency between the two states. Using Equations \ref{Eq: General scatter rate} and \ref{Eq: Decay rate expression}, the total scattering rate for \( ^{137}\mathrm{Ba}^+ \) can be derived to be
\begin{align}
    \label{Eq: General total scatter rate}
    R_{total,Q} &= \frac{\lvert \langle 6P, J = 1/2 \lvert \lvert \hat{d} \rvert \rvert 6S, J = 1/2\rangle \rvert^2 \xi_{Q}^2}{4 \hbar^2} \times 
    \nonumber\\
    &\left[ \gamma_{P_{1/2} \rightarrow S_{1/2}} \frac{\omega_R^3}{\omega_{P_{1/2} \rightarrow S_{1/2}}^3} \left( \frac{\alpha_{P_{1/2},Q}}{\Delta_{1/2}^2} + \frac{\beta_{P_{1/2},Q}}{\Delta_{3/2}^2} \right) + \gamma_{P_{1/2} \rightarrow D_{3/2}} \frac{(\omega_R-\omega_{D_{3/2}})^3}{\omega_{P_{1/2} \rightarrow D_{3/2}}^3} \left( \frac{\alpha_{D_{3/2},Q}}{\Delta_{1/2}^2} \right) \right.
    \nonumber\\
    &\left. + \gamma_{P_{3/2} \rightarrow D_{3/2}} \frac{(\omega_R-\omega_{D_{3/2}})^3}{\omega_{P_{3/2} \rightarrow D_{3/2}}^3} \left( \frac{\beta_{D_{3/2},Q}}{\Delta_{3/2}^2} \right) + \gamma_{P_{3/2} \rightarrow D_{5/2}} \frac{(\omega-\omega_{D_{5/2}})^3}{\omega_{P_{3/2} \rightarrow D_{5/2}}^3} \left( \frac{\beta_{D_{5/2},Q}}{\Delta_{3/2}^2} \right) \right],
\end{align}
where \( \alpha_{i,Q} = \sum_{f \in i} \left( \sum_{p \in J = 1/2} \left( C_{fp}C_{gp,Q} \right) \right)^2 \) and \( \beta_{i,Q} = \sum_{f \in i} \left( \sum_{p \in J = 3/2} \left( C_{fp}C_{gp,Q} \right) \right)^2\), with \( i \in \{S_{1/2}, D_{3/2}, D_{5/2}\} \).

The Rayleigh scattering rate is derived to be
\begin{align}
    \label{Eq: General Rayleigh scatter rate}
    R_{g,Q} &= \gamma_{P_{1/2} \rightarrow S_{1/2}}\frac{\lvert \langle 6P, J = 1/2 \lvert \lvert \hat{d} \rvert \rvert 6S, J = 1/2\rangle \rvert^2 \xi_{Q}^2}{4 \hbar^2} \frac{(\omega_R-\omega_f)^3}{\omega_{P_{1/2} \rightarrow S_{1/2}}^3}\times
        \nonumber\\
        &\left( \frac{\sum_{p \in J = 1/2} \left( C_{gp}C_{gp,Q} \right)}{\Delta_{1/2}} + \frac{\sum_{p \in J = 3/2} \left( C_{gp}C_{gp,Q} \right)}{\Delta_{3/2}} \right)^2.
\end{align}
\end{widetext}

The Rabi frequency for each of the Raman transitions in Figure 5(a) (main text) is derived to be \cite{Wineland-et-al-2003}
\begin{align}
    \label{Eq:Rabi frequencies Raman 3-level}
    \Omega_0 &= \frac{1}{2\sqrt{12} \hbar^2} \left( b_0 r_+ + b_- r_0 \right) \left( -\frac{1}{\Delta_{1/2}} + \frac{1}{\Delta_{3/2}} \right) 
        \nonumber\\
        &\times \lvert \langle 6P, J = 1/2 \lvert \lvert \hat{d} \rvert \rvert 6S, J = 1/2 \rangle \rvert^2 \xi_r \xi_b
        \nonumber\\
        \Omega_1 &= \frac{1}{2\sqrt{12} \hbar^2} \left( b_0 r_- + b_+ r_0 \right) \left( \frac{1}{\Delta_{1/2}} - \frac{1}{\Delta_{3/2}} \right) 
        \nonumber\\
        &\times \lvert \langle 6P, J = 1/2 \lvert \lvert \hat{d} \rvert \rvert 6S, J = 1/2 \rangle \rvert^2 \xi_r \xi_b,
\end{align}
where \(\Omega_l\) is the Rabi frequency for the transition from \(\ket{l}\) to \(\ket{l+1}\), \(r_i\) and \(b_i\) are components of the red and blue electric fields of the Raman beams polarized in the \(i\) direction respectively, i.e.
\begin{align}
    \vec{\xi}_{r}&=\xi_{r} \left(r_+ \hat{\epsilon}_+ + r_0 \hat{\epsilon}_0 + r_- \hat{\epsilon}_-\right)
    \nonumber\\
    \vec{\xi}_{b}&=\xi_{b} \left(b_+ \hat{\epsilon}_+ + b_0 \hat{\epsilon}_0 + b_- \hat{\epsilon}_-\right).
\end{align}
Similarly, the Rabi frequency for each of the Raman transitions in Figure 5(b) can be derived to be
\begin{widetext}
\begin{align}
    \label{Eq:Rabi frequencies Raman 5-level}
    \Omega_0 &= \frac{1}{2\hbar^2}\frac{1}{\sqrt{6}}(r_0b_-+r_+b_0)\left( -\frac{1}{\Delta_{1/2}} + \frac{1}{\Delta_{3/2}} \right) \left|\bra{ 6P, J=1/2}|d|\ket{6S,J=1/2}\right|^2 \xi_r \xi_b
    \nonumber\\
    \Omega_1 &= \frac{1}{2\hbar^2}\frac{1}{6}(b_0r_-+b_+r_0)\left( \frac{1}{\Delta_{1/2}} - \frac{1}{\Delta_{3/2}} \right) \left|\bra{ 6P, J=1/2}|d|\ket{6S,J=1/2}\right|^2 \xi_r \xi_b
    \nonumber\\
    \Omega_2 &= \frac{1}{2\hbar^2}\frac{1}{6}(r_0b_-+r_+b_0)\left( -\frac{1}{\Delta_{1/2}} + \frac{1}{\Delta_{3/2}} \right) \left|\bra{ 6P, J=1/2}|d|\ket{6S,J=1/2}\right|^2 \xi_r \xi_b
    \nonumber\\
    \Omega_3 &= \frac{1}{2\hbar^2}\frac{1}{\sqrt{6}}(b_0r_-+b_+r_0)\left( \frac{1}{\Delta_{1/2}} - \frac{1}{\Delta_{3/2}} \right) \left|\bra{ 6P, J=1/2}|d|\ket{6S,J=1/2}\right|^2 \xi_r \xi_b,
\end{align}
\end{widetext}

For single qudit gates, two laser frequencies are needed for each 2-level transition. For the zig-zag configuration as shown in Figure 5, one laser has to be pi-polarized and the other is sigma-polarized. Assuming that the electric field amplitudes for the two laser frequencies are equal, the total and Rayleigh scattering rates that give the largest Raman scattering rate for 3-level qudits is the state \( \ket{0} \) with one sigma-plus and one pi-polarized laser frequencies. Using Equations \ref{Eq: General total scatter rate}, \ref{Eq: General Rayleigh scatter rate} and \ref{Eq:Rabi frequencies Raman 3-level}, the expressions for the total and Rayleigh scattering rates given in Equations 13 and 14 in the main article can be derived. For 5-level qudits, with the same assumption that the electric field amplitudes for both frequencies are equal, the state with the largest Raman scattering rate is state \( \ket{1} \) with a sigma-minus and a pi-polarized laser frequencies. Using Equations \ref{Eq: General total scatter rate}, \ref{Eq: General Rayleigh scatter rate} and \ref{Eq:Rabi frequencies Raman 5-level}, the derived expressions for the total and Rayleigh scattering rates are given in Equations 15 and 16 in the main article.

For the qudit MS gate, \( 2d-1 \) laser frequencies are required for the zig-zag encoding scheme that we used. For the case of 3-level qudit, we assume that \(\xi_r=\xi_b\) for all transitions. We also assume pure polarization of Raman beams, i.e. \( \lvert r_+ \rvert = 1\) for Raman beams 1 and 2, \( \lvert r_- \rvert = 1\) for Raman beams 3 and 4, \(\lvert b_0 \rvert =1\) for Raman beam 0 as indexed in Figure 5(a) (main text). With Equations \ref{Eq: General total scatter rate}, \ref{Eq: General Rayleigh scatter rate} \ref{Eq:Rabi frequencies Raman 3-level}, we arrive at Equations 46 and 47 in the main article. 
For the case of 5-level qudit, we assume that \(\xi_r=\xi_b\) for the transitions \( \ket{1} \leftrightarrow \ket{2} \) and \( \ket{2} \leftrightarrow \ket{3} \). The electric amplitudes for the transitions \( \ket{0} \leftrightarrow \ket{1} \) and \( \ket{3} \leftrightarrow \ket{4} \) are fully constrained with Equation 32 (main text) and the aforementioned assumption. We also assume pure Raman beam polarizations. Equations \ref{Eq: General total scatter rate}, \ref{Eq: General Rayleigh scatter rate}, \ref{Eq:Rabi frequencies Raman 5-level}, gives Equations 48 and 49.

\section{Magnetic Field Noise Threshold Estimation}
\label{app:Magnetic_Field_Threshold}
\begin{figure}
    \centering
    \includegraphics[width=\linewidth]{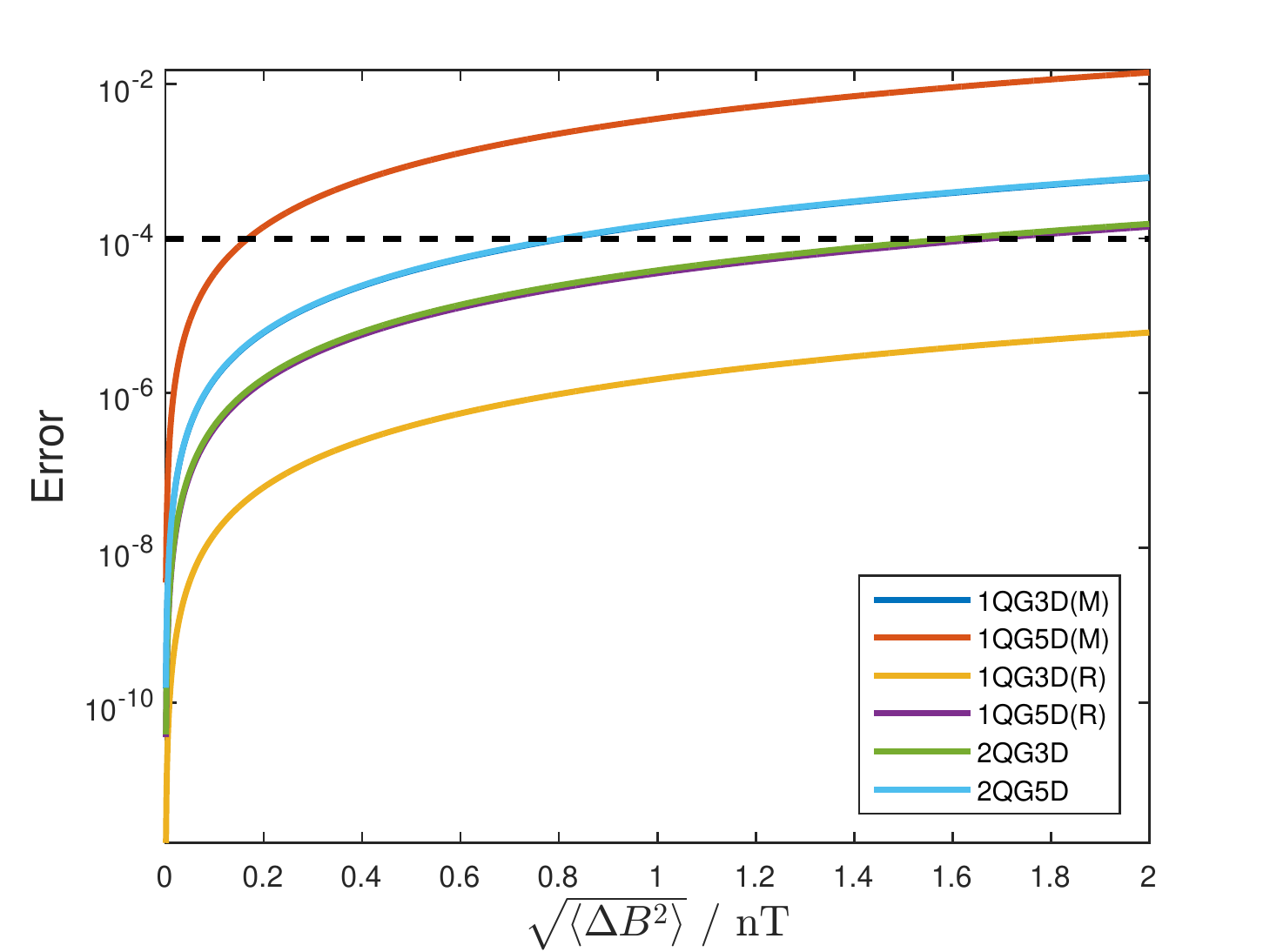}
    \caption{Estimations of error values from magnetic field noise for single and two qudit gates as a function of magnetic field standard deviation. Legend: 1QG3D(M) - abbreviation for 1-qudit gate 3-dimensional (microwave transition), 1QG5D(R) - 1-qudit gate 5-dimensional (Raman transition). The rest of the abbreviations of the legends can be extrapolated from the information presented. Black dashed line marks the error value of \(10^{-4}\).}
    \label{fig:magnetic field noise threshold estimations}
\end{figure}
In Sections \ref{Sec:Single-Qudit} and \ref{Sec:Two-Qudit}, the extent of relaxation of the magnetic field noise requirement such that this error remains below \(10^{-4}\) for qudit gates were presented. The thresholds are estimated by extending the formula for state decoherence in a static Hamiltonian as shown in Equation \ref{Eq:Decoherence-Fidelity} to ions encoding a qudit evolving under qudit gates. To be conservative, the fidelities are computed for a state in equal superposition of encoded states, where their difference in angular momenta, \(|m_l-m_{l'}|\), is the largest. Equation \ref{Eq:Decoherence-Fidelity} then simplifies to
\begin{equation}
    \mathcal{F}(t) = \frac{1}{2} + \frac{1}{2}e^{-\frac{\mu_B^2}{2 \hbar^2} g_F^2 max\left( |m_l-m_{l'}|^2 \right) \langle \Delta B^2 \rangle t^2}.
\end{equation}
For \( d=3 \), \(max\left( |m_l-m_{l'}|^2 \right)=2\). For \( d=5 \), \(max\left( |m_l-m_{l'}|^2 \right)=4\). Time \( t \) is set to the gate times for the respective gates. The error for single qudit gates is obtained by \(\varepsilon = 1 - \mathcal{F}(t)\) and the error for two qudit gates, accounting for two ions, is \(\varepsilon = 1 - \mathcal{F}^2(t)\). The results of the calculations are presented in Figure \ref{fig:magnetic field noise threshold estimations}.

\bibliographystyle{apsrmp4-2.bst}
\bibliography{bib}
\bibliographystyle{unsrt}
\end{document}